\documentclass{jfm}
\nonstopmode%
\usepackage{graphicx}
\usepackage[utf8]{inputenc}

\usepackage{upgreek}
\newlength{\figwidth}
\shorttitle{Laminar horseshoe vortex around emerging obstacles.}
\shortauthor{G. Launay, E. Mignot, N. Riviere, R. Perkins}

\title{An experimental investigation of the laminar horseshoe vortex around an emerging obstacle.}

\author{Gaby Launay\aff{1}
  E. Mignot\aff{1}
  \corresp{\email{emmanuel.mignot@insa-lyon.fr}}
  N. Riviere\aff{1}
  R. Perkins\aff{2}}
\affiliation{%
  \aff{1}LMFA, CNRS-Universite de Lyon, INSA de Lyon, Bat. Joseph Jacquard, 20 avenue A. Einstein, 69621 Villeurbanne Cedex, France
  \aff{2}LMFA, CNRS-Universite de Lyon, ECL de Lyon, 36, avenue Guy de Collongue, 69134 Ecully Cedes, France
}

\begin{document}

\maketitle

\begin{abstract}
  An emerging long obstacle placed in a boundary layer developing under a free-surface generates a complex horseshoe vortex (HSV) system, which is composed of a set of vortices exhibiting a rich variety of dynamics.
  The present experimental study examines such flow structure and characterizes precisely, using PIV measurements, the evolution of the HSV geometrical and dynamical properties over a wide range of dimensionless parameters (Reynolds number $Re_h \in [750, 8300]$, boundary layer development ratio $h/\delta \in [1.25, 4.25]$ and obstacle aspect ratio $W/h \in [0.67, 2.33]$).

  The dynamical study of the HSV is based on the categorization of the HSV vortices motion into an enhanced specific bi-dimensional typology, separating a coherent (due to vortex-vortex interactions) and an irregular evolution (due to appearance of small-scale instabilities).
  This precise categorization is made possible thanks to the use of vortex tracking methods applied on PIV measurements,
  A semi-empirical model for the HSV vortices motion is then proposed to highlight some important mechanisms of the HSV dynamics, as (i) the influence of the surrounding vortices on a vortex motion and (ii) the presence of a phase shift between the motion of all vortices.
  The study of the HSV geometrical properties (vortex position and characteristic lengths and frequencies) evolution with the flow parameters shows that strong dependencies exist between the streamwise extension of the HSV and the obstacle width, and between the HSV vortex number and its elongation.
  Comparison of these data with prior studies for immersed obstacles reveals that emerging obstacles lead to greater adverse pressure gradients and down-flows in front of the obstacle.
\end{abstract}


\section{Introduction}
\label{sec:introduction}

\subsection{Context}
\label{sec:context}
\begin{figure}
  \centering
  \includegraphics[width=.75\figwidth]{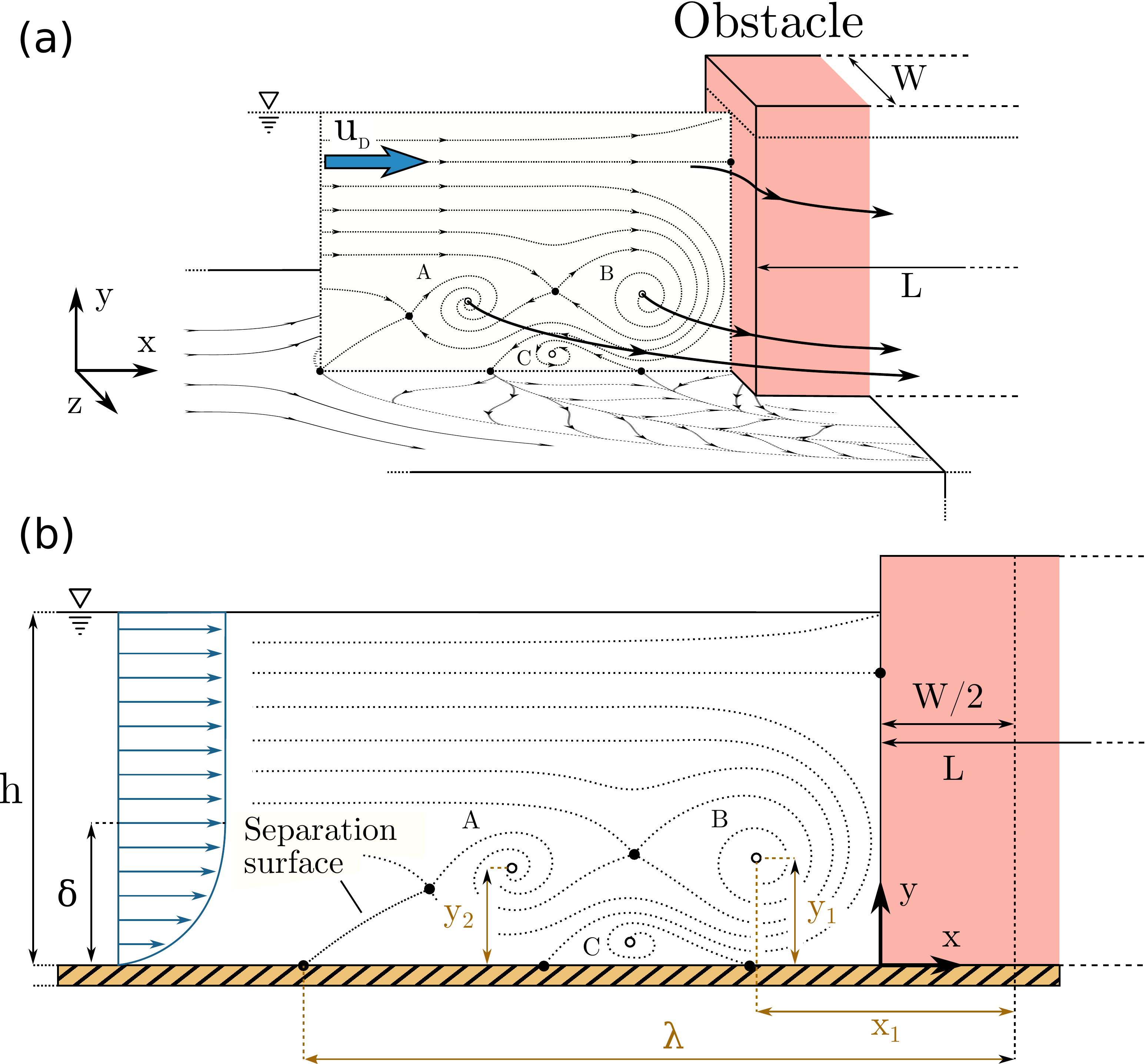}
  \caption{%
    Schematic representation of a laminar horse-shoe vortex (HSV), created by the interaction of a free-surface flow and an emerging obstacle of width $W$:
    (a) 3D illustration of the HSV system showing 3 vortices (A, B, C).
    (b) Side mid-plane view of the HSV system showing the control parameters (boundary layer thickness $\delta$, water level $h$, bulk velocity $u_D$ and obstacle width $W$ and length $L$) along with the definition of the HSV main geometrical characteristics (streamwise elongation $\lambda$ and vortex center positions $x_i$, $y_i$).
  }
  \label{fig:INTRO-Hsv-diagram}
\end{figure}

An obstacle placed in a developing boundary layer over a flat plate creates an adverse pressure gradient which, if sufficiently strong, makes the boundary layer detach.
The boundary layer separation creates a shear-layer, separating the main (upper) flow and the back (bottom) flow, which can contains a succession of vortices \citep{greco_flow_1990}.
Those vortices do not appear in the upstream-most part of the shear layer, called the separation surface in figure~\ref{fig:INTRO-Hsv-diagram} \citep{younis_topological_2014}, but can exhibit complex dynamics (oscillating motion, merging by pairs, diffusion, turbulent behavior) in the downstream part \citep{lin_characteristics_2008}.
The resulting set of vortices wraps around the obstacle with a particular shape, explaining the name given to the whole structure: the horseshoe vortex (HSV, as illustrated on figure \ref{fig:INTRO-Hsv-diagram}a).
Depending on the obstacle shape and on the flow velocity, recirculation zones can also appear at the sides of the obstacle and behind it \citep{larousse_flow_1993}.

The HSV has been extensively studied since $1962$ \citep{schwind_three_1962} for its numerous applications:
(i) The HSV influences the amount of turbulence released in the downstream boundary layer, impacting the aerodynamic properties.
(ii) The shear stress at the bottom wall and at the obstacle are affected by the HSV, modifying both the thermal exchanges \citep{sabatino_boundary_2008} and the scouring process in hydraulic with mobile bed \citep{euler_controls_2012}.
(iii) HSV appear in transverse jets interacting with boundary layers \citep{kelso_horseshoe_1995} in flow control issues.
(iv) Finally, the force applied by the flow on the obstacle, of interest in turbo-machinery \citep{eckerle_horseshoe_1987} and hydraulics, is also affected by the HSV.
Flow configurations vary in those studies, with different obstacle shapes (cylinders, prisms, foils) and emerging, immersed or traversing obstacles.
The present work focuses on emerging long rectangular prisms (see figure~\ref{fig:INTRO-Hsv-diagram}).

\subsection{State-of-the-art}
\label{sec:state-art}
In the case of a laminar boundary layer interacting with an (immersed or emerging) obstacle, different HSV dynamics typologies have been reported, based either on the number of vortices in the HSV and/or on their dynamics \citep{schwind_three_1962, baker_laminar_1978, greco_flow_1990, lin_characteristics_2002}.
The most complete and generally accepted typology is the one from~\citet{greco_flow_1990}, obtained through flow visualizations, who defined five HSV regimes based on the dynamics of the vortices, namely:
(i)~Steady vortex system (stationary HSV),
(ii)~Oscillating vortex system (periodical HSV, with streamwise vortex position oscillation),
(iii)~Amalgamating vortex system (vortex creation in the upstream part of the HSV and periodical vortex merging by pairs in its downstream part),
(iv)~Breakaway vortex system (periodical vortex shedding from the HSV, disappearing by diffusion near the obstacle foot), and
(iv)~Transitional vortex system (aperiodical vortex dynamics).
This typology was later confirmed, partially or completely, by~\citet{khan_dynamics_1995, lin_simultaneous_2003, khan_topological_2005, lin_characteristics_2008} for immersed obstacles and by~\citet{seal_dynamics_1997, escauriaza_reynolds_2011} for emerging obstacles, using either flow visualization, PIV, numerical simulations and/or pointwise velocity measurements. However, studies devoted to understand how those regimes evolve with the flow and obstacle parameters are rare in this context: for an immersed obstacle configuration, ~\citet{lin_characteristics_2008} recently showed that the HSV regimes evolution mainly depends on the Reynolds number based on the obstacle width $\Rey_W$ and on the ratio of the boundary layer thickness over the obstacle width $\delta/W$.

Turbulent HSV are characterized by temporally non-coherent vortices and therefore do not allow a typology definition.
The HSV is rather characterized by a bi-modal phenomenon: the time alternation between the so-called ``zero-flow'' and ``back-flow'', first described by~\citet{devenport_time-depeiident_1990} and later confirmed by~\citet{agui_experimental_1992, larousse_flow_1993, doligalski_vortex_1994, escauriaza_reynolds_2011}.
\citet{paik_bimodal_2007} showed, using numerical simulations, that this bi-modal phenomenon was actually three-dimensional and linked to the G\"ortler instability developing under the downstream-most vortex.
As the transition to turbulence of the HSV has not been extensively studied, it is not known if turbulent HSV can arise from laminar boundary layers.

Regarding \emph{immsersed} obstacles, \citet{simpson_junction_2001} summarized the state-of-the-art concerning flows around blunt or streamlined obstacles.
For laminar HSV, he compared the HSV dynamics regimes observed by different authors.
For turbulent HSV, he summarized the different descriptions of the bimodality phenomenon.
He also discussed the effect of the HSV on the scouring process and the ``bluntness factor'' that allows to take the obstacle geometry into account in the dimensional analysis.
\citet{ballio_survey_1998} collected existing data on the evolution of the main HSV geometrical characteristics (such as the separation distance $\lambda$ and the vortices position, see figure \ref{fig:INTRO-Hsv-diagram}) for both laminar and turbulent HSV, using, \textit{inter alia}, the works of~\citet{baker_laminar_1978, baker_vortex_1979, baker_turbulent_1980, baker_position_1985, baker_oscillation_1991}.
They indicated that the obstacle width $W$ is the main parameter for the HSV evolution, the obstacle height $\xi$ being significant only with low ratios of $\xi/W$.
They also concluded that the HSV increases significantly the bottom wall shear stress, making the HSV an important structure for the scouring process and thermal transfers.

The literature dedicated to \emph{emerging} obstacles is far less exhaustive than for \emph{immersed} obstacles.
The specificity of the \emph{emerging} obstacle configuration is that the flow cannot pass over the obstacle and is forced to skirt the obstacle by its sides.
When studying the effect of the obstacle submergence (with varying obstacle heights from immersed to emerging) on turbulent HSV,~\citet{sadeque_flow_2008} indicated that the separation distance $\lambda$ and the shear stress below the HSV are more important in the case of emerging obstacles.
In fact, most studies with emerging obstacles are dedicated to turbulent HSV~\citep{dargahi_turbulent_1989, graf_experiments_1998, johnson_measurements_2003, roulund_numerical_2005, ozturk_flow_2008, sadeque_flow_2008}, and draw similar qualitative conclusions regarding the turbulent HSV dynamics than in the immersed obstacle configuration.
\citet{seal_dynamics_1997} examined the laminar HSV upstream of an emerging obstacles, but they focused only on the breakaway vortex system regime.
Indeed, to the authors knowledge, only a few studies in laminar HSV around an emerging obstacle provide a comprehensive description of the HSV dynamics and their geometrical properties as a function of the flow and obstacle parameters.

To summarize, the HSV in immersed obstacle configuration is well-documented thanks to numerous studies (see \citealt{ballio_survey_1998} survey).
This is, nonetheless, not the case for the emerging obstacle configuration subjected to a laminar boundary layer.
While some characteristics, such as the HSV regimes or some parametric dependencies, seem to be qualitatively similar for both \emph{emerging} and \emph{immersed} configurations,
the evolution of the HSV properties with the dimensionless parameters of the flow remains poorly known for emerging obstacles.
In addition, it is not clear yet how the confinement of the free surface, that should strongly influence the HSV, affects its geometrical and dynamical properties.

In this context, this work aims at characterizing the evolution of the HSV with the dimensionless flow parameters, in the case of a long obstacle emerging from a laminar free-surface flow.
The studied HSV characteristics are separated in two main parts in the sequel:
(i) the HSV vortices dynamics, whose study is based on a dynamics typology, and
(ii) the HSV geometrical properties (size, number of vortex and vortex average position).

\subsection{Dimensional analysis}
\label{sec:dimensional-analysis}
Any property of the HSV that forms in a laminar boundary layer facing an emerging rectangular obstacle can be expressed as a function of the fluid, flow and geometrical parameters, as:
\begin{equation}
  \label{eq:AD:full-dim}
  X = f(\nu, \rho, \sigma, u_D, \delta, H, W, L, h, k_s, g)
\end{equation}
with $\nu$ the kinematic viscosity, $\rho$ the fluid density, $\sigma$ the surface tension, $u_D$ the bulk velocity, $\delta$ and $H=\delta^*/\theta$ respectively the boundary layer thickness and shape factor at the obstacle face location before introducing it (with $\delta^*$ the boundary layer displacement thickness and $\theta$ the boundary layer momentum thickness), $W$ the obstacle width (along $z$), $L$ the obstacle length (along $x$), $h$ the water level at the obstacle location before introducing it, $k_s$ the bed and obstacle roughness, and $g$ the gravity acceleration (see figures~\ref{fig:INTRO-Hsv-diagram} and~\ref{fig:EXPE-Disp-diagram}).
Those 11 parameters involve 3 scales.
Using $h$ as length scale, $h/u_D$ as time scale and $\rho h^3$ as mass scale, Vaschy-Buckingham $\Pi$-Theorem then allows to reduce the dependency to the following $8$ dimensionless parameters:
\begin{equation}
  \label{eq:AD:full-adim}
  X^* = f\left(~\Rey_h=\frac{4 u_D h}{\nu},~\frac{W}{h},~\frac{h}{\delta},~Fr=\frac{u_D}{\sqrt{gh}}, \frac{W}{L},~H,~We=\frac{\rho u_D^2 h}{\sigma},~\frac{k_s}{W} \right)
\end{equation}
with $X^*$ any flow property made dimensionless using the appropriate scale, $\Rey_h$ the Reynolds number based on the hydraulic diameter $D_h=4bh/(b+2h) \approx 4h$ herein, so that $\Rey_h \approx 4Q/b$ with $Q$ the total discharge and $b$ the channel width, $Fr$ the Froude number and $We$ the Weber number.

Previous works suggested that the obstacle length $L$ has no impact on the HSV when studied in the vertical upstream plane of symmetry~\citep{dargahi_turbulent_1989, ballio_survey_1998}.
However, preliminary velocity measurements with increasing obstacle lengths showed that the boundary layer separation average position and its transverse oscillation (along $z$) are affected by this parameter.
In order to avoid the effect of the wake and to neglect the influence of the aspect ratio parameter $W/L$, all obstacles considered in the sequel will be chosen sufficiently long with respect to the obstacle width ($W/L < 0.3$).
The Froude number remains small enough throughout the present work ($Fr < 0.3$) to neglect its influence.
Preliminary water level measurements showed that the flow remains in the hydraulic smooth regime for the studied flow parameter domain, allowing to neglect the influence of the roughness parameter $k_s/W$.
The shape factor $H$ remains fairly constant around the value of $2.68$ in this study.
The measured free-surface deformations are small ($\Delta h/h < 0.13$) throughout the experiment, and so, surface tension effects can be safely disregarded.
Therefore, in the present work, any HSV dimensionless characteristic $X^*$ should depend only on $3$ dimensionless parameters as
\begin{equation}
  \label{eq:AD:reduced-adim}
  X^* = f\left(\Rey_h,~\frac{h}{\delta},~\frac{W}{h}\right)
\end{equation}

\section{Experimental methods}
\label{sec:exper-appar}

\subsection{Experimental set-up}
\label{sec:experimental-set-up}
\begin{figure}
  \centering
  \includegraphics[width=1.1\figwidth]{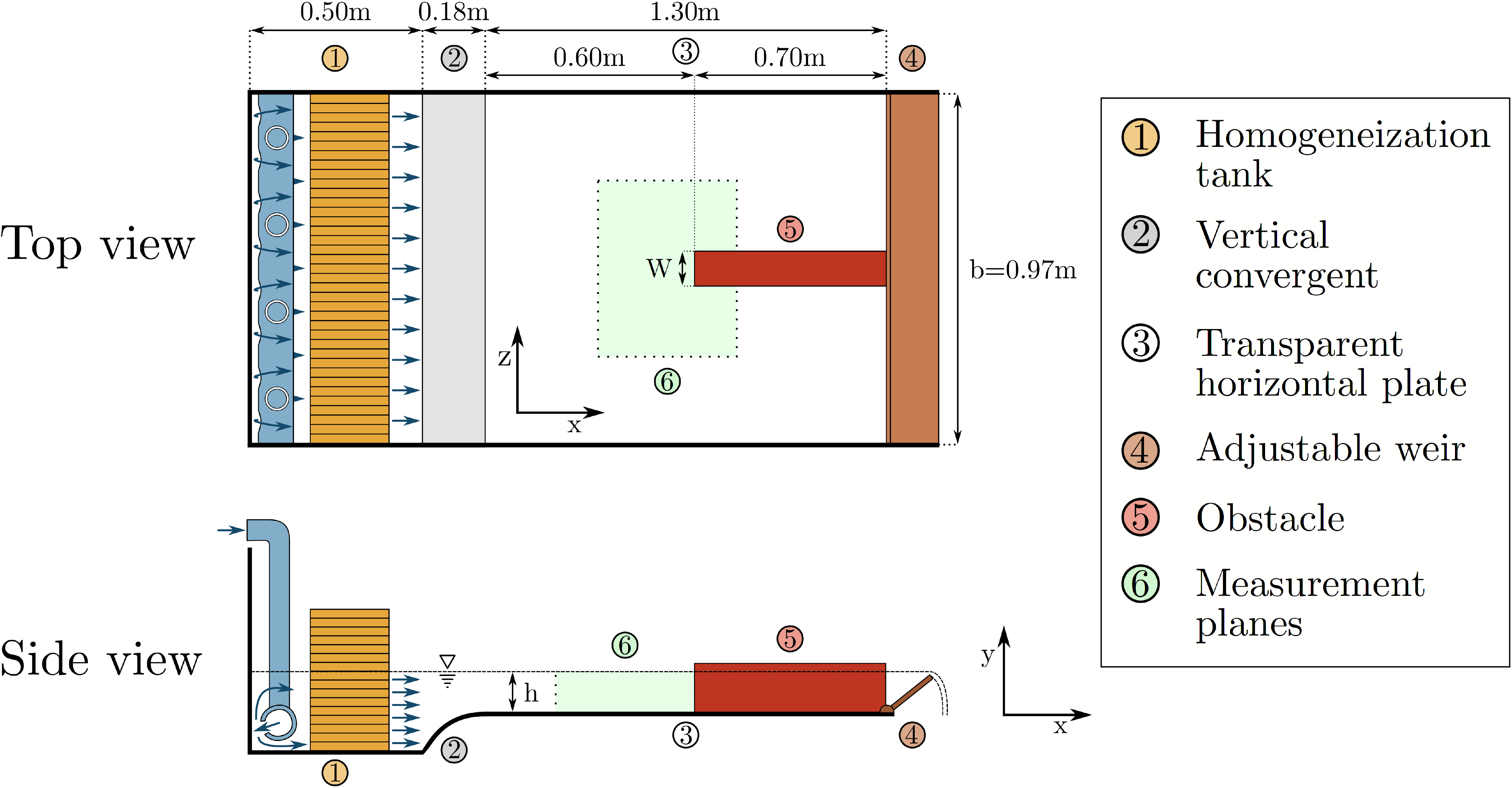}
  \caption{%
    Experimental set-up consists of a feeding loop and an adjustable weir to control the flow parameters ($u_D$ and $h$) and a transparent water tank to visualize and measure, in vertical and horizontal planes, HSV around long emerging obstacle (width $W$ and length $L$).
  }
  \label{fig:EXPE-Disp-diagram}
\end{figure}

The water table schematized in figure~\ref{fig:EXPE-Disp-diagram} is used to generate a horseshoe vortex at the foot of long, emerging, rectangular obstacles with varying dimensionless parameters $\Rey_h$, $h/\delta$ and $W/h$.
The water tank is fed by a pumping loop which includes a valve for discharge control, an electromagnetic flowmeter (Promag W, of Endress+Hauser, uncertainty of $0.01$L/s, \textit{i.e} a precision of $0.5\%$ to $4\%$ depending on the discharge value), a homogenization tank composed of several grids and honeycombs (1) and a vertical convergent (2) to compress the boundary layer.
The water then flows on a horizontal smooth plate (width $b$=0.97m and length 1.3m) made of glass to allow optical access from the side and bottom walls (3).
Water level can be controlled by an adjustable weir (4) and is measured by a mechanical water-depth probe with digital display (uncertainty of $0.15$mm, according to \citealt{riviere_supercritical_2011}).
The obstacle (5) with adjustable width $W$ is placed at a distance of $0.6$m downstream the convergent end.

\subsection{Measurement techniques}
\label{sec:meas-techn}
HSV measurements are obtained using either particle image velocimetry (PIV) or trajectographies in the vertical plane of symmetry ($z=0$) and, for section~\ref{sec:horizontal-view-hsv}, in a horizontal plane near the bed at the elevation $y/h=0.01$ (marker 6 on figure \ref{fig:EXPE-Disp-diagram}).
For both techniques, a 532nm, 4W continuous laser with a Powell lens is used to illuminate 10$\upmu$m hollow glass spheres included as tracers in the flow.
The displacement of these particles is recorded with a mono-chromatic, 12bit, $2048 \times 1088$ pixels camera.
For trajectographies, frames with adapted time-exposure are taken at frequencies ranging from 1 to 25Hz.
For PIV measurements, double frames are taken at frequencies from 1 to 2Hz, depending on the typical flow velocities and the frames spatial resolution.
Image processing and PIV computations are realized under DaVis software (Lavision) and further velocity field analyzes are performed using Python.
Image processing includes ortho-rectification, background removal, intensity capping~\citep{shavit_intensity_2006} and/or moving average.
PIV computations are realized using cross-correlations with $50\%$ overlapping and adaptive interrogation windows size decreasing from $64$ to $16$ pixels, leading to a spatial resolution of approximately $0.01h$.
Measurement quality is ensured by following recommendations from~\citet{adrian_particle_2011} and others:
(i)~a thin laser sheet (1mm thick), ensuring that particles displacements in the direction normal to the measurement plane do not influence the measured velocity,
(ii)~a small Stokes number for the seeding particles (maximum $St=0.012$, leading to less than $1\%$ error according to~\citealt{tropea_springer_2007}),
(iii)~a low sedimentation ratio (ratio between typical velocity and sedimentation velocity) of $0.004$ ensuring that sedimentation velocity of the particles is negligible in regard to the advection velocity,
(iv)~ortho-rectification of the obtained frames to avoid velocity errors due to optical deformation,
(v)~suitable particle concentration (at least 10 particles per interrogation windows),
(vi)~sufficiently large particle displacement between two frames (at least 10 pixels), leading to a roughly approximated uncertainty on the velocity of $1\%$,
(vii)~filtering of the obtained vectors in regard to the cross-correlation peak-ratio (with a minimum of $1.5$) to remove possibly wrong vectors (generally around $5\%$ of the vectors in the present case).

The measurement protocol for each case is the following:
(i)~values for the dimensionless parameters and associated dimensional parameters are selected and the experimental set-up is tuned accordingly, without obstacle, setting the desired bulk velocity $u_D$ along with the water level $h$ at the $x$ future position of the obstacle upstream face.
(ii)~The boundary layer profile at the future position of the obstacle face is measured using PIV to access the boundary layer properties.
(iii)~An obstacle of given width $W$ is placed so that its upstream face is located at $x=60$cm from the convergent end, and its lateral faces are parallel to $x$ axis (figure \ref{fig:EXPE-Disp-diagram}).
This creates the adverse pressure gradient responsible for the boundary layer separation and the HSV appearance.
(iv)~PIV measurements or trajectographies are performed, once the flow is established, in the vertical plane ($x, y$) of symmetry upstream from the obstacle (or in a horizontal plane $x, z$ upstream from the obstacle for section~\ref{sec:horizontal-view-hsv}).

\subsection{Characteristics of the flow without the obstacle}
\label{sec:char-flow-with}
The state of the laminar boundary layer as it interacts with the obstacle is a key parameter for the formation and the evolution of the HSV.\@
It is then essential to ensure that the boundary layer is not polluted by experimental set-up biases.
Without obstacle, the boundary layer freely develops from the end of the convergent, where the velocity profile is uniform along a vertical profile.
The vertical velocity profile at a given $x$ value can be fully characterized by:
(i) the boundary layer thickness $\delta$, defined as the vertical position where the velocity reaches $99\%$ of the maximum velocity,
and (ii) the shape factor $H$.
The boundary layer thickness at the future position of the obstacle face depends on the bulk velocity $u_D$, on the distance from the end of the convergent, but also, on the vertical confinement imposed by the free-surface, \textit{i.e.} on the water depth $h$.
The vertical profile of streamwise velocity in the boundary layer is expected to fit the laminar Blasius solution for high $h/\delta$ (unconfined situation) and the half-parabolic Poiseuille profile for low $h/\delta$ (highly confined situation), by analogy with closed channel flows.

\begin{figure}
  \centering
  \includegraphics[width=.75\figwidth]{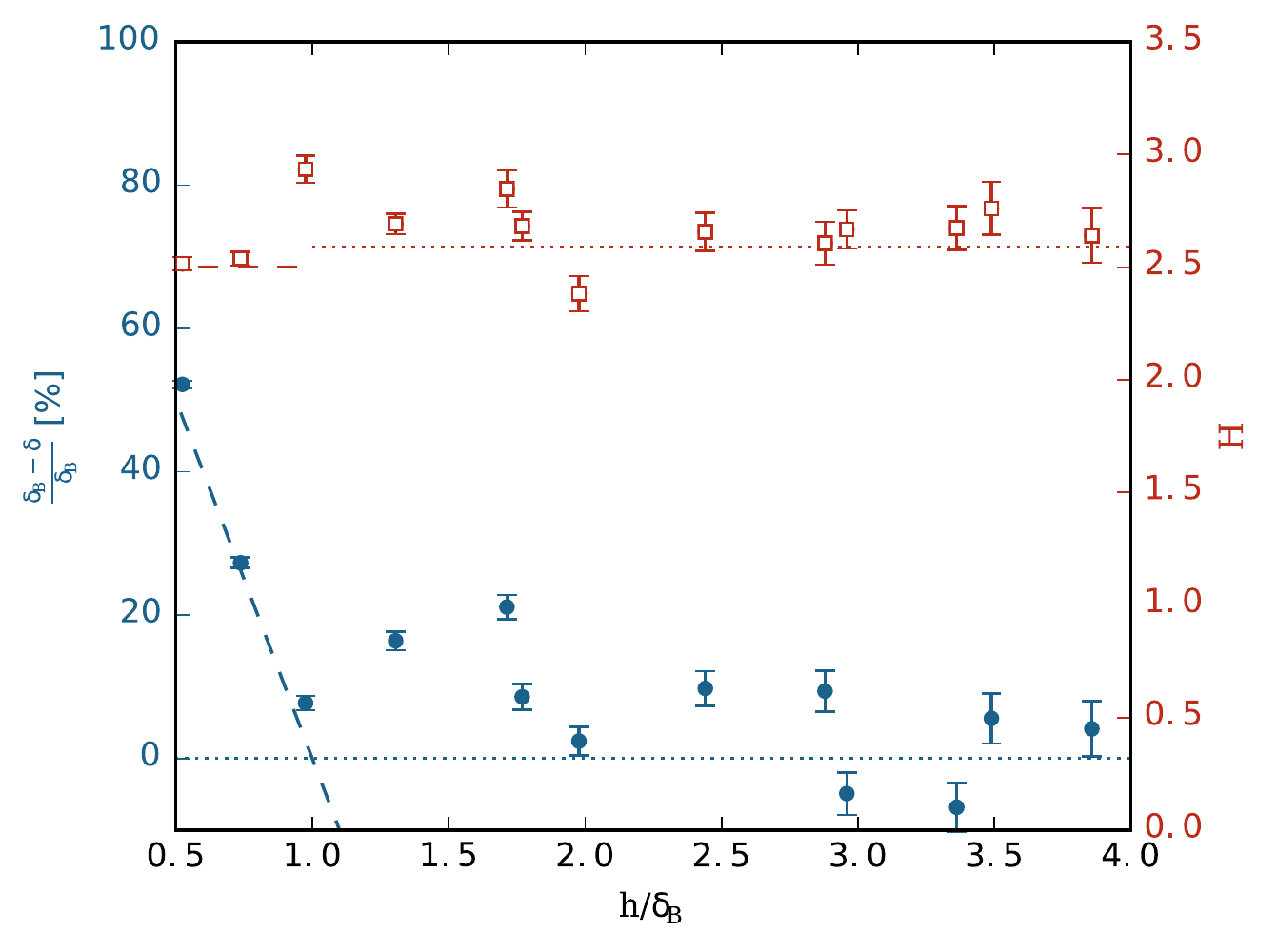}
  \caption{%
    Characteristics of the boundary layer before introducing the obstacle. Left axis: measured boundary layer thickness $\delta$ deviation from the Blasius solution $\delta_B$ (blue filled circles) and analytic solution for a parabolic profile (dash line). Right axis: evolution of the measured shape factor $H$ (red open square symbol), with dotted and dashed lines representing respectively the Blasius shape factor ($H_B = 2.59$) and the parabolic shape factor ($H_P = 2.5$). uncertainties on $(\delta_B - \delta)/\delta_B$ and $H$ are calculated with an estimated uncertainty of $0.03h$ on the measured values of $\delta$.
    As expected, the boundary layer fits a laminar Blasius profile and tends to a Poiseuille profile for important free-surface confinements.
  }
  \label{fig:EXPE-Bl-fact}
\end{figure}

To confirm this statement, figure~\ref{fig:EXPE-Bl-fact} shows the measured boundary layer thicknesses and shape factors for all boundary layers used in the sequel, compared to the corresponding values expected for Blasius ($\delta_B$ and $H_B = 2.59$) and Poiseuille ($\delta_P$ and $H_P = 2.5$) profiles.
This figure confirms that the boundary layers match with Poiseuille profiles for highly confined flows and approach Blasius-like profiles as the confinement decreases.
It should be noted that the shape factor $H$ remains fairly constant around an average value of $2.68$, ensuring that the boundary layer does not undergo a turbulent transition in the present cases.
This is also confirmed by the measured turbulent intensity (not shown here), remaining lower than $<3\%$.

These results show that the experimental apparatus is able to produce laminar boundary layers (despite the high Reynolds number of up to $\Rey_h=8000$) that are affected by the vertical free-surface confinement for $h/\delta_B < 2$.

\subsection{Experiment plan}
\label{sec:experiment-plan}
\begin{figure}
  \centering
  \includegraphics[width=.75\figwidth]{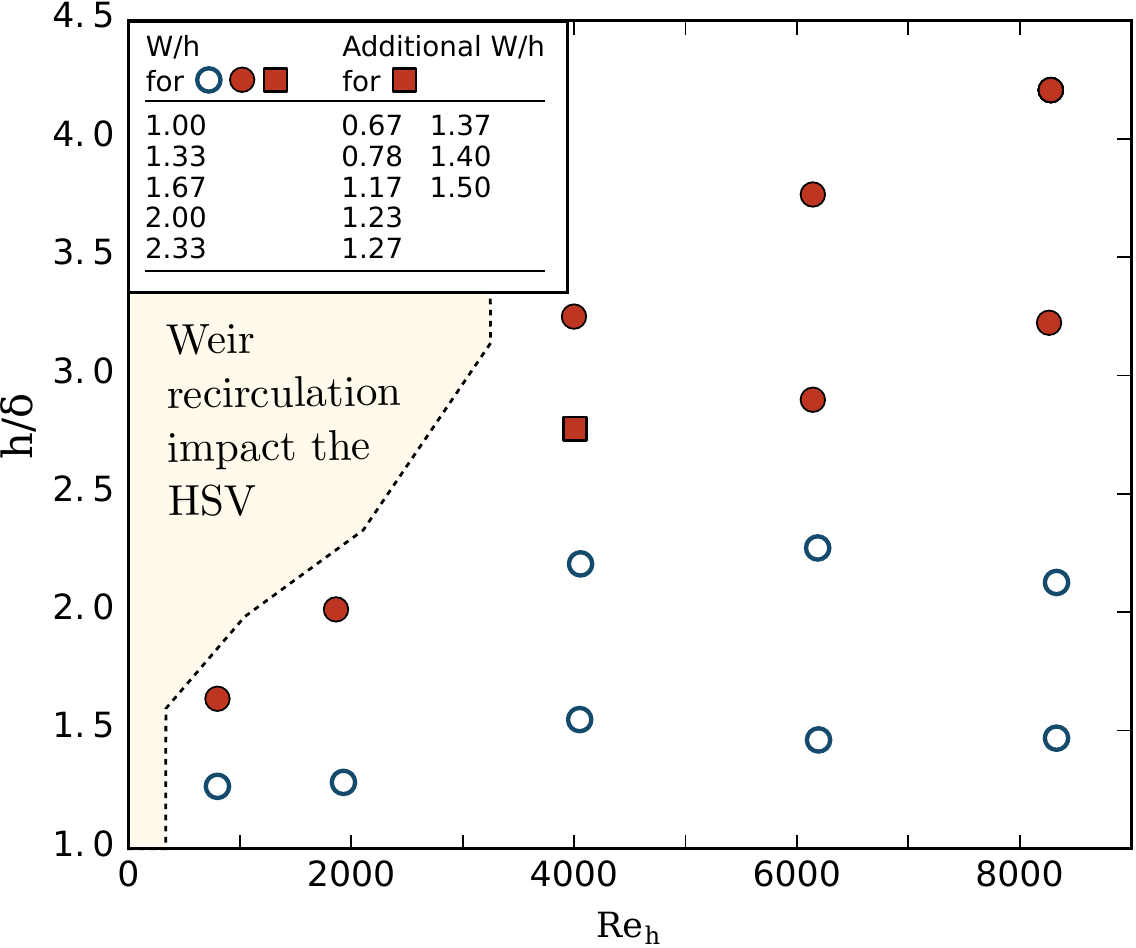}
  \caption{%
    Experiment plan used to study the HSV geometrical and dynamical properties evolution.
    Blue open symbols represent cases investigated by trajectography and red filled symbols cases investigated by PIV.\@ For each circle symbol (hollow or filled), 5 different values of $W/h$ are considered.
    For the red square symbol, 13 values of $W/h$ have been measured in order to study the HSV regime transitions in details.
    The left zone is inaccessible to measurement, as the experimental set-up weir creates a recirculation impacting the HSV.
  }
  \label{fig:EXPE-Expe-plan}
\end{figure}

In order to have a good insight on how the three dimensionless flow parameters affect the HSV structure and dynamics, the dimensionless parameter space ($\Rey_h$, $h/\delta$, $W/h$) is mapped such as presented in figure~\ref{fig:EXPE-Expe-plan}.
$\Rey_h$ and $W/h$ are well-controlled (respectively by the discharge $Q$ and the obstacle width $W$) and allow homogeneous mapping, avoiding inter-dependencies.
Tuning $h/\delta$, however, is more difficult, as the boundary layer thickness $\delta$ depends on the bulk velocity $u_D$ the water level $h$ and the distance between the convergent and the obstacle, which has a very limited variation range in the present experimental set-up.

Measurements duration always exceeds at least $20$ periods (in case of periodic HSV behavior) for the $75$ cases of the parametric study, and at least $200$ periods for the $13$ cases of the detailed transition study (square symbol in figure~\ref{fig:EXPE-Expe-plan}).
Time resolutions of the measurements ensure at least $15$ velocity fields per period.

\subsection{Horseshoe vortex properties}
\label{sec:hors-vort-prop-1}
This section presents the methods used to measure the HSV geometrical and dynamical properties from the acquired PIV velocity fields and trajectographies.

\subsubsection{Horseshoe properties measurement}
\label{sec:hors-prop-meas}
One first main characteristic of interest is the vortex center positions (see figure~\ref{fig:INTRO-Hsv-diagram}b).
The vortices sharing the boundary layer vorticity sign are designated as $V_i$, with $i$ the number of the vortex, starting with the downstream-most one, and ($x_i,y_i$) their positions.
In the sequel, $V_1$ and $V_2$ will also be referred as ``main vortex'' and ``secondary vortex''.
The counter-rotating vortices, located near the bed and just upstream from the previously defined vortices are designated as $V_{ci}$ with $i$ the number of the associated vortex $V_i$.
A second characteristic of the HSV is the location of the boundary layer separation, defined as the upstream-most $x$ position where the shear-stress along $x$ equals zero.
Its distance to the obstacle is noted $\lambda$ (see figure~\ref{fig:INTRO-Hsv-diagram}b).

As proposed by~\citet{ballio_survey_1998} in order to enhance the comparison between obstacles of different shapes, the reference points for streamwise distances ($\lambda$ and $x_i$) is not the obstacle face but rather a point located at $W/2$ behind it (see figure~\ref{fig:INTRO-Hsv-diagram}b).
Uncertainties regarding those distances are estimated as about $0.05h$ for trajectographies and $0.02h$ for PIV measurements (as the measurements spatial resolution depends on $h$).

The vortex circulation, which is also an interesting property, is quite challenging to define in the vicinity of a strong shear layer (the boundary layer in this case).
The classical method to estimate the vortex radius, consisting of defining a vorticity threshold, can be rendered inoperable by the strong boundary layer vorticity due to the shear.
The residual vorticity~\citep{kolar_vortex_2007}, known to be the vorticity associated to rotation, is in place used herein to get the vortex region.

\subsubsection{HSV dynamics characterization}
\label{sec:hsv-dynam-char}
In order to have quantitative data on the HSV dynamics and to establish a clear typology, vortex positions need to be followed in time.
To do so, critical points of the velocity field in the vertical plane of symmetry are detected and tracked in time (for PIV measurements), summarizing efficiently the HSV structure evolution.

Critical points are Lagrangian dependant and, as such, are unable to detect vortices advected at high velocities. In those cases, gradient-based criteria such as the $\lambda_2$-criterion~\citep{jeong_identification_1995} or the residual vorticity~\citep{kolar_vortex_2007} are far more efficient in detecting and tracking vortices. However, as the present vortex advection velocities (velocities of their centers) remain small compared to the velocities they induce, and regarding the valuable additional topological information brought by the critical points, they are used in the sequel to characterize the HSV dynamics.

The method of detection and tracking, inspired from~\citet{graftieaux_combining_2001},~\citet{depardon_automated_2007} and~\citet{effenberger_finding_2010}, consists of six steps:
(i)~Pre-filtering of the time-resolved velocity fields, using POD reconstruction on a truncated modal base.
POD modes are not filtered using the classical energy criterion \citep{peltier_meandering_2014}, but according to the dispersion of their spatial spectra, which is representative of the presence of large-scale structures.
This step aims at reducing the measurements noise, removing the small-scale fluctuations to promote the large-scale structures and replacing the missing velocity vectors by spatially and timely interpolated ones.
(ii)~Detection, on each instantaneous velocity field, of the measured vector grid cells susceptible of containing a critical point using a scan of the Poincarré-Bendixson index~\citep[see][for more details on this index]{hunt_kinematical_1978}.
(iii)~Detection of the sub-grid position of the critical points, using~\citet{effenberger_finding_2010} method.
(iv)~Determination of the critical points type (saddles points, stable or unstable nodes, stable rotating or counter-rotating vortex centers) based on the local Jacobian matrix eigenvalues.
(v)~Optional topological simplification using~\citet{graftieaux_combining_2001} $\Gamma$ criterion.
This step allows to select only large-scale bounded critical points.
(vi)~Trajectory reconstruction using distance sum minimization.

For a more synthetic visualization of the HSV vortices motion for a particular configuration, the vortex centers trajectories are averaged by group according to their similarities:
(i)~For each couple of trajectories, the normalized integral of the squared difference is computed:
\begin{equation}
  \epsilon_{mn} = \frac{1}{2} \frac{\int\left[ x_m(t) - x_n(t)\right]^2 dt}{\frac{1}{2} \int\left[ x_m(t) + x_n(t)\right]^2 dt}
  + \frac{1}{2} \frac{\int\left[ y_m(t) - y_n(t)\right]^2 dt}{\frac{1}{2} \int\left[ y_m(t) + y_n(t)\right]^2 dt}
  \label{eq:diff-number}
\end{equation}
where $(x_m(t), y_m(t))$ is the position of the trajectory $m$ at time $t$ and $\epsilon$ is representative of the difference between the two trajectories (small for close trajectories, large for different trajectories).
(ii)~The maximum difference between two trajectories considered similar is arbitrarily defined (here to $\epsilon_{crit} = 0.07$, based on the obtained results).
(iii)~Trajectories are distributed in groups, ensuring that no group includes couples of trajectories with $\epsilon > \epsilon_{crit}$.
(iv)~Groups of similar trajectories are averaged to obtain mean trajectories.

One can finally associate instantaneous velocity fields to each of those mean trajectories and so perform a conditional averaging of the velocity fields on the vortex center position.

\subsection{Horizontal view of the HSV}
\label{sec:horizontal-view-hsv}
Main HSV geometrical and dynamical properties can be observed from measurements in the vertical plane of symmetry.
Nonetheless, the HSV does not remain necessarily symmetric with respect to this plane at all time, and \citet{eckerle_horseshoe_1987} pointed out that the vortex filament maximal radii and intensities can be located outside of the symmetry plane.
In this context, measurements in a horizontal plane near the bed are mandatory to ensure that the HSV driven phenomena can be deducted from 2D PIV measurements in the vertical plane of symmetry.

\begin{figure}
  \centering
  \includegraphics[width=1.\figwidth]{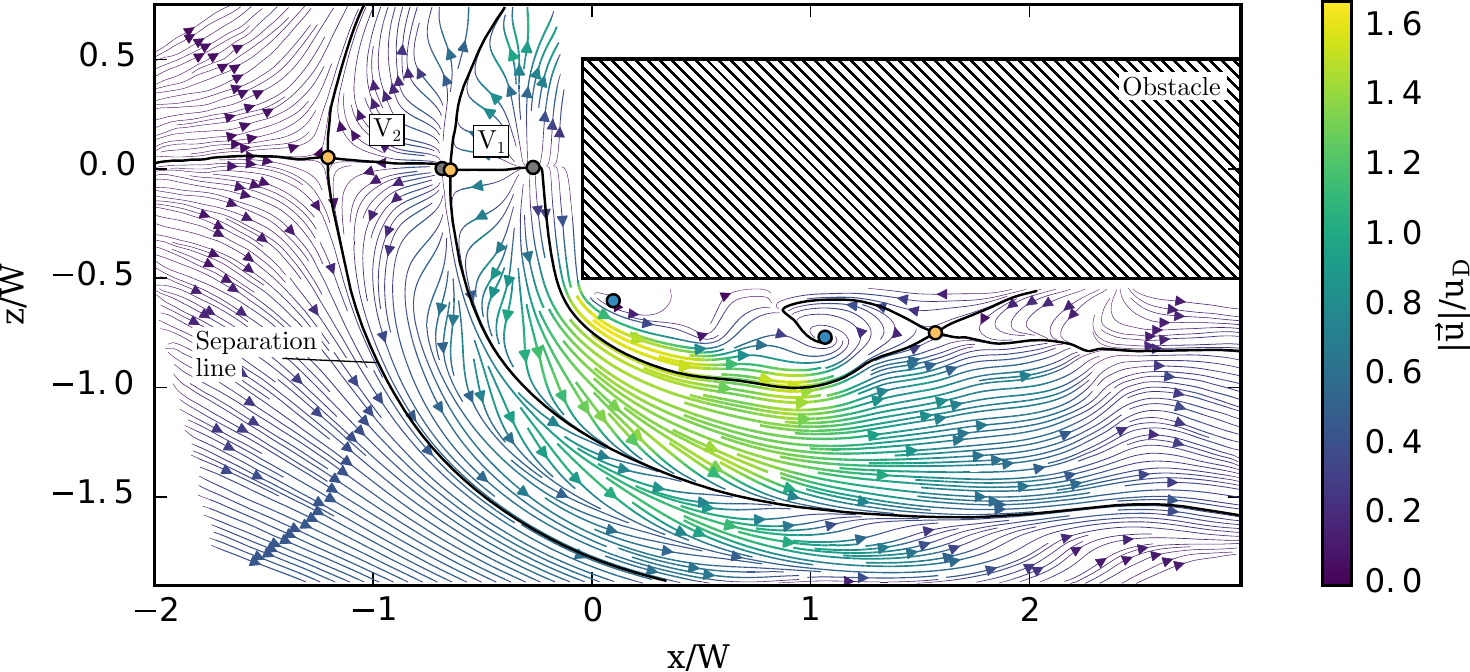}
  \caption{%
    Instantaneous velocity field in the horizontal plane near the bed ($y=0.01h$) for a laminar flow around an obstacle for $\Rey_h=4272,~h/\delta=2.70,~W/h=1.67$.
    Streamlines are coloured with velocity magnitude.
    Circles represent detected critical points: saddle points (yellow), unstable nodes (grey) and vortex centers (blue).
    Plain thick lines are streamlines coming from and going to saddle points and delimit the vortices within the HSV.\@ $V_1$ and $V_2$ are the main and secondary vortices.
    The topology of the flow and the velocity distribution show that the HSV vortices properties do not evolve drastically when rolling around the obstacle in the upstream part of the HSV (before $x=0$).
  }
  \label{fig:EXPE-Horizontal}
\end{figure}

Figure~\ref{fig:EXPE-Horizontal} presents a velocity field in the horizontal plane near the bed ($y/h = 0.01$).
The evolution of the main and secondary vortex ($V_1$ and $V_2$) filaments location while bypassing the obstacle can be evaluated on this figure with the help of the critical points and their associated lines.
Vortices do not undergo strong modifications in size and velocity while wrapping around the obstacle, until they interact with the complex lateral separation bubble on the sides of the obstacle.
Time-resolved measurements show that the position of the horizontal separation line (see figure~\ref{fig:EXPE-Horizontal}) remains constant with time (not shown here), ensuring that shedding from the lateral recirculation bubbles are not strong enough to disrupt the HSV and the boundary layer separation.
Those measurements also reveal that the dynamics of the HSV vortices does not evolve significantly along the filaments (while rolling around the obstacle).

Conclusion can be made, on this configuration, that:
(i)~the instantaneous HSV symmetry plane coincides at each time with the obstacle symmetry plane,
(ii)~the vortex filaments remain approximately of the same size and keep the same dynamics while wrapping around the obstacle (contrary to the observations of~\citealt{eckerle_effect_1991}), and consequently,
(iii)~a measurement on the vertical symmetry plane of the HSV ($z=0$) is an adapted and sufficient approach to characterize the HSV behavior.

The same conclusions apply for all flows investigated with horizontal measurements, allowing the present work to be based solely on the study of measurements in the vertical plane of symmetry.

\subsection{Comparison between various obstacle shape and submergence}
\label{sec:comp-betw-vari}
Most previous works concerning the HSV, in both immersed and emerged configuration, considered cylindrical obstacles, making the comparison with the present results with rectangular obstacles challenging.

\citet{baker_oscillation_1991} proposed, in order to solve this problem, to estimate that HSV from obstacles of different shapes are comparable if they have the same separation distance (method that requires to known the separation distance) while~\citet{ballio_survey_1998} considered comparable HSV from square obstacles of width $W$ and HSV from cylindrical obstacles of diameter $W$.

A more systematic method based on the estimated adverse pressure gradient is proposed herein.
Because the whole HSV and notably the boundary layer separation is governed by the streamwise adverse pressure gradient, one can assume that two obstacles of different shapes creating the same adverse pressure gradient in the upstream symmetry plane should generate similar HSV.\@
Based on this assumption, and using pressure profiles from potential flow computation, an equivalent obstacle width $W_{eq}$ can be computed for cylindrical obstacles of diameter $D$, so that a quadrilateral obstacle of width $W_{eq}$ and infinite length and a cylindrical one with equivalent width $W_{eq}$ lead to nearly the same upstream pressure gradient and consequently, comparable HSV.\@
This method can further be applied to obstacles with other shapes (such as quadrilateral obstacles with non-infinite length, foil, bevelled quadrilateral, oblong obstacles, $\ldots$).
As the separation distance is linked to the adverse pressure gradient, this method is very similar to the one of~\citet{baker_oscillation_1991}, but does not necessitate prior knowledge of the separation distance $\lambda$.

\begin{figure}
  \centering
  \includegraphics[width=.75\figwidth]{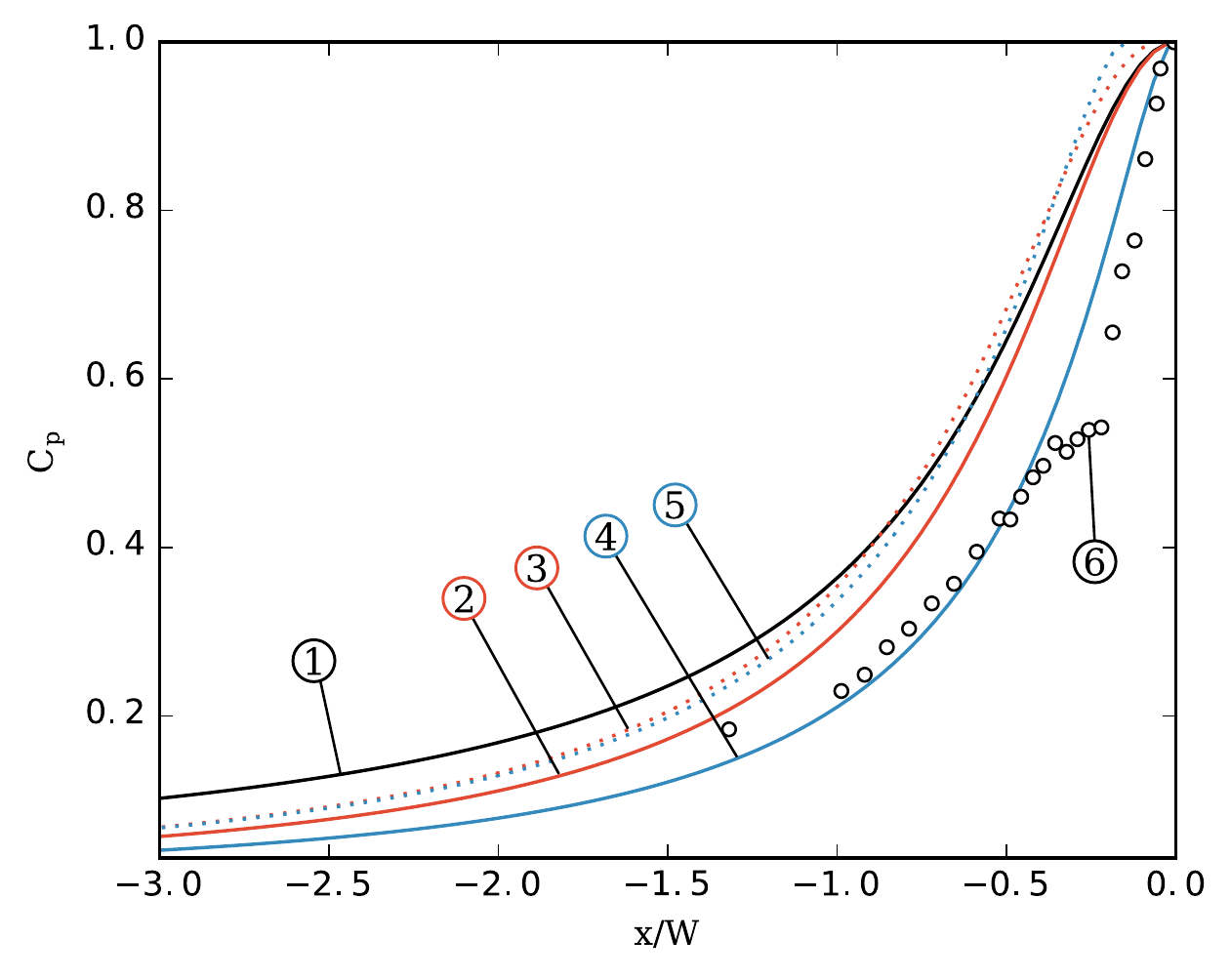}
  \caption{%
    Pressure coefficients $C_p$ obtained using 2D (in the horizontal plane) potential flow theory computations upstream of different obstacles:
    (1) quadrilateral obstacle of width $W_{eq}$ and infinite length,
    (2) squared-shaped obstacles of widths $W_s=W_{eq}$,
    (3) squared-shaped obstacles of width $W_s=1.093W_{eq}$,
    (4) cylindrical obstacles of diameters $D=W_{eq}$, and
    (5) cylindrical obstacles of diameters $D=1.29W_{eq}$.
    (6) measurements from~\citet{baker_laminar_1978} at the bed in front of a cylindrical immersed obstacle.
    These results show that the pressure gradients for different obstacles can be aggregated by modifying their characteristic sizes.
  }
  \label{fig:EXPE-Deq}
\end{figure}

Figure~\ref{fig:EXPE-Deq} shows that pressure coefficients $C_p$ (from 2D potential flow computation) upstream of cylindrical, squared-shaped and infinitely long rectangular obstacles can effectively be aggregated by adjusting their size (diameter $D$ or width $W$).
The optimal size coefficients, in a least squares sense, have been found to be:
\begin{equation}
  W_{eq} = W
\end{equation}
\begin{equation}
  W_{eq} = \frac{W_s}{1.093} = 0.915 W_s
\end{equation}
\begin{equation}
  W_{eq} = \frac{D}{1.29} = 0.775 D
  \label{eq:EXPE-Equivalent-diameter}
\end{equation}
with $W$ the width of an infinitely long rectangular obstacle and $W_s$ the width of a square-shaped obstacle.
Moreover, the fair agreement between the potential flow computation and measurements from~\citet{baker_laminar_1978} shows that potential flow is effectively able to predict the pressure distribution.

This equivalence will be used in the sequel to modify empirical correlations from the literature so that they use $W_{eq}$ instead of $D$ and $W_s$ (counting respectively $1.29W_{eq}$ and $1.093W_{eq}$).

To compare immersed and emerging obstacle data, the obstacle height $\xi$ for immersed obstacles will be said analogous to the water depth $h$ for emerging obstacles, as proposed by~\citet{ballio_survey_1998}, thus keeping the same wet portion of obstacle shape ratio ($\xi/W$ for immersed obstacles and $h/W$ for emerging ones).

\section{Dynamic typology}
\label{sec:dynamic-typology}

Defining a typology categorizing the different observed behaviors is a convenient way of studying the HSV dynamics.
Authors such as~\citet{baker_laminar_1978} establish typologies using the number of vortices in the HSV.\@
Nevertheless, as in~\citet{greco_flow_1990} and~\citet{lin_characteristics_2008}, the present typology is rather based on the vortex dynamics.
The number of vortices will be discussed further on, in the section dedicated to the HSV geometrical properties.

For a given flow configuration, the HSV shows either large-scale, coherent and well-defined vortices (like in figure~\ref{fig:INTRO-Hsv-diagram}), and will be said in ``coherent regime'', or a non-stationary and aperiodical behavior, with presence of small-scale non-coherent structures and will be said in ``irregular regime''.

In coherent regime, the HSV vortices can either be steady or follow a ``pseudo-periodic'' motion, showing an alternation of elementary processes of same duration $T$ (HSV period) called ``phases'' in the sequel.
For one flow configuration, successive phases can be substantially different but always bring back the HSV to its initial state (and consequently, all phases share the same initial state).

The different observed phases are described in section~\ref{sec:phases-definition} and are the base of the definition of coherent ``sub-regimes'', defined in section~\ref{sec:regimes-definition}, and presented in details in section~\ref{sec:coherent-regimes}.
The evolution from coherent to irregular regimes is presented in section~\ref{sec:irregular-regimes}.
Finally, the evolution of all those regimes (coherent regimes and irregular regime) with the flow parameters is detailed in section~\ref{sec:regim-evol-with}.

\subsection{Coherent regimes definition}
\label{sec:coher-regim-defin}
\subsubsection{Phases (elementary processes) definition}
\label{sec:phases-definition}
All observed phases can be classified in four categories:
(i)~The ``oscillating phase'', where the displacement of each vortex is a horizontal ellipse that brings it back near its initial position at the end of the phase.
(ii)~The ``merging phase'', where the main vortex ($V_1$) follows the same pattern as in the oscillating phase, but merges with the secondary vortex ($V_2$) at the end of the phase.
The vortex merging is defined, from a ``critical points'' view, as a bifurcation from a saddle point and two vortex centers to a single vortex center.
As it disappears in the merging, $V_2$ is replaced by the third vortex $V_3$ while a new vortex is created near the boundary layer separation point.
(iii)~The ``diffusing phase'', analogue to the merging phase, but for which the main vortex circulation and size decrease along the phase, reaching the merging position with low circulation compared to the secondary vortex.
A phase will be considered to be ``diffusing'' if the main vortex circulation is less than $25\%$ of the secondary vortex circulation (this ratio can reach $0\%$ if the vortex fully diffuses before reaching the merging).
(iv)~The ``breaking phase'', where the main vortex breaks (escapes) from the HSV, is advected further downstream and finally diffuses near the obstacle.
The main vortex is considered to break from the HSV if the distance between $V_1$ and $V_2$ exceeds two times the diameter of the bigger vortex.
From this time on, the breaking vortex is considered outside of the HSV, $V_2$ replaces $V_1$ and a new vortex arises near the boundary layer separation point.
\subsubsection{Regimes definition}
\label{sec:regimes-definition}
\begin{figure}
  \centering
  \includegraphics[width=.8\figwidth]{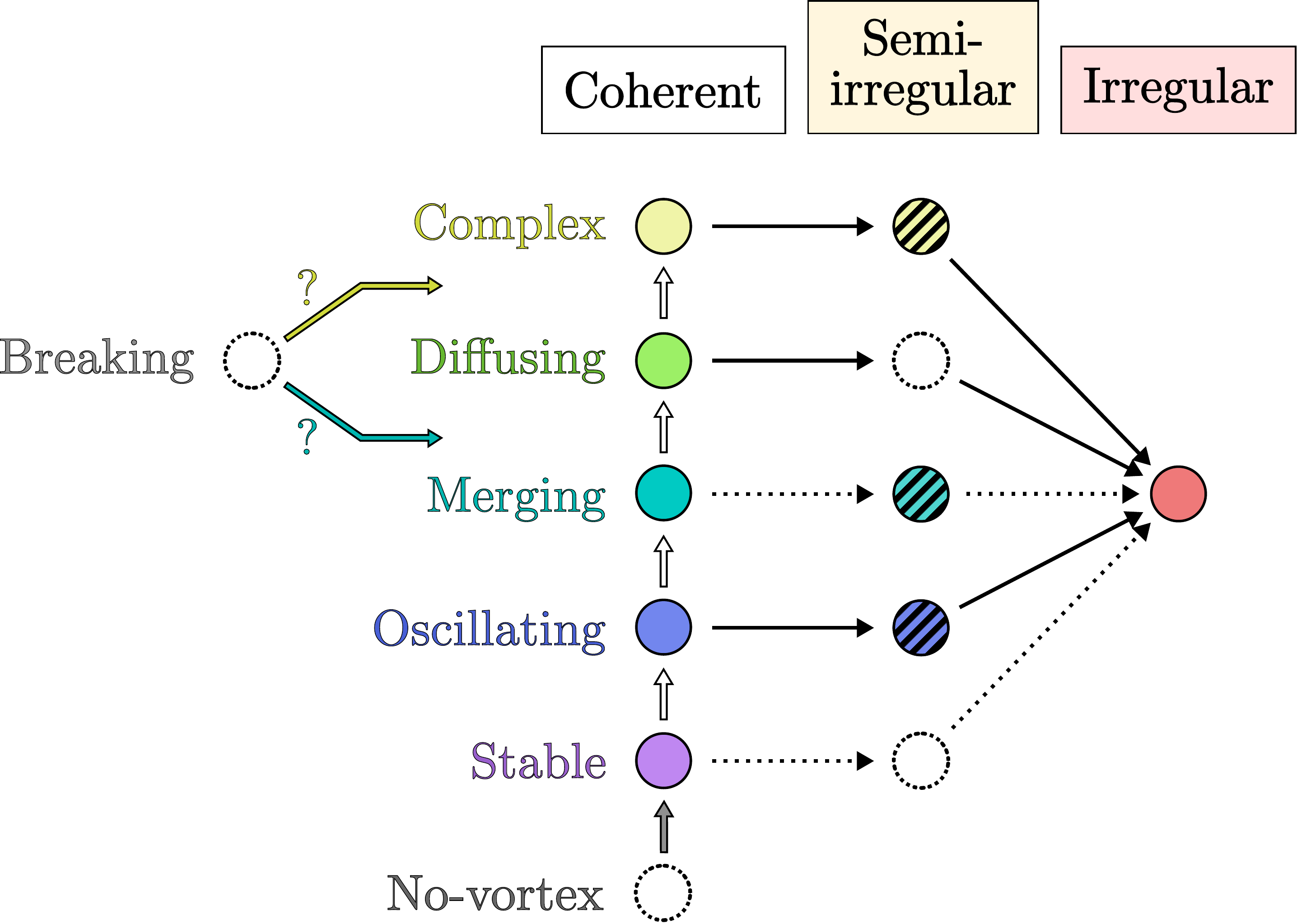}
  \caption{%
    HSV regimes organization for a laminar flow around a long rectangular obstacle.
    Each circle represents an observed dynamics, categorized along the coherent evolution (vertically) and the irregular evolution (horizontally).
    White arrows represent the coherent transitions and black arrows the irregular transitions.
    Dotted circles and arrows correspond to regimes and transitions not observed herein, but expected to exist.
  }
  \label{fig:TYPO-Regimes}
\end{figure}

The definition of the four phases, observed in the experiments, allow to classify the flow configurations in seven coherent regimes, represented on figure~\ref{fig:TYPO-Regimes}, plus two transitional regimes:
(i)~The ``no-vortex regime'', where no vortex appears on the shear layer.
This regime was not observed in the present study, but was reported by~\citet{schwind_three_1962} for immersed obstacles, and is expected to exist in the present emerging obstacle configuration for $\Rey_h$ lower than those considered herein.
(ii)~The ``stable regime'', where the location and the circulation of the vortices remain constant with time.
In practice, a HSV is considered in stable regime if the mean amplitude of the vortices displacements does not exceed $0.03W$.
(iii)~The ``oscillating regime'', (iv)~the ``merging regime'' and (v)~the ``diffusing regime'', composed by a succession of similar associated phases (in practice at least $95\%$ of all phases).
(vi)~The ``oscillating-merging transitional regime'', composed by non-regular alternations of oscillating and merging phases.
(vii)~The ``merging-diffusing transitional regime'', composed by non-regular alternation of merging and diffusing phases.
(viii)~The ``breaking regime'', composed by a succession of breaking phases.
This regime was not observed in this study, but was reported by~\citet{thomas_unsteady_1987} and \citet{greco_flow_1990} for the immersed obstacle configuration.
Where this regime should be placed on the 2D typology (figure~\ref{fig:TYPO-Regimes}) remains unclear.
(ix) The ``complex regime'', composed by an apparently chaotic succession of oscillating, merging, diffusing and/or breaking phases.
This regime is characterized by an important phase dispersion, \textit{i.e.}\ successive phases notably differ from each others even if they have the same phase type, contrary to the previously defined transitional regimes.
These regimes will be detailed in the next section.

The typology defined here is partially similar to the one presented by~\citet{greco_flow_1990} for the immersed obstacle configuration: the present stable, oscillating, merging, breaking and irregular regimes can be, respectively, assimilated to Greco's ``steady vortex system'', ``oscillating vortex system'', ``amalgamating vortex system'', ``breakaway vortex system'' and ``transitional vortex system''.
However, as the typology of~\citet{greco_flow_1990} is based on flow visualization and does not precisely define the regimes boundaries, this terminology is not used herein.
Complex regime was not reported (to the authors knowledge) in the immersed nor emerging obstacle literature.

\subsection{Coherent Regimes}
\label{sec:coherent-regimes}
\subsubsection{Stable regime}
\label{sec:stable-regime}
\begin{figure}
  \centering
  \includegraphics[width=1.\figwidth]{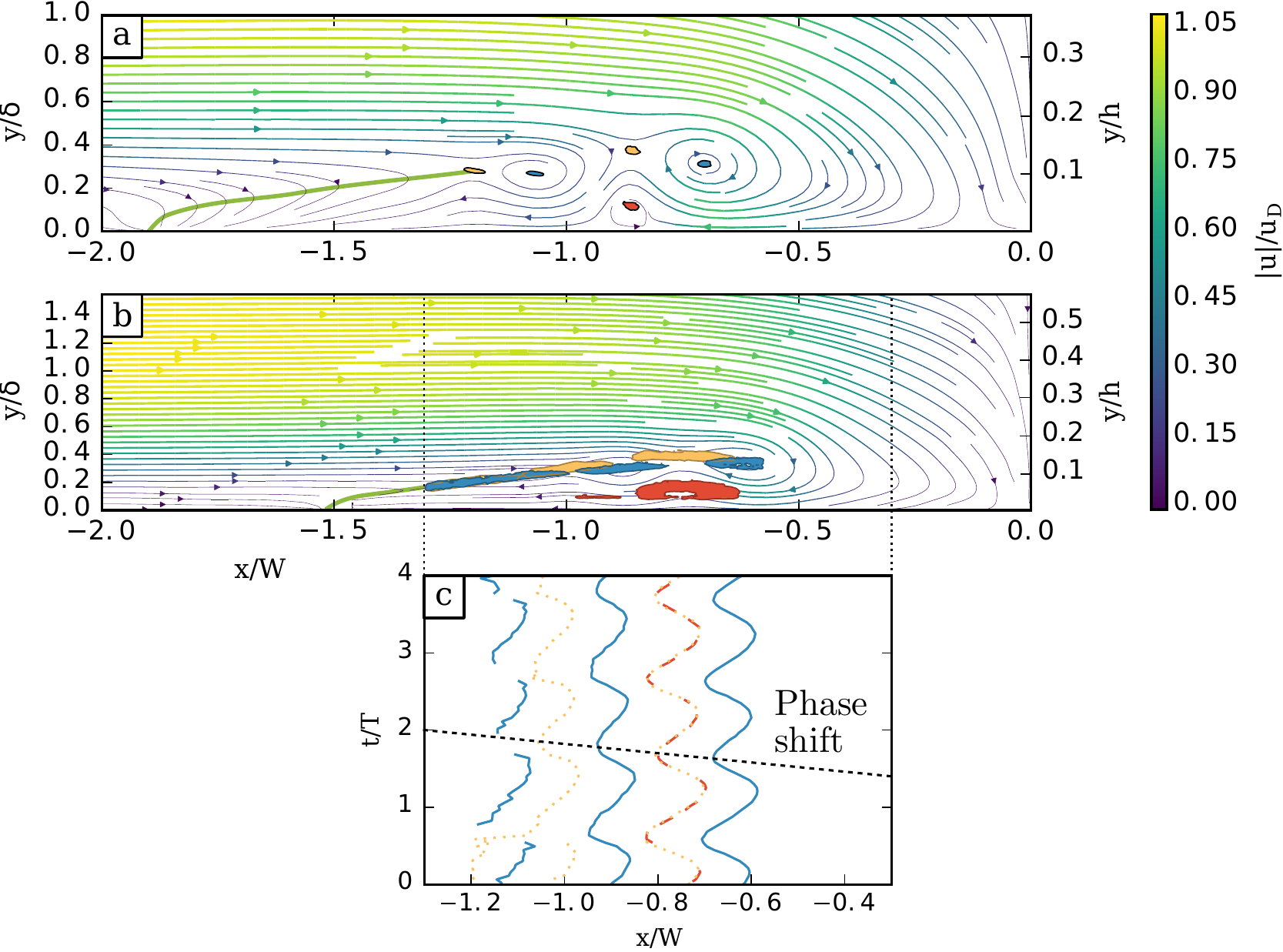}
  \caption{%
    Mean velocity field (magnitude-colored streamlines) in the vertical plane of symmetry for (a): a stable flow case ($\Rey_h=4271,~h/\delta=2.70,~W/h=0.79$) and (b): an oscillating flow case ($\Rey_h=4271,~h/\delta=2.70,~W/h=1.23,~T=21.01s$).
    Zones where 99\% of the critical points are found are calculated using kernel density estimator and are shown as filled contour (blue for vortex centers, red for counter-rotating vortex centers and yellow for saddle points).
    Green lines are the separation surface section on the symmetry plane.
    (c) Time evolution of the critical points position along $x/W$ for the oscillating regime (b).
    Plain (blue) lines stand for vortices, dashed (red) lines for counter-rotating vortices and dotted (yellow) lines for saddles points.
    Critical points displacement exhibit a phase shift from the main vortex toward upstream at a celerity of $0.29 u_D$.
    Aspect ratio is conserved on velocity fields despite the use of dimensionless axes.
  }
  \label{fig:TYPO-Stable-Oscillating}
\end{figure}

In stable regime, each vortex remains at the same spatial location at all time.
Figure~\ref{fig:TYPO-Stable-Oscillating}a shows an example of a HSV in stable regime where critical points detection indicates the presence of two vortices, confirmed by the streamlines pattern.
The steadiness of the flow dynamics is visible through the small size of the critical points presence zones.

The classical stable HSV topology, as discussed for instance in~\citet{younis_topological_2014}, is well represented, with: (i)~a boundary layer separation at $x/W=-1.9$, (ii)~a separation surface evolving from $x/W=-1.9$ to $x/W=-1.2$, (iii)~a succession of clockwise-rotating vortices (two in the present case), separated by saddle points, (iv)~counter-clockwise rotating vortices between the clockwise rotating ones (one in the present case) and (v)~the down-flow, visible through the streamlines curvature, that makes the flow re-attach near the obstacle foot at $x/W \approx -0.1$.

The three-dimensionality of the flow is clearly visible as the upper flow slows down while approaching the obstacle and as the HSV vortices streamlines are spiralling toward the vortex centers.
\subsubsection{Oscillating regime}
\label{sec:oscillating-regime}
The oscillating regime is illustrated for one flow configuration in figures~\ref{fig:TYPO-Stable-Oscillating}b-c.
The topology of the mean flow remains the same as for the stable regime (with an additional vortex in the present case).
The vortex centers presence zones are elongated but do not collide with each other: vortices are sustainable in time.
It is to be noted that vortex centers remain fairly on the shear layer originating from the separation point and ending at the main vortex $V_1$ position, and separating the upper flow (going towards downstream), and the backflow (going back upstream).
Saddle points presence zones present the same behavior as vortex centers, showing that the oscillating motion is shared by the whole HSV structure.

Figure~\ref{fig:TYPO-Stable-Oscillating}c shows the evolution of each critical point streamwise location over the time during 4 consecutive periods.
The periodic, quasi-sinusoidal streamwise displacement behavior is shared by all critical points of the HSV.\@
A phase-shift is nevertheless present between those oscillations, and reveals that the oscillating dynamics source of the HSV is the main vortex $V_1$, while the other vortices follow its motion.
\subsubsection{Merging regime}
\label{sec:merging-regime}
\begin{figure}
  \centering
  \includegraphics[width=1\figwidth]{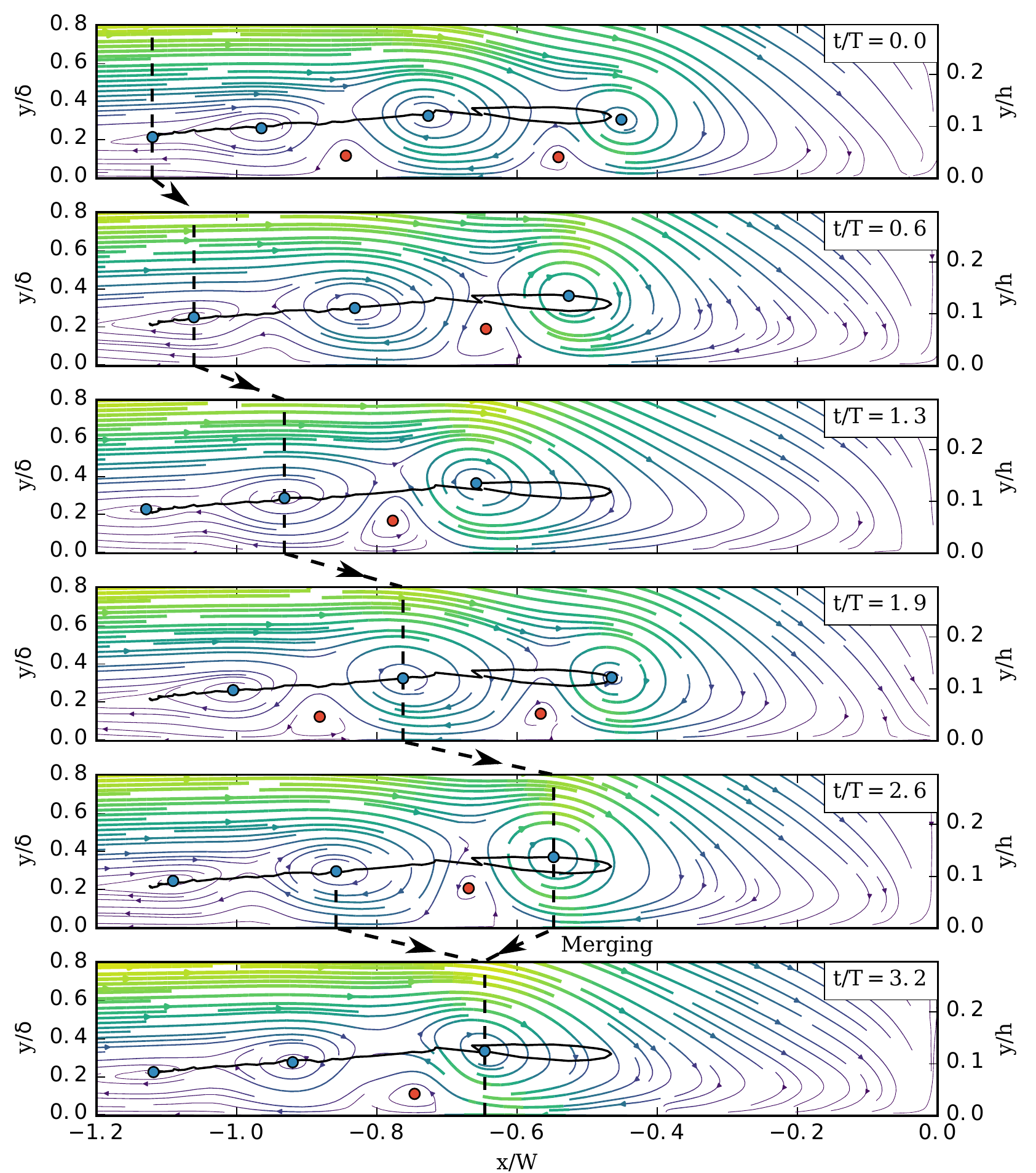}
  \caption{%
    Successive instantaneous velocity fields in the vertical plane of symmetry for a merging regime ($\Rey_h=6397,~h/\delta=3.69,~W/h=1.67,~T=20.02s$).
    Coloured symbols are detected critical points (same colours scheme as figure~\ref{fig:TYPO-Stable-Oscillating}).
    Plain line is the highlighted vortex trajectory and dashed lines help follow the vortex of interest through presented instantaneous fields.
    Aspect ratio is conserved despite the use of dimensionless axes.
  }
  \label{fig:TYPO-Merging-bd}
\end{figure}

Figure~\ref{fig:TYPO-Merging-bd} illustrates the evolution of the so-called merging regime during $3$ periods.
The global topology, already mentioned for the stable (figure~\ref{fig:TYPO-Stable-Oscillating}a) and oscillating regimes (figure~\ref{fig:TYPO-Stable-Oscillating}b) is also valid for the merging regime.
The main difference is that new vortices appear periodically at the end of the separation surface to replace those disappearing by merging.
Consequently, vortices have a life cycle:
the highlighted vortex in figure~\ref{fig:TYPO-Merging-bd} appears at $x/W=-1.15$ (and $y/\delta=0.22$) at $t/T=0$, is advected toward downstream while gaining in radius and circulation until $t/T=2.6$, then slows down as its radius decreases and ends up going back upstream and merging with the previous vortex ($V_2$) at $t/T=3.2$.
\subsubsection{Diffusing regime}
\label{sec:diffusing-regime}
\begin{figure}
  \centering
  \includegraphics[width=1.25\figwidth]{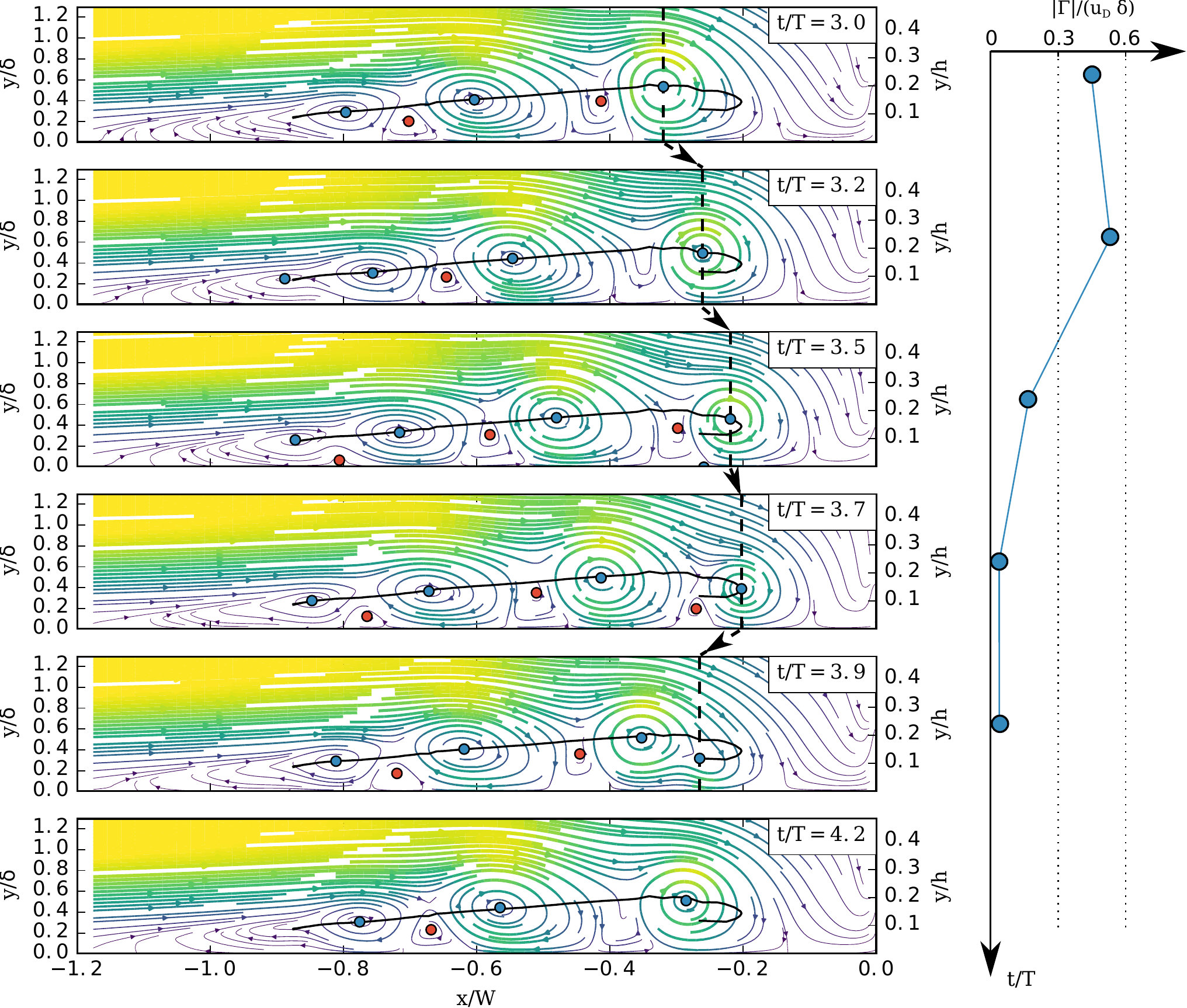}
  \caption{%
    Successive instantaneous fields in the vertical plane of symmetry for a diffusing regime ($\Rey_h=4250 ,~h/\delta=2.7,~W/h=2.36,~T=16.5s$).
    $t/T=0$ corresponds to the birth of the vortex of interest at the end of the separation surface.
    See legend from figure~\ref{fig:TYPO-Merging-bd} for the velocity fields.
    Right plot shows the evolution of the vortex of interest circulation with time.
  }
  \label{fig:TYPO-diffusing-bd}
\end{figure}

The diffusing regime is quite similar to the merging regime, but the main vortex radius and circulation drop before reaching the merging.
The end of the life cycle of a vortex in this flow regime is shown on figure~\ref{fig:TYPO-diffusing-bd}, from the moment when the main vortex $V_1$ starts to lose its circulation ($t/T=3.0$) to its disappearance ($t/T=3.9$).
At the merging, the main vortex has a very low circulation and is absorbed by the secondary one without changing $V_2$ properties nor trajectory.

The evolution of the circulation and radius of the vortices is similar for all diffusing phases (see circulation evolution on figure~\ref{fig:TYPO-diffusing-bd}): the vortices increase in size in the upstream part of the shear layer and decrease in size in the downstream part.
This can be explained as vortices size and circulation evolution is the result of the balance between:
(i)~The vertical gradient of streamwise velocity due to the boundary layer separation, strong in the upstream part of the HSV, feeding the vortices.
(ii)~The opposite-sign vorticity generated at the wall that rolls around the vortices and decreases their circulation by vorticity diffusion \citep{seal_quantitative_1995}.
It is to be noted that the main vortex disappearance (by diffusion) and the vortex creation at the end of the separation surface are not necessarily simultaneous, resulting in a varying instantaneous number of vortices (for instance $3$ at $t/T=3.0$ or $4$ at $t/T=3.5$ for instance in figure~\ref{fig:TYPO-diffusing-bd}).

\subsubsection{Note on the vortex merging}
\label{sec:vortex-fusion}
Vortex merging appears in the merging and diffusing phases and is thus an important mechanism for the HSV dynamics.
Co-rotating vortex merging in uniform flows was described by~\citet{dritschel_quantification_1992,trieling_monopolar_1998,meunier_physics_2005,josserand_merging_2007}.
Those studies showed that, for 2 identical vortices, vortex merging occurs when the ratio of the vortices radius over their separation distance exceeds a threshold value.
For non-identical vortices, a general merging criterion is not established, as the merging has been shown to be governed by different processes (some of them leading to the destruction of the smaller/weaker vortex without increasing the larger vortex circulation, see~\citealp{dritschel_quantification_1992}).

The situation is more complicated in the present study, as the vortices are not two-dimensional, and are surrounded by a complex flow.
The vortex-vortex interaction match an elastic-like behavior:
two vortices start to interact as the distance separating them $\Delta x$ roughly equals the sum of their radii $\Delta x \approx R_1 + R_2$.
When getting closer, they repel each other, as can be seen in the oscillating phases when $V_1$ travels upstream and pushes $V_2$ in figure~\ref{fig:TYPO-Stable-Oscillating}c.
If they succeed in getting even closer, the two vortices merge, resulting in a briefly (regarding the time-scale of the HSV oscillations) disrupted vortex.
However, for vortices with very different sizes, as in the diffusing phases (see figure~\ref{fig:TYPO-diffusing-bd}), the weaker vortex is simply absorbed by the larger one, without any elastic-like behavior.

The appearance of this elastic-like behavior can be attributed to the restriction  of the vortex position along $y$ (due to the presence of the bed and the boundary layer) that prevents the vortices from rolling around each other, which is the typical behavior for vortices in uniform flows (see~\citealp{dritschel_quantification_1992} for instance).

\subsubsection{Breaking phase}
\label{sec:breaking-phase}
\begin{figure}
  \centering
  \includegraphics[width=1\figwidth]{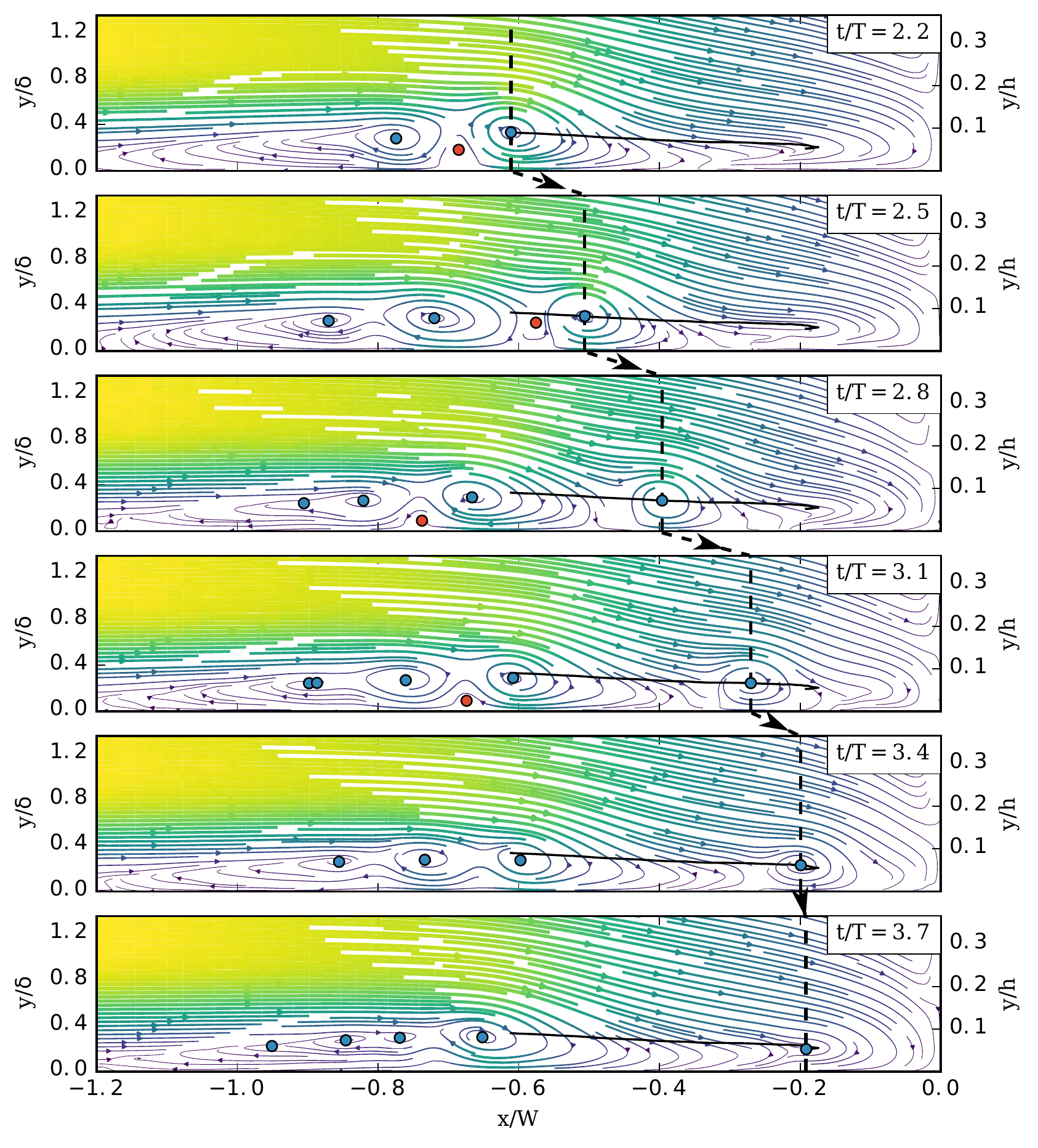}
  \caption{%
    Successive instantaneous velocity fields in the vertical plane of symmetry for a breaking phase in a complex regime ($\Rey_h=6397,~h/\delta=3.69,~W/h=1.67,~T=17.22s$).
    $t/T=0$ (not shown here) corresponds to the birth of the vortex of interest at the end of the separation surface at $x/W=-0.9$.
    See legend from figure~\ref{fig:TYPO-Merging-bd} for the velocity fields.
  }
  \label{fig:TYPO-Breaking}
\end{figure}

As no breaking regimes have been observed in the present work, the breaking phase presented in figure~\ref{fig:TYPO-Breaking} is part of a complex regime configuration, explaining why the flow pattern differs between $t/T=2.2$ and $t/T=3.1$.
This phase differs from a merging phase by the fact that the main vortex escapes from the HSV, travels toward downstream and diffuses near the obstacle.
Figure~\ref{fig:TYPO-Breaking} shows the end of a vortex life cycle in a breaking phase.
The highlighted vortex (main vortex at $t/T>2.2$) breaks from the HSV at $t/T \approx 2.8$ and is advected downstream at high velocity while loosing in radius and circulation until disappearing at $t/T=3.7$.
Unlike observed by~\citet{doligalski_vortex_1994}, in this case no velocity eruptions is at the origin of the breaking.
\subsubsection{Oscillating - merging regime transition}
\label{sec:stable-merg-regime}
\begin{figure}
  \centering
  \includegraphics[width=.75\figwidth]{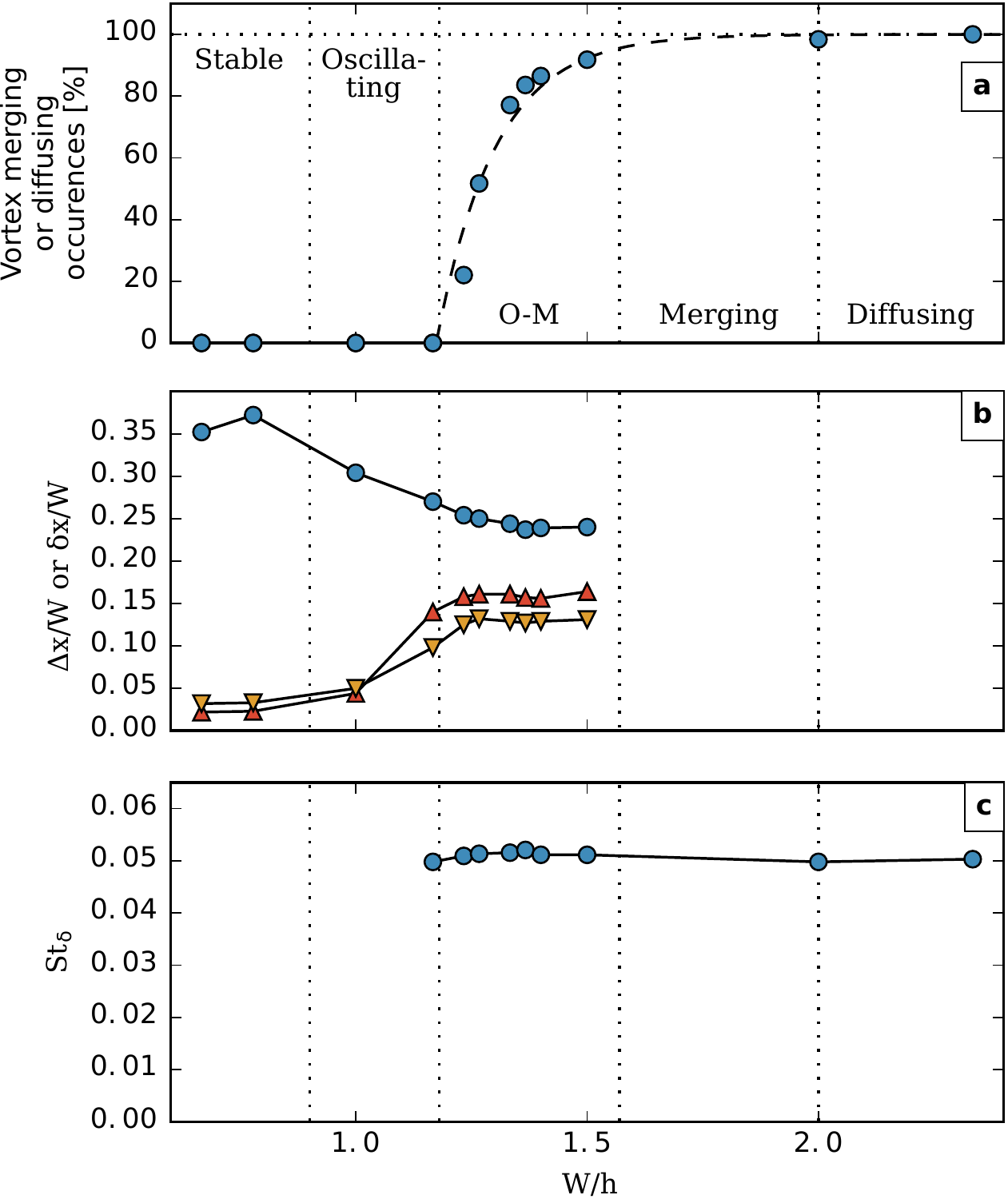}
  \caption{%
    Evolution of some HSV characteristics for $\Rey_h=4272,~h/\delta=2.7$ and increasing $W/h$ values, corresponding to the square symbol on figure \ref{fig:EXPE-Expe-plan}.
    (a) Percentage of merging or diffusing phase.
    (b) blue circles: mean distance between the main and the second vortices $\Delta x_{avg}$, red triangles: mean spatial oscillation amplitude of the main vortex $\delta x_1$, yellow unversed triangles: mean spatial oscillation amplitude of the secondary vortex $\delta x_2$.
    Those values are only defined, and consequently computed, for stable regimes and oscillating phases.
    (c) Strouhal number based on the boundary layer thickness $\delta$.
    These plots illustrate well the continuous transition from the oscillating to the merging regimes.
  }
  \label{fig:TRANS-O-M}
\end{figure}

While the transition from the stable to the oscillating regime is simply a continuous increase in vortex motion amplitude, the transition from the oscillating regime to the merging regime is more complex.
This transition is investigated more in details on figure~\ref{fig:TRANS-O-M} for different cases with increasing $W/h$ but constant $\Rey_h$ and $h/\delta$ values (corresponding to the square symbol on figure~\ref{fig:EXPE-Expe-plan}).

For $W/h < 0.9$, the main and secondary vortices are far from each other and remain steady, the HSV being in stable regime.
As $W/h$ increases (from $0.9$ to $1.2$), the main vortex starts to oscillate and the average distance between the main and secondary vortices decreases.
After a critical value of $W/h = 1.2$, $V_1$ and $V_2$ begin to merge for some periods \textit{i.e.}\ merging phases appear.
Occurrences of merging phases increase with $W/h$ until reaching the merging regime ($>95\%$ merging phases) at $W/h=1.6$.
The transition from oscillating to merging regimes (from $W/h=1.2$ to $W/h=1.6$) is asymmetrical:
The number of oscillation periods decreases rapidly for $W/h \approx 1.2$ and more slowly for $W/h \approx 1.6$ (see fitting in figure~\ref{fig:TRANS-O-M}a).
Oscillations amplitudes can only be measured for oscillating phases, which explain the saturation of the vortex spatial amplitude (for $W/h = 1.2$ to $1.5$ on figure~\ref{fig:TRANS-O-M}b):
this saturation value $\delta x/W = 0.16$ can be understood as the maximum possible oscillation amplitude before appearance of merging between $V_1$ and $V_2$ (Note that the critical value $\delta x/W=0.16$ is expected to differ regarding $h/\delta$ and $\Rey_h$ values).
A general parameter governing the vortex merging occurrences is, in this situation, quite challenging to define.

It is to be noted that the Strouhal number, presented in figure~\ref{fig:TRANS-O-M}c, is directly proportional to the oscillation frequency $f$ ($\delta$ and $u_D$ being kept constant as $W/h$ increase).
Consequently, figure~\ref{fig:TRANS-O-M}c only indicates that the obstacle aspect ratio $W/h$ and the HSV regime have no influence on the HSV oscillation frequency.
\subsubsection{Merging to diffusing regimes transition}
\label{sec:merg-diff-regim}
Contrary to the stable/oscillating/merging transition that is continuous, the merging to diffusing transition is more complex.

\begin{figure}
  \centering
  \includegraphics[width=.75\figwidth]{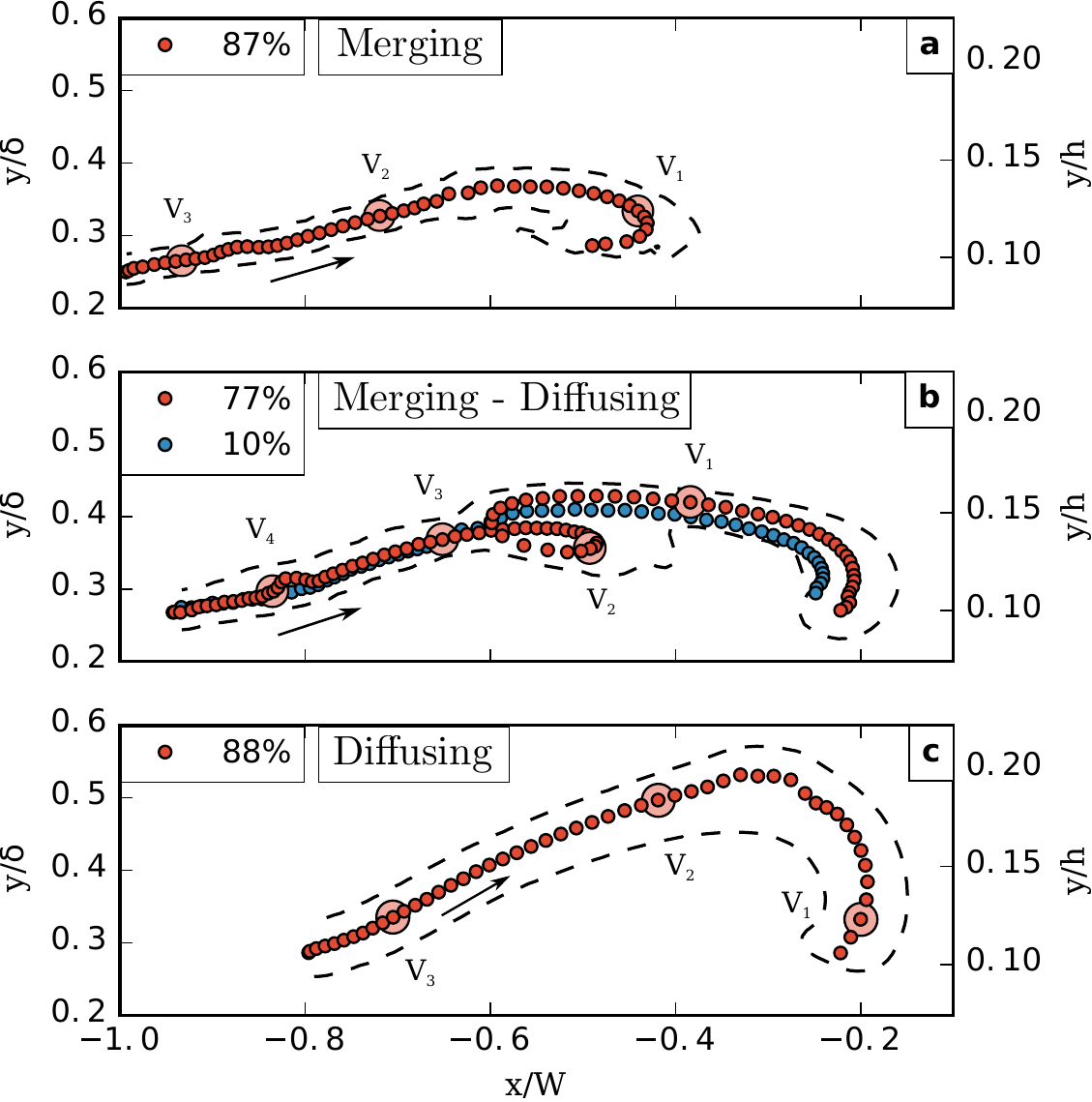}
  \caption{%
    Main vortex centers mean trajectories on the $x, y$ plane of symmetry for three cases with $\Rey_h=4272,~h/\delta=2.7$ and increasing $W/h$ (a: 1.51, b: 2.01, c: 2.45), corresponding to the square symbol on figure \ref{fig:EXPE-Expe-plan}.
    Dashed contours represent detected vortex centers envelopes.
    Each mean trajectory begins at the leftmost point.
    Percentages represent the percentage of trajectories used to compute each mean trajectories, the rest is not considered as it differs too much.
    Large circles are simultaneous vortex positions at a given arbitrary time.
    For visibility, aspect ratios are not conserved (trajectories representation is stretched along the $y/\delta$ axis).
    Mean trajectories are computed on approximately $200$ single trajectories.
    The discontinuous transition from merging to diffusing regimes is clearly visible on these plots with the increase of $W/h$.
  }
  \label{fig:TRANS-merging-diffusing}
\end{figure}

Figure~\ref{fig:TRANS-merging-diffusing} shows the evolution of mean trajectories with increasing values of $W/h$, while passing from a merging regime to a diffusing one.

The mean trajectory shown in figure~\ref{fig:TRANS-merging-diffusing}a is characteristic of a merging regime (see figure~\ref{fig:TYPO-Merging-bd}):
the vortex appears at the end of the separation surface ($x/W \approx -1$), is advected downstream while gaining in size and ends up going back upstream and merging with the secondary vortex.
This last part of the trajectory is not well reconstructed, because of the main vortex high velocity.

Oppositely, the trajectory shown in figure~\ref{fig:TRANS-merging-diffusing}c is characteristic of a diffusing regime (see figure~\ref{fig:TYPO-diffusing-bd}):
the vortex is advected downstream, but diffuses at $x/W \approx -0.2$ instead of going back upstream and merging with the previous vortex.

For the transitional case in figure~\ref{fig:TRANS-merging-diffusing}b, the majority ($77\%$, red trajectory) of the vortices appearing at the end of the separation surface merge with the previous vortices at $x/W \approx -0.6$ (merging phases).
However, after this merging, these vortices are advected further downstream and diffuse at $x/W = 0.25$ (diffusing phases).
Therefore, the HSV dynamics is ($77\%$ of the time) a regular alternation between merging and diffusing phases.
$10\%$ of the vortex (blue trajectory) are simply diffusing-like.
The remaining trajectories ($13\%$) differ from those mean trajectories.

Nevertheless, the observation of this bi-modal transition does not exclude the possibility of a more simple and continuous transition:
the main vortex being simply more and more diffused during a phase, as the transition from merging to diffusing regimes occurs.

\begin{figure}
  \centering
  \includegraphics[width=.5\textwidth]{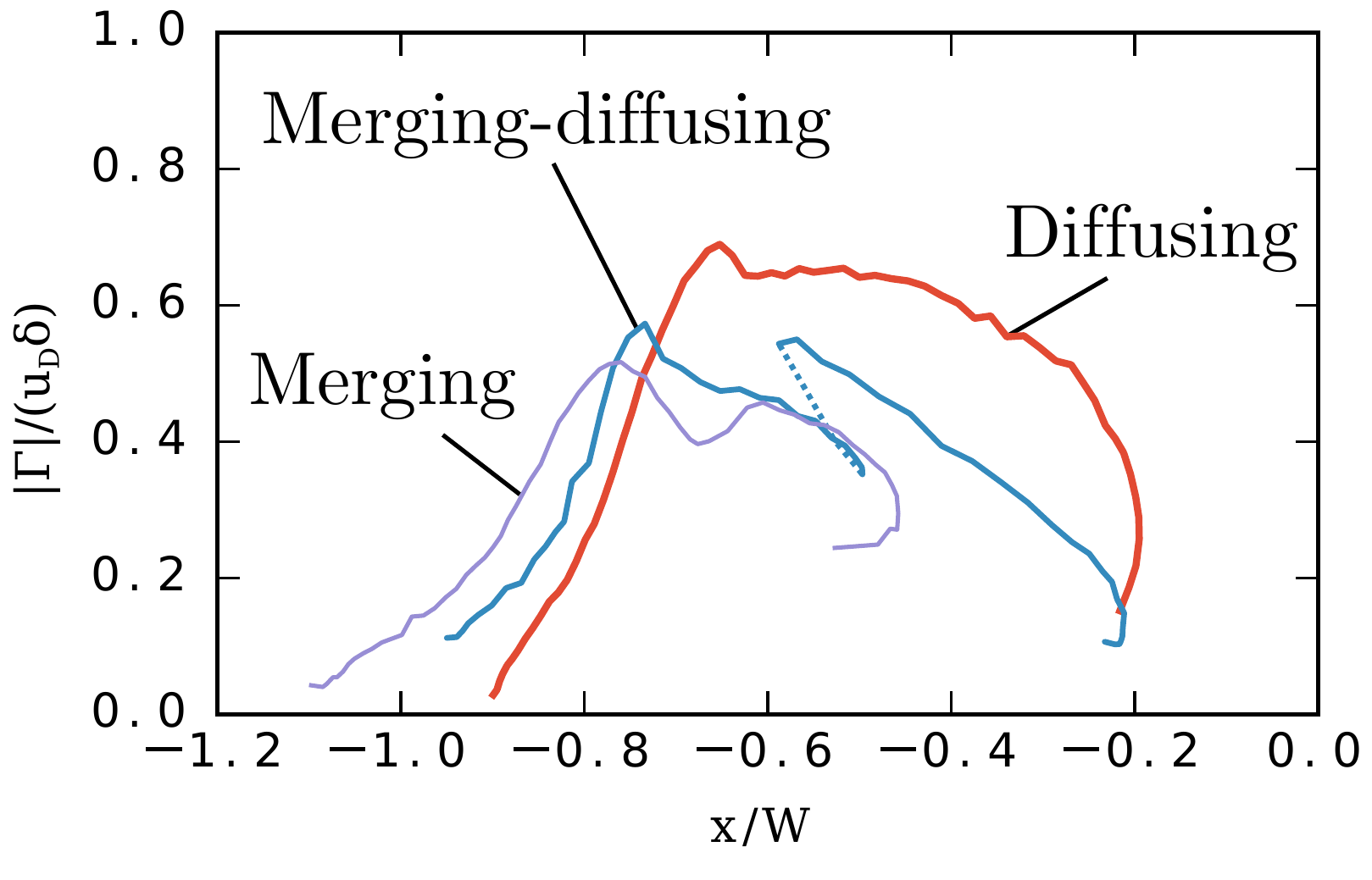}
  \caption{%
    Dimensionless vortex circulation evolution along the streamwise direction for the mean trajectories of figure~\ref{fig:TRANS-merging-diffusing}.
    These evolutions show another aspect of the merging-diffusing transition linked to the increase of the vortex circulation.
  }
  \label{fig:TYPO-merging-diffusing-circ}
\end{figure}

Circulation evolution for those three cases are presented on figure~\ref{fig:TYPO-merging-diffusing-circ}.
For the merging-diffusing case, the abrupt circulation increase at $x/W=-0.6$ corresponds to the merging of the main vortex with the secondary one, and separates the vortex trajectory in two parts:
(i) The first part, quite similar to the merging case: the vortex gains in circulation from $x/W \approx -1$ to $x/W\approx -0.8$, and looses in circulation further downstream.
(ii) The second part, quite similar to the diffusing case: the vortex circulation decreases while approaching the obstacle, reaching a very low circulation ($\Gamma/(u_D \delta) \approx 0.1$) at $x/W\approx-0.2$.
At the beginning of the second part, the circulation of the vortex is the same as at $x/W=-0.75$, showing that the diffusing behavior occurrences are not only related to high circulations.
\subsubsection{Complex regime}
\label{sec:complex-regime}
\begin{figure}
  \centering
  \includegraphics[width=.75\figwidth]{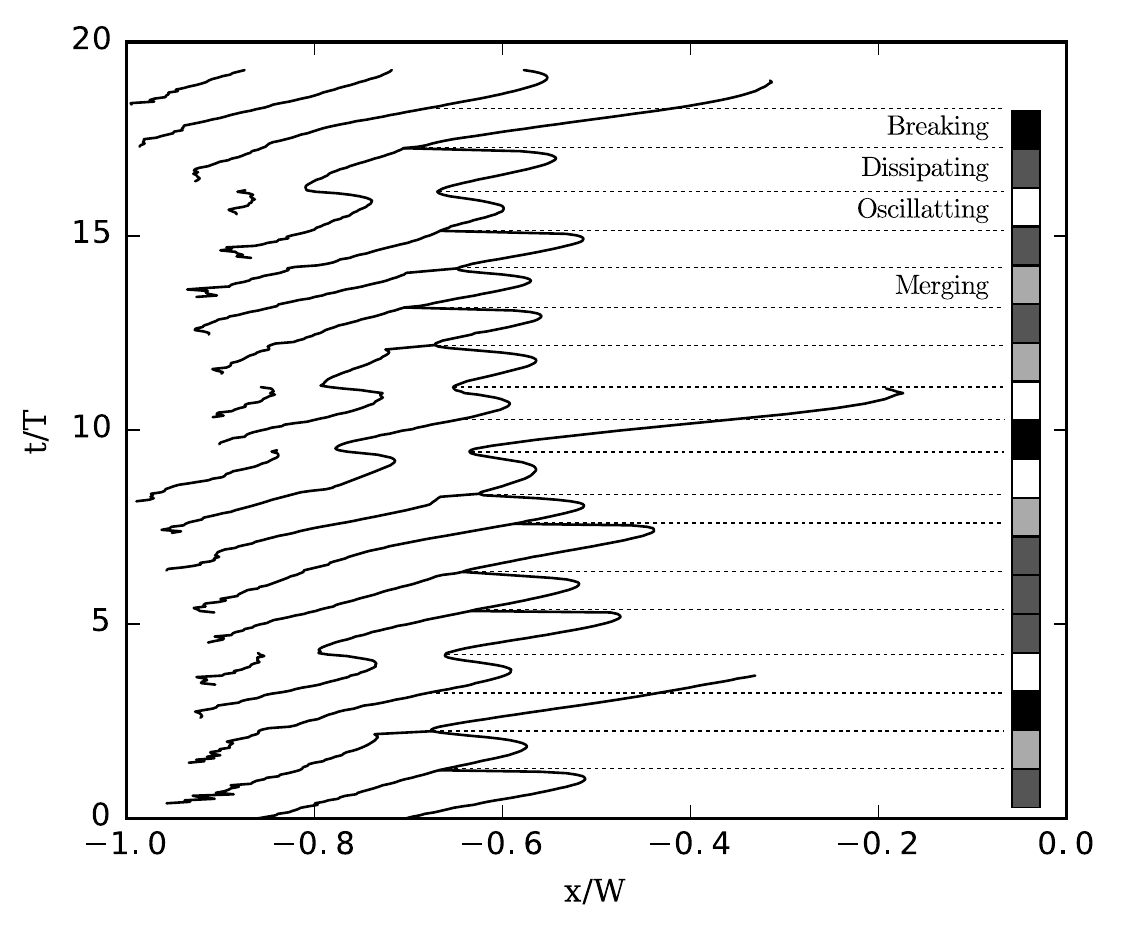}
  \caption{%
    Time evolution of the streamwise vortex center positions for a complex regime ($\Rey_h=6397,~W/h=1.67,~h/\delta=3.69,~T=17.22s$).
    Identified phase types are presented on the right panel.
    Dashed lines highlight the initials states, here the upstream most position of the main vortex $V_1$.
    The unordered phase succession illustrates the chaotic behavior of the complex regime.
  }
  \label{fig:TYPO-Complex-col}
\end{figure}

An example of vortex center trajectory for a complex regime is presented on figure~\ref{fig:TYPO-Complex-col}, where oscillating, merging, diffusing and breaking phases alternate during $20$ consecutive periods.
It has been verified by additional measurements upstream of the obstacle (not shown here) that the phase alternation is not provoked by perturbations from outside the HSV.\@
No particular order can be seen in the phases organizations, or in the main vortex maximum and/or minimal positions.

In the case of oscillating, merging or diffusing regimes, the initial and final states of each phase are very close, \textit{i.e.\ } each phase brings the HSV back in its initial state.
The phase behavior being governed by the initial state, successive phases have no particular reasons to differ from each other.
In the case of the complex regime, each phase final state slightly differs from its initial state, leading to an alternation of different phase types.

With this point of view, the HSV dynamics can be considered as a dynamic system linking the next initial state to the current one:
\begin{equation}
  \phi_{n+1} = F(\phi_{n})
  \label{eq:chaotic-dyn-syst}
\end{equation}
with $\phi_n$ the $n\textsuperscript{th}$ initial state and $F$ the function representing the system dynamics.
This system is expected to admit stable equilibrium positions for stable, oscillating, merging, diffusing and breaking regimes, to be chaotic for the complex regime and to undergo dynamic bifurcations leading to complex transitional behaviors, as it is the case for the transitional regime between the fusion and the diffusing regimes (figure~\ref{fig:TRANS-merging-diffusing}b).
Further, studying this dynamical system properties would require measuring at least $1000$ consecutive phases, which represents approximately $20000$ instantaneous velocity fields and is out of the scope of the present study.

\subsection{Irregular Regimes}
\label{sec:irregular-regimes}
\subsubsection{Coherent to irregular transition}
\label{sec:coher-irreg-trans}
From the previously defined coherent regimes, the HSV can evolve to an irregular, aperiodical, turbulent-like state, despite the boundary layer remaining in a laminar state.
This transition is characterized by the appearance of small-scale non-coherent structures.
Transitional cases (between coherent and irregular regimes) will be said in ``semi-irregular regime'', and consist of time periods of coherent regime, punctually disturbed by eruptions of small-scale perturbations.
Such transitions have been observed for oscillating, merging and complex regimes, as shown in figure~\ref{fig:TYPO-Regimes}.
The fact that the irregular transition occurs for different coherent regimes make necessary the two-dimensional typology presented on figure~\ref{fig:TYPO-Regimes}, separating the coherent and the irregular evolutions.
This distinction was not included in any previous work~\citep{schwind_three_1962, baker_laminar_1978, greco_flow_1990, lin_characteristics_2008}, where typologies remained strictly one-dimensional.

\begin{figure}
  \centering
  \includegraphics[width=1\figwidth]{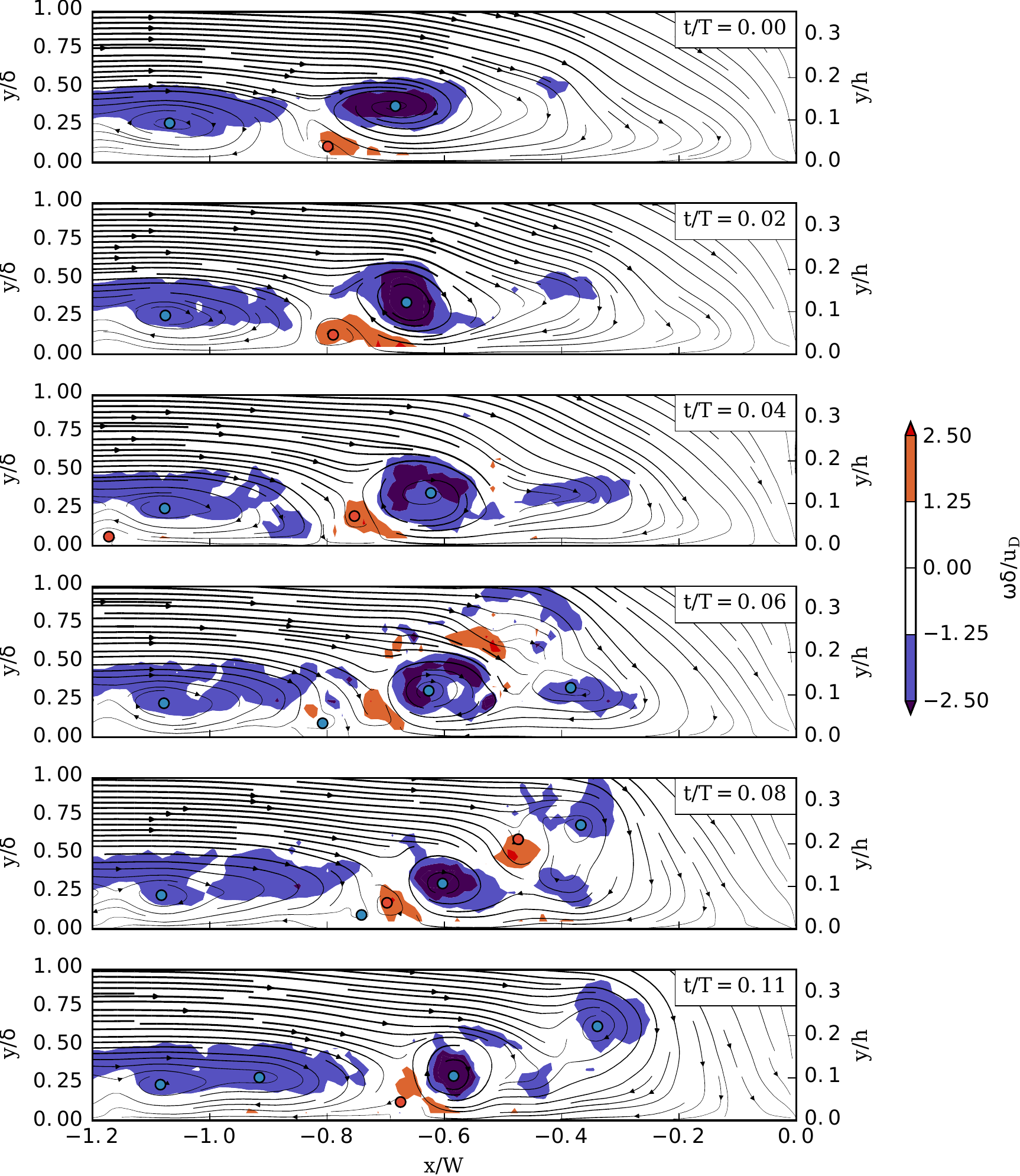}
  \caption{%
    Successive instantaneous streamlines and vorticity contours in the vertical plane of symmetry for a semi-irregular complex regime ($\Rey_h=6395$, $h/\delta=2.81$, $W/h=1.36$, $T=16.37s$), illustrating the first type eruption.
    Circles represent detected critical points.
    This set of instantaneous flows illustrate well the first type eruption, characterized by the apparition of negative vorticity above and downstream of the main vortex.
  }
  \label{fig:TYPO-Irregular-transition}
\end{figure}

The eruptions of small scale non-coherent structures are of two types.
First type eruptions are linked to the appearance of positive vorticity above the main vortex $V_1$.
Figure~\ref{fig:TYPO-Irregular-transition} illustrates this phenomenon and shows small-scale structures appearing above and downstream of the main vortex at $t/T=0.06$, and provoking the appearance of positive vorticity of the same order of magnitude (but opposite sign) as the main vortex vorticity.
This positive vorticity is then advected around the main vortex from $t/T=0.06$ to $0.08$ and ends up creating a new vortex downstream from $V_1$ at $t/T = 0.11$.
Second type eruptions are linked to the merging between two vortices, resulting in a destabilized main vortex (not shown here).

The bi-modal behavior~\citep{devenport_time-depeiident_1990}, previously reported for fully turbulent HSV, has not been clearly identified for irregular nor semi-irregular regimes, even if zero-flow and back-flow occurrences can be seen episodically.
The conclusion may be that the bi-modal behavior mainly occurs for higher Reynolds numbers, such as those used in studies devoted to this phenomenon ($\Rey_W=115000$ for~\citet{devenport_time-depeiident_1990} and~\citet{paik_bimodal_2007}, $\Rey_W=20000$ and $39000$ for~\citet{escauriaza_reynolds_2011}, while $\Rey_W < 4950$ in the present study).

\subsubsection{Instabilities origins}
\label{sec:inst-orig}
Several hypotheses can be put forward to explain the appearance of turbulent-like (small-scale and non-coherent) structures in the HSV and, consequently, its transition to an irregular regime.
The first possible origin can be the boundary layer transition to turbulence.
Turbulence bursts coming from the upstream boundary layer can certainly destabilize the HSV and break its periodicity.
Yet, no perturbation in the upstream boundary layer could be observed herein, thus excluding this hypothesis.

Second, the interaction between proximate vortices can lead to the appearance of an elliptical instability~\citep{kerswell_elliptical_2002} in the core region of the vortices. This interaction cannot be the origin of the first type eruption (figure~\ref{fig:TYPO-Irregular-transition}), where the instability originates from above the vortex, but is the best candidate to explain instabilities resulting from merging vortices (second type eruption).

Third,~\citet{escauriaza_reynolds_2011} show that eruptions of vorticity from the wall, responsible for the bi-modal behavior of fully turbulent HSV, can be understood as thee-dimensional Görtler instabilities~\citep{floryan_gortler_1986}.
In such scenario, the vorticity of the first counter-rotating vortex $V_{c1}$ punctually wraps around the main vortex $V_1$ and destabilizes it.
While the first type eruption in figure~\ref{fig:TYPO-Irregular-transition} could agree with this definition, there is no evidence that the positive vorticity, appearing at $t/T=0.06$, originates from the counter-rotating vortex.
A definitive conclusion on the first type eruption origin cannot, unfortunately, be drawn without proper 3D information on the flow, out of the scope of the present work.

\subsection{Regimes evolution with flow parameters}
\label{sec:regim-evol-with}
\begin{figure}
  \centering
  \includegraphics[width=1.\textwidth]{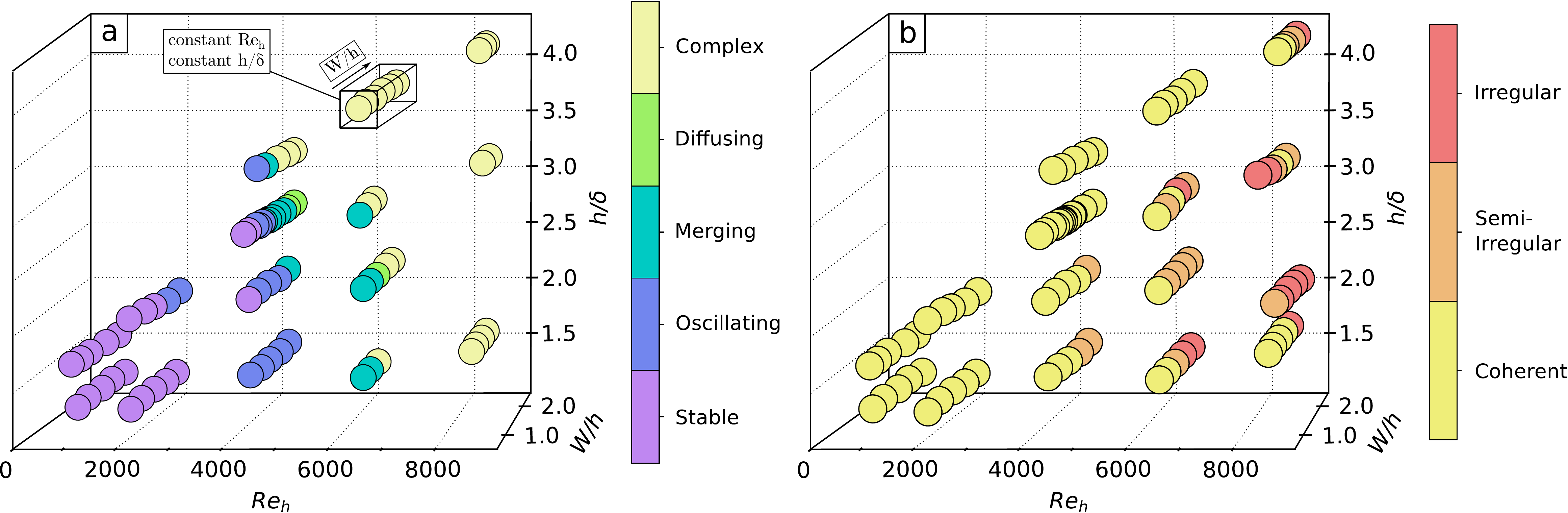}
  \caption{%
    (a) Coherent regimes evolution as a function of the three dimensionless parameters.
    Each circle, representing a measured flow, is coloured according to the observed HSV regime.
    (b) Irregular regimes evolution.
    As coherent regimes cannot be defined on irregular regimes configurations, irregular HSV flows are not represented on (a).
    These evolutions establish well the dependence of the HSV coherent regimes to the three dimensionless parameters, and the main dependence of the irregular regimes to the Reynolds number.
  }
  \label{fig:TYPO-stab_all}
\end{figure}

The evolution of the flow regimes with the dimensionless flow parameters is presented on figure~\ref{fig:TYPO-stab_all}, separately for the coherent (a) and the irregular (b) evolutions.

Figure~\ref{fig:TYPO-stab_all}a clearly shows that the coherent regimes depend on the three dimensionless parameters $h/\delta$, $W/h$ and $\Rey_h$:
an increase of any one of them leads to a destabilization of the HSV, potentially modifying the regime toward the complex one (toward the top in figure~\ref{fig:TYPO-Regimes}).
The influence of $\Rey_h$ is the most obvious for the studied domain.

The interpretation of the parametric dependencies of the irregular evolution (figure~\ref{fig:TYPO-stab_all}b) is more challenging.
Influence of the Reynolds number $\Rey_h$ and the aspect ratio $W/h$ is clearly visible, but the influence of $h/\delta$ remains unclear.

\begin{figure}
  \centering
  \includegraphics[width=.75\figwidth]{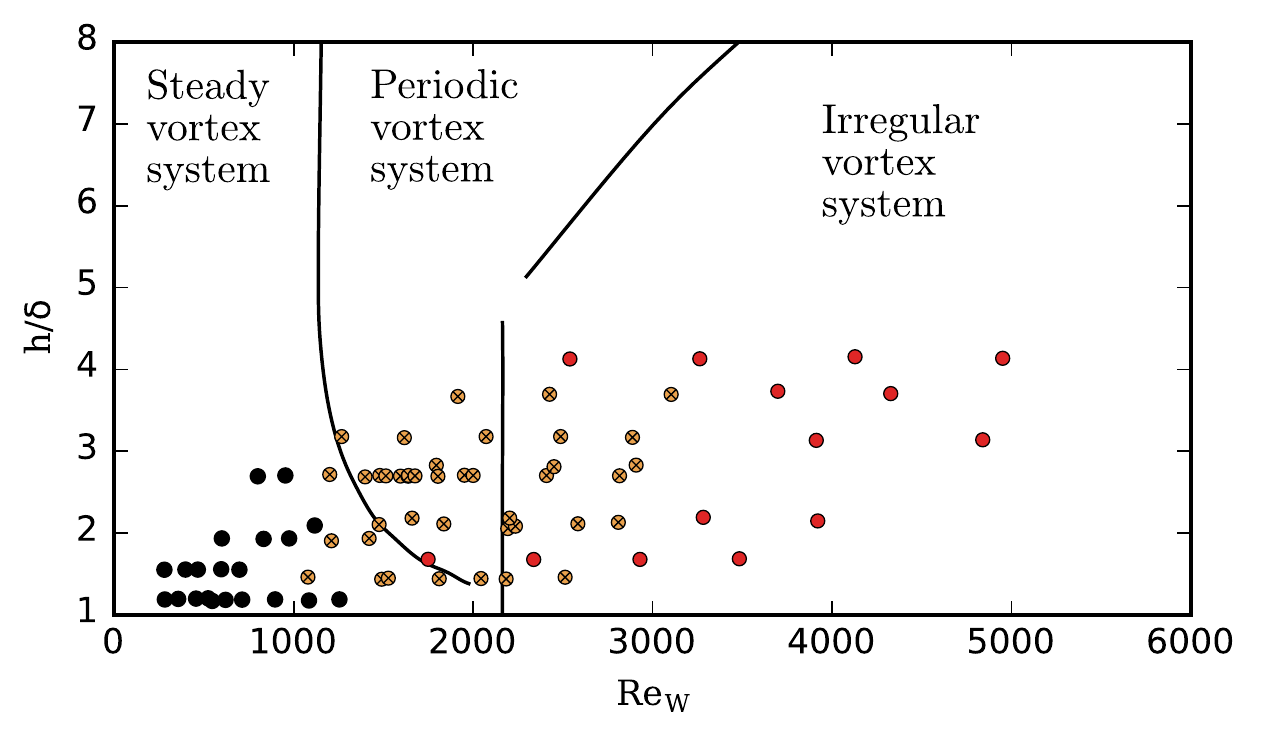}
  \caption{%
    HSV regimes evolution comparison with typology from~\citet{lin_characteristics_2008}.
    Each point represents a flow of the present study, classified according to~\citet{lin_characteristics_2008} typology: filled symbols for the steady vortex system (equivalent to the stable regime), cross-filled (yellow) symbols for periodic vortex system (equivalent to the oscillating, merging and diffusing regimes) and hollow symbols (red) for irregular vortex system (equivalent to the irregular regime).
    Lines represent regime boundaries reported by~\citet{lin_characteristics_2008} for immersed obstacles, and adapted with the equivalent diameter method (section~\ref{sec:comp-betw-vari}).
  }
  \label{fig:TYPO-Comp-lin}
\end{figure}

Figure~\ref{fig:TYPO-Comp-lin} compares the evolution of the HSV regimes between an immersed obstacle configuration from~\citet{lin_characteristics_2008} and the present emerging obstacle configuration.
This figure shows an overall agreement with the regimes boundaries of~\citet{lin_characteristics_2008}, with a main dependency to $\Rey_W$.
Nevertheless, the boundary between steady and periodic vortex system is not well reproduced, and the transition from periodic to irregular regimes appears at higher $\Rey_W$ in the emerging obstacle configuration.
As the periodic and irregular regimes overlap, the couple of dimensionless parameters used by~\citet{lin_characteristics_2008} ($\Rey_W$ and $h/\delta$) should not be the leading parameters of the regime evolution for the emerging obstacle configuration.

\section{Mechanisms behind the horseshoe vortex dynamics}
\label{sec:mech-behind-hors}
The previous section presented the dependence of the HSV dynamics to the flow dimensionless parameters, but few is known concerning the physical mechanisms at the origin of these transitions.
This section proposes a model which purpose is to identify the main mechanisms behind the HSV periodic motion detailed in the previous section.
After a brief note on the HSV vortex appearance in section~\ref{sec:vortex-creation}, a semi-empirical correlation for the main vortex velocity is proposed in section~\ref{sec:vortex-motion} and on top of this, a model for the HSV dynamics is proposed in section~\ref{sec:vort-dynam-repr}.
Results obtained with this model are presented in section~\ref{sec:numer-simul-results} and allow to draw conclusions on the leading mechanisms of the HSV dynamics.
The reader is reminded that the presented model is not an attempt to obtain a predictive model for the HSV dynamics, but is rather to gain information on the HSV dynamics main mechanisms.

\subsection{Vortex creation}
\label{sec:vortex-creation}
\begin{figure}
  \centering
  \includegraphics[width=.75\figwidth]{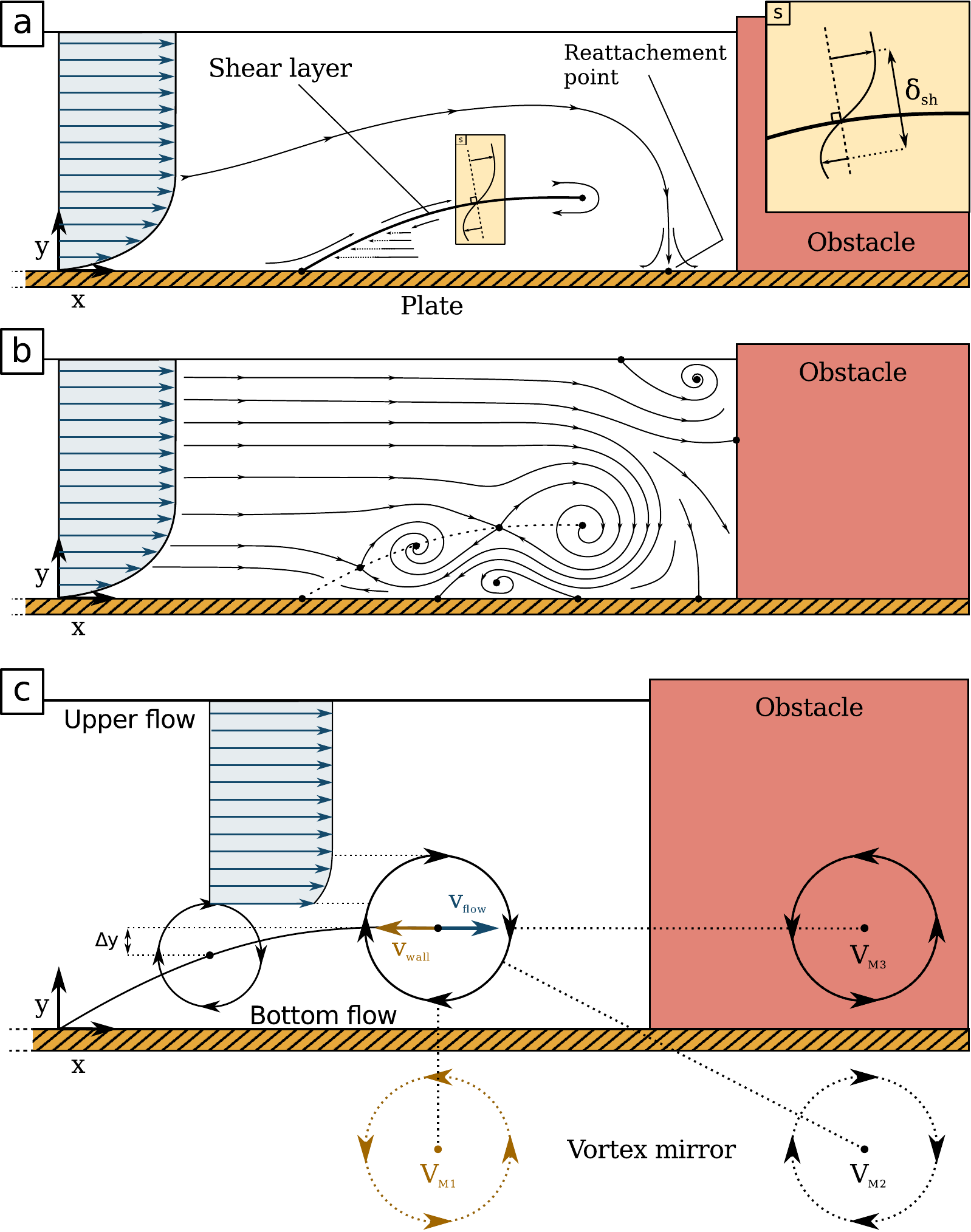}
  \caption{%
    (a) Schematic representation of the boundary layer separation in the vertical plane of symmetry and the resulting shear layer at low $\Rey_{sh}$ (\textit{i.e.}\ in no-vortex regime).
    (b) Diagram showing the vortex formation on the shear layer at moderate $\Rey_{sh}$ (here in stable regime).
    (c) Illustrative diagram for the model on the main vortex velocity $v_{adv}$, detailed in section~\ref{sec:vort-dynam-repr}.
  }
  \label{fig:TRANS-Vortex-formation}
\end{figure}

Because of its curvature and the strong three-dimensionality of the flow, the shear layer differs from the classical straight shear layer (see for instance~\citealt{wygnanski_two-dimensional_1970}).
Assumption can be made that it still behaves qualitatively like a classical shear layer and that its stability is governed by the shear layer Reynolds number:
\begin{equation}
  \Rey_{sh} = \frac{\Delta U \delta_{sh}}{\nu}
  \label{eq:Re_cis_a}
\end{equation}
with $\Delta U$ the velocity difference between the outer flow on both sides of the shear layer and $\delta_{sh}$ the shear layer thickness.
For the present shear layer (see figure~\ref{fig:TRANS-Vortex-formation}a), one can estimate an upper bound for the associated Reynolds number as:
\begin{equation}
  \Rey_{sh, \max} = \frac{2 u_D y_{sh}}{\nu}
  \label{eq:Re_cis_b}
\end{equation}
where $y_{sh}$ is the elevation of the downstream end of the shear layer.
In the present study, $\Rey_{sh, \max}$ ranges from $105$ to $1363$.

By analogy with classical shear layers, different dynamics behavior are expected depending on the $\Rey_{sh}$ values:
(i)~For low $\Rey_{sh}$ ($\Rey_{sh} < 55$ for classical shear layers), the shear layer should be stable and laminar~\citep{bhattacharya_critical_2006}, and the HSV should exhibit no vortex, leading to the no-vortex regime (see figure~\ref{fig:TYPO-Regimes} and figure~\ref{fig:TRANS-Vortex-formation}a).
\citet{schwind_three_1962} observed this no-vortex regime for $\Rey_{sh,  \max} = 78$.
(ii)~For high Reynolds numbers ($\Rey_{sh} > 10^4$ for classical shear layers), the shear layer and the HSV should be fully turbulent~\citep{dimotakis_mixing_2000}.
This has not been observed in this study, as other instabilities appear before reaching such $\Rey_{sh}$ values.
(iii)~For moderate Reynolds numbers ($55 < \Rey_{sh} < 10^4$ for classical shear layers), coherent, large-scale vortices should be generated periodically in the shear layer and advected downstream (see~\citet{loucks_velocity_2012} and~\citet{mignot_coherent_2016} for $\Rey_{sh}$ up to respectively $9700$ and $5000$).
This range is in fair agreement with the present measured configurations ($Re_{sh, max} \in [105, 1363]$).
One main difference between the classical shear layer and this curved shear layer is the fact that HSV can show steady vortices (stable regime) instead of continuously advected downstream vortices.

Nevertheless, no correlation could be obtained between the estimated $\Rey_{sh, max}$ and the HSV typology, promoting the hypothesis that the vortex dynamics and periodical behavior are not directly linked to the shear layer vortex shedding.

\subsection{Vortex motion}~\\
\label{sec:vortex-motion}
The velocity of a vortex along the shear layer depends on the surrounding flow.
For the main vortex $V_1$, which is supposed to govern the HSV dynamics, its advection velocity will depend on:
(i)~The wall influence, whose induced velocity on $V_1$ can be estimated using the vortex mirror concept~(\citealt{doligalski_vortex_1994}, see figure \ref{fig:TRANS-Vortex-formation}c) as:
\begin{equation}
  v_{wall} = - \frac{\Gamma}{4\upi \overline{y_{1}} u_D}
  \label{eq:vortex_model_wall}
\end{equation}
with $\Gamma$ the main vortex circulation and $\overline{y_1}$ the average location of the main vortex along $y$.
Note that only the bed mirror vortex ($V_{M1}$ on figure~\ref{fig:TRANS-Vortex-formation}c) is taken into account, as the two others ($V_{M2}$ and $V_{M3}$) are much further and have negligible influences on $V_1$.
(ii)~The secondary vortex $V_2$ influence, which depends on the relative position of the two vortices ($\Delta x$, $\Delta y$).
(iii)~The global state of the flow induced by the boundary layer separation, which should be constant for a given configuration and depend only of the dimensionless parameters $\Rey_h$, $h/\delta$ and $W/h$.

An empirical correlation for the main vortex velocity $v_{adv}$ can finally be found by using measurements of all these parameters instantaneous values over stable and oscillating regimes (using mean trajectories and their associated velocity fields):
\begin{equation}
  \frac{v_{adv}}{u_D} =
  \underbrace{- \frac{\Gamma}{4\upi \overline{y_{1}} u_D}}_{v_{wall}} +
  \underbrace{\left[1.19\frac{\Delta y}{\overline{y_{1}}}  - 1.15\left(\frac{\Delta y}{\overline{y_{1}}}\right)^2\right] \left(\frac{h}{\delta}\right)^{3.73} \left(\frac{\overline{y_{1}}}{h}\right)^2}_{v_{flow}}
  \label{eq:vortex_model_conv}
\end{equation}
with $R^2=0.89$ and $v_{flow}$ the influence of the flow, aggregating points (ii) and (iii).
The effect of $\overline{y_{1}}/h$ on the main vortex velocity can be explained by the upper flow contraction caused by the HSV, which provokes an increased $v_{adv}$.
The effect of $h/\delta$ accounts for the impact of the shear layer shape (governed by $\delta$ and the down-flow).
Note that for a given flow configuration, the only time-varying parameters are $\Delta y$ and $\Gamma$.

\begin{figure}
  \centering
  \includegraphics[width=.85\textwidth]{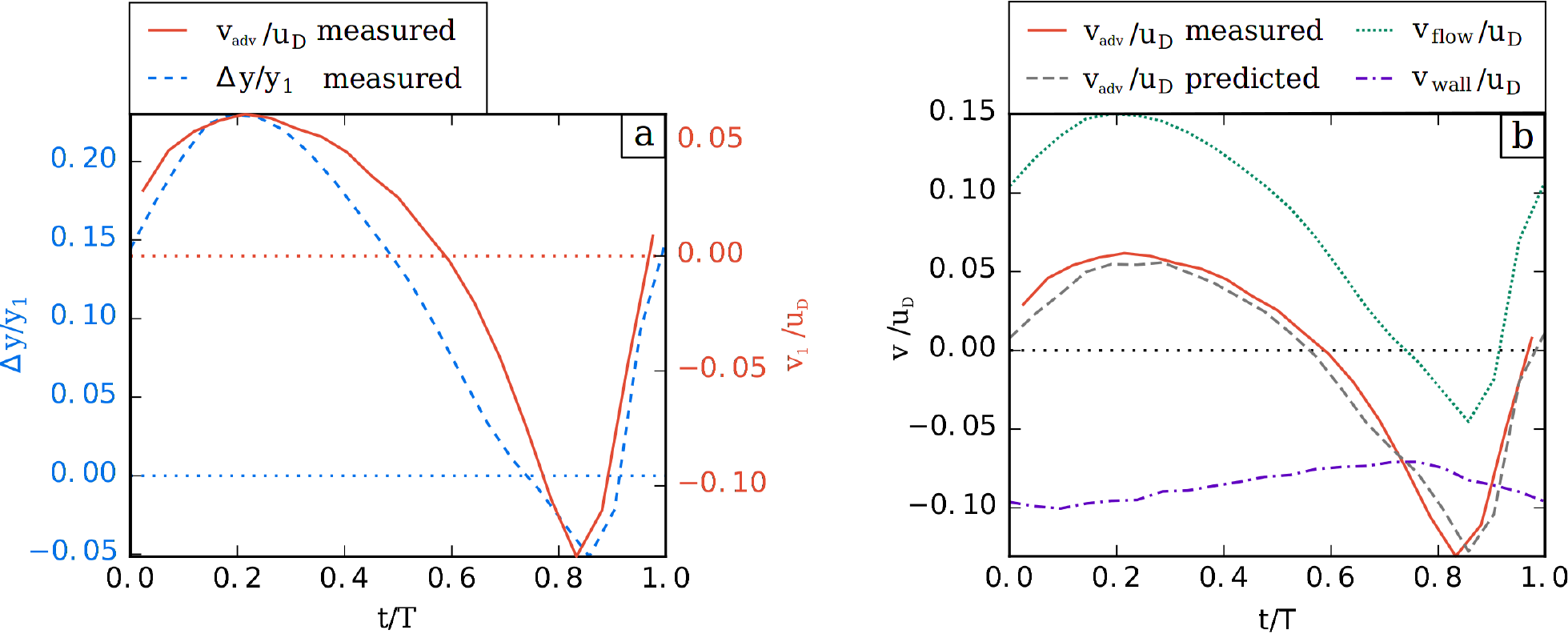}
  \caption{%
    Evolution of different components of the correlation~\ref{eq:vortex_model_conv} on one selected period of the oscillating case presented in figure~\ref{fig:TYPO-Stable-Oscillating}.
    (a) Comparison between the measured main vortex velocity $v_{adv}$ and the measured difference of altitude between the main and the secondary vortex $\Delta y$, showing the strong correlation existing between these two parameters.
    (b) Comparison between the two components of correlation~\ref{eq:vortex_model_conv} ($v_{flow}$ and $v_{wall}$) and the main vortex measured velocity $v_{conv}$, showing the good agreement obtained for the main vortex velocity correlation.
    It is to be noted that for a given flow, the sole non-constant parameters in the correlation~\ref{eq:vortex_model_conv} are $\Delta y$ and $\Gamma$.
  }
  \label{fig:MOD-params-evol-on-M290}
\end{figure}

This correlation is illustrated in figure~\ref{fig:MOD-params-evol-on-M290} for the oscillating regime previously presented in section~\ref{sec:oscillating-regime}.
Figure~\ref{fig:MOD-params-evol-on-M290}a presents the strong correlation between the main vortex velocity $v_{conv}$ and the difference of altitude between the main and the secondary vortex $\Delta y$, confirming the necessity to take this parameter into account in relation~\ref{eq:vortex_model_conv}.
Figure~\ref{fig:MOD-params-evol-on-M290}b presents the comparison between the measured main vortex velocity $v_{conv}$ and its prediction using the correlation~\ref{eq:vortex_model_conv}, along with the two components of the predicted main vortex velocity: $v_{flow}$ and $v_{wall}$.
The good agreement allows to pass to the next step.

\subsection{Vortex dynamics reproduction}
\label{sec:vort-dynam-repr}
To identify the physical phenomenon at the origin of the HSV dynamics, a numerical model is established based on the followings hypotheses:
(i)~The vortices can travel upstream and downstream but remain along the shear layer, which shape is taken from measurements.
(ii)~The main vortex instantaneous velocity $v_{adv}$ is estimated by equation~\ref{eq:vortex_model_conv}.
(iii)~The secondary vortex velocity $v_{2,adv}$ is equal to the velocity of the main one $v_{adv}$.
(iv)~The initial vortex locations are their measured equilibrium or mean positions plus a perturbation.
(v)~The vortices velocity are initially set to zero.

Numerical simulations using this model succeed in replicating the vortex equilibrium position for stable regime configurations but do not exhibit a periodic behavior for the oscillating regime configurations (not shown here).
One missing characteristic of the HSV dynamics that may explain the lack of periodicity is the delay $\Delta t$ between $V_1$ and $V_2$ motion (previously discussed in section~\ref{sec:oscillating-regime}, and presented on figure~\ref{fig:TYPO-Stable-Oscillating}).

\begin{figure}
  \centering
  \includegraphics[width=.75\figwidth]{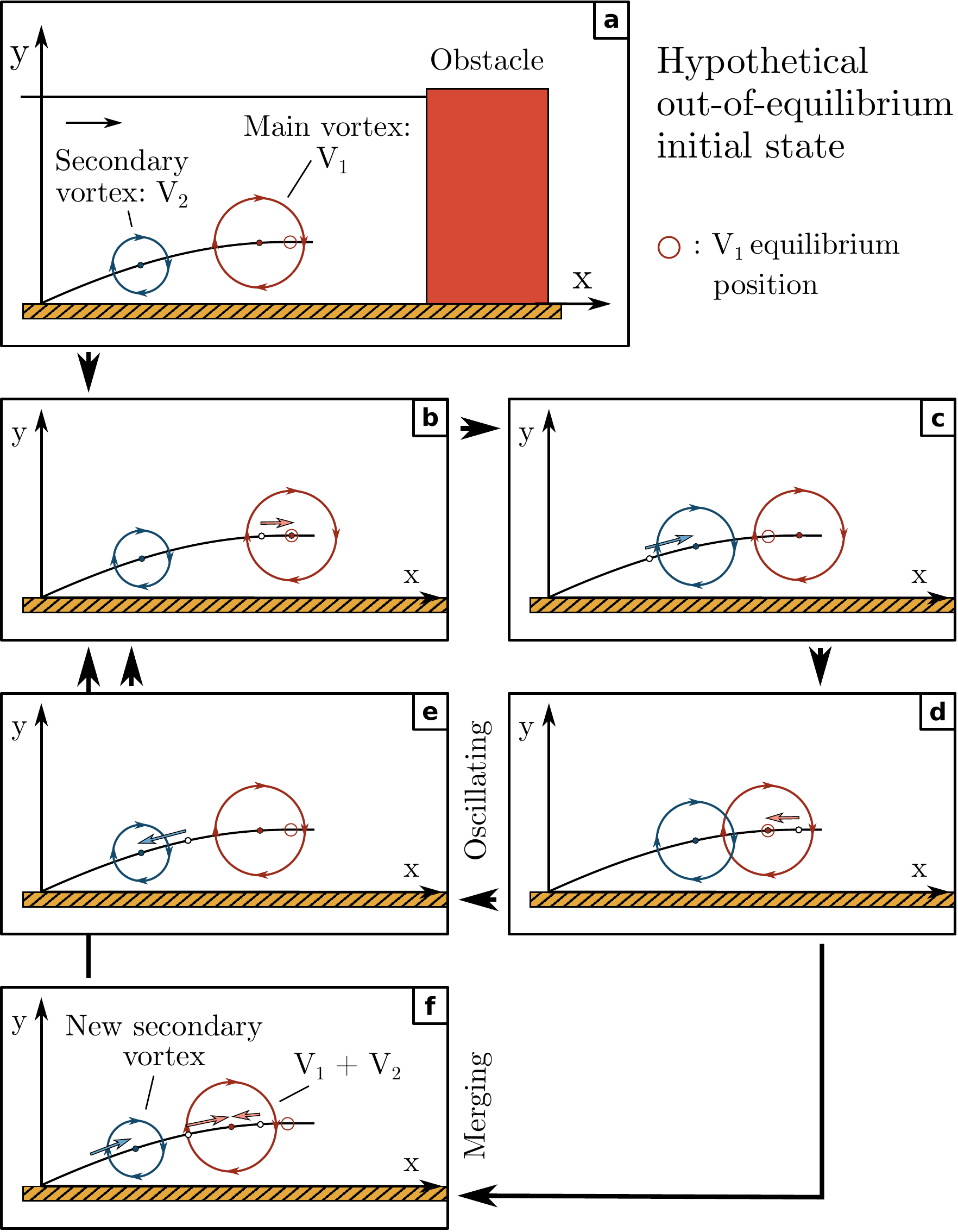}
  \caption{%
    Schematic illustration of how the motion delay between the main and the secondary vortices can lead to an oscillating behavior (explained step by step in section~\ref{sec:vort-dynam-repr}).
  }
  \label{fig:TRANS-Model-dyn}
\end{figure}

Figure~\ref{fig:TRANS-Model-dyn} shows a scenario that illustrates how the complexity added by this delay can lead to a periodic behavior:
(a) Vortices are placed on the shear layer, the main vortex being placed upstream of its equilibrium position.
(b) The main vortex naturally goes towards its position of equilibrium, according to the equation~\ref{eq:vortex_model_conv} while the secondary vortex remains in place, due to the delay (exaggerated for the sake of this demonstration).
(c) The secondary vortex, after a delay of $\Delta t$, moves downstream, decreasing the $\Delta y$ value and thus moving upstream the main vortex equilibrium position.
(d) The main vortex moves towards its new equilibrium position while the secondary vortex stays still due to the delay.
(e) The secondary vortex, after a delay $\Delta t$, moves upstream, increasing the $\Delta y$ value, pushing downstream the equilibrium position and bringing the vortex system back to its initial state (similar to (a)).
(f) If the main and the secondary vortices come close enough to each other in (d), they merge, the secondary vortex is replaced by a new one from upstream, and the vortex system recovers its initial state (similar to (a)).

\begin{figure}
  \centering
  \includegraphics[width=.5\textwidth]{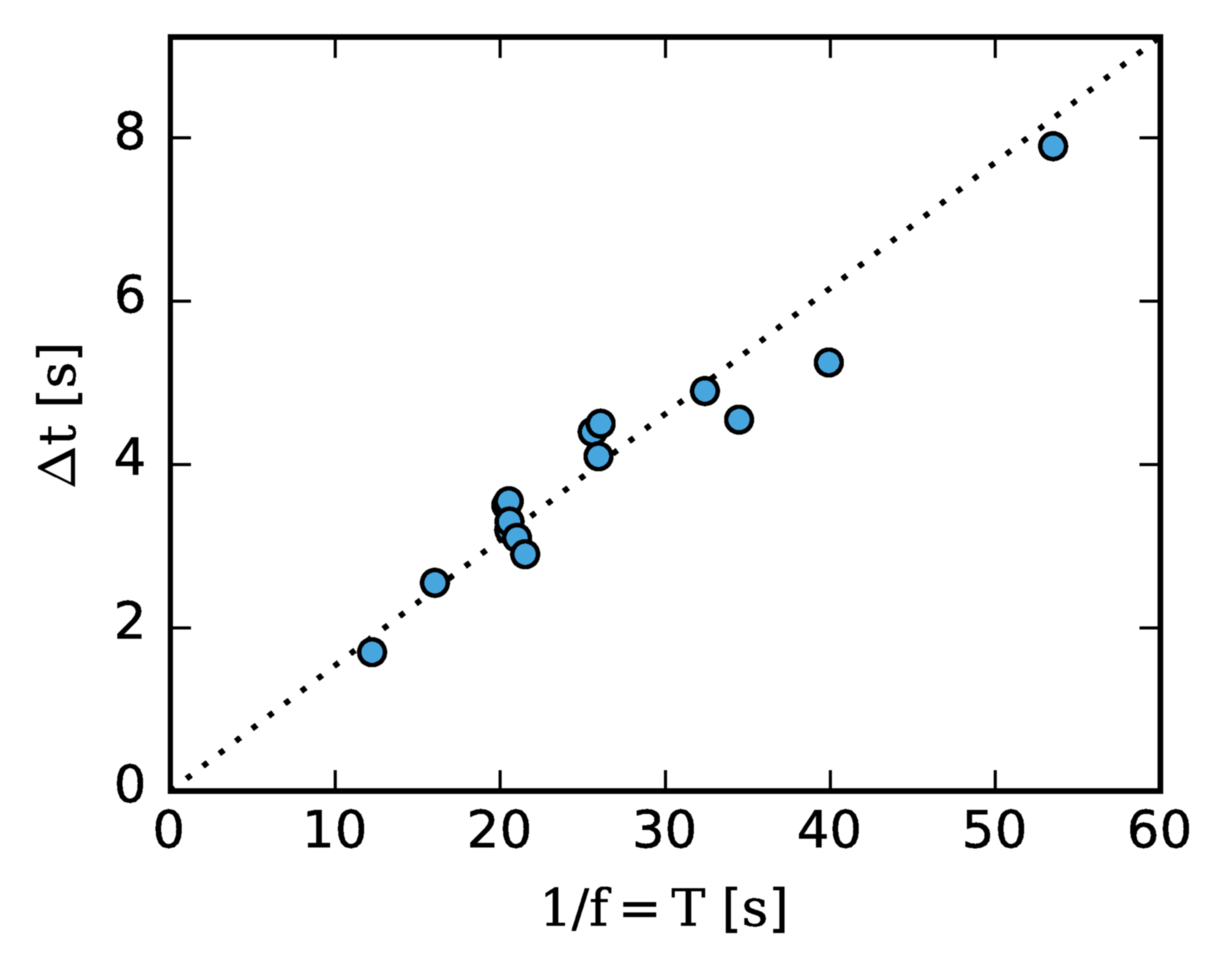}
  \caption{%
    Evolution of the averaged measured delay $\Delta t$ between the main and the secondary vortices motions according to the measured oscillation period $T$ for $15$ configurations ($\Rey_h \in [2113, 6406]$, $h/\delta \in [1.9, 3.8]$, $W/h \in [0.46, 1.00]$, in oscillating, merging or complex regimes).
    Dotted line is the best linear correlation (equation~\ref{eq:corr-delay-freq}), with a $R^2=0.92$, indicating a strong link between the delay and the periodic behavior of the HSV.
  }
  \label{fig:MOD-delay}
\end{figure}

The fact that the delay is at the origin of the oscillation process if further ensured by the strong correlation between measured delays $\Delta t$ and oscillation frequencies $f$:
\begin{equation}
  f = \frac{0.154}{\Delta t}
  \label{eq:corr-delay-freq}
\end{equation}
with $R^2 = 0.92$ on the dimensional correlation (see on figure~\ref{fig:MOD-delay}).
Unfortunately, no correlation could be obtained between the delay $\Delta t$ and the dimensionless parameters $\Rey_h$, $h/\delta$ and $W/h$.
The delay is expected to result from vortex-vortex interactions, and so to depend on vortex properties (position, radius, circulation), themselves depending on the boundary layer shape, which depends on the dimensionless parameters.
This dependency chain explains the difficulty to build direct correlations between the delay $\Delta t$ and the dimensionless parameters.

The following section aims at checking if taking this delay into account in the model is sufficient to retrieve a self-sustainable periodic behavior  for the selected oscillating regime flow \textit{i.e.}\ if a small perturbation applied on the main vortex position while at its equilibrium position can lead to a stabilized oscillation amplitude.
It is to be noted that the delay does not have to be constant but, as a result of the interaction between the main and the secondary vortices, should depend on the distance between the main and the secondary vortex and on their circulations.

This delay is added in the model by considering that $V_2$ velocity equals $V_1$ velocity, with a measured constant average delay $\Delta t$:
\begin{equation}
  v_{2, adv}(t) = v_{adv}\left(t - \Delta t\right).
\end{equation}
As a lot of hypotheses have been made, an adjustment variable is necessary for the model to be able to reproduce the observed HSV behavior.
The addition of an empirical factor of $3$ on the delay $\Delta t$ fairly close the system:
\begin{equation}
  v_{2, adv}(t) = v_{adv}\left(t - 3 \Delta t\right).
\end{equation}

\subsection{Numerical simulations results}
\label{sec:numer-simul-results}
\begin{figure}
  \centering
  \includegraphics[width=1\figwidth]{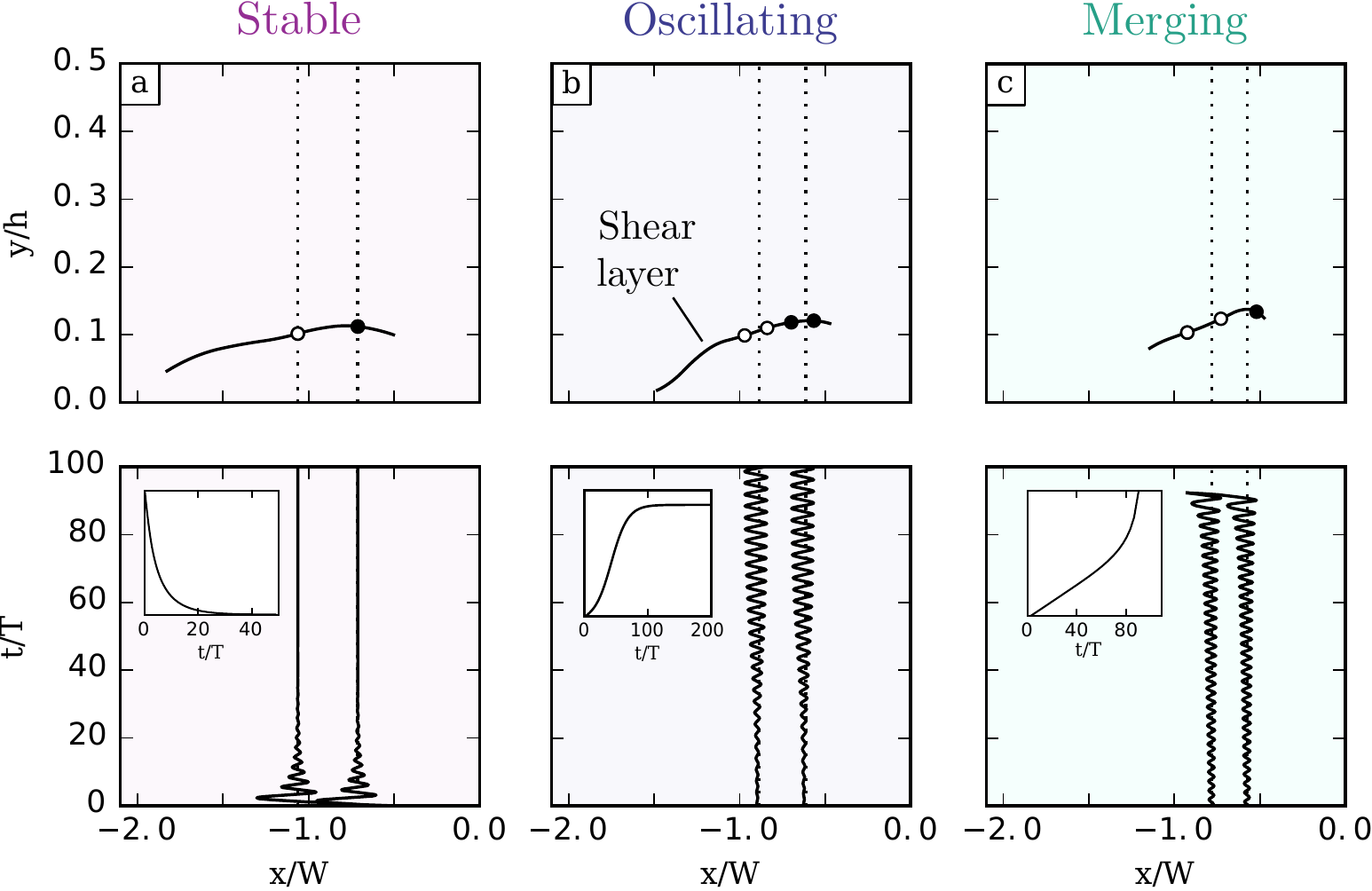}
  \caption{%
    Numerical simulation results for the model presented in section~\ref{sec:vort-dynam-repr}, for:
    (a) a stable regime configuration ($\Rey_h=4271$, $h/\delta=2.70$, $W/h=0.79$),
    (b) an oscillating regime configuration ($\Rey_h=4271$, $h/\delta=2.70$, $W/h=1.23$) and
    (c) a merging regime configuration ($\Rey_h=4271$, $h/\delta=2.70$, $W/h=1.37$).
    Upper figures show the shear layer measured shapes (black lines), the average measured vortex positions (dashed lines) and the computed extreme positions for the main vortex (black circles) and the secondary vortex (white circles).
    Bottom figures show the vortices streamwise trajectories with time (black lines) and the average measured vortex positions (dashed lines, hidden by the black lines for the left case).
    Vignettes show the time evolution of the main vortex oscillation amplitude.
    Those results are in good agreement with the observations, indicating that the mechanisms taken into account in the model are sufficient to reproduce the HSV dynamics.
  }
  \label{fig:MOD-Recap}
\end{figure}

Figure~\ref{fig:MOD-Recap} shows the results of simulations using the model presented in section~\ref{sec:vort-dynam-repr} for a stable, an oscillating and a merging regime configurations.
Firstly, regarding the simulation of the stable regime configuration, $V_1$ is initially introduced away from its equilibrium position ($x/W=-0.5$ instead of $x/W=-0.7$).
Vortices indeed appear to rapidly reach their equilibrium positions at $t/T\approx 30$ (with $T$ the vortex position oscillation period), after a transitional damped oscillation.
Then, regarding the oscillating regime configuration (figure~\ref{fig:MOD-Recap}b), a small perturbation is applied on the position of the main vortex ($x/W=-0.62$ instead of $x/W=-0.61$).
The oscillation amplitude increases with time and reaches a stable value after $t/T \approx 100$.
Finally, regarding the merging regime configuration, the same small perturbation is applied on the position of the main vortex.
The oscillation amplitude increases but does not reach a stable position.
The main and secondary vortices end up close to each other, in which case the simulation is stopped, as no vortex-vortex interaction model was implemented.

Despite the delay adjustment, the Strouhal number based on the delay $St_{\Delta t}$, shown to be nearly constant on experiments ($St_{\Delta t} = 0.154$), reaches a similar value of $St_{\Delta t}=0.146$ in the simulation.
This shows that the relation between the delay $\Delta t$ and the periodic behavior at frequency $f$ is well simulated by this model and that the proposed model is able to recreate the HSV dynamics for the stable, oscillating and merging regimes.

Conclusions can be made regarding the origin of the periodic motion of the HSV that:
(i) the secondary vortex position has a strong influence on the velocity of the main one, represented by the parameter $\Delta y$ in the correlation~\ref{eq:vortex_model_conv}.
This effect may be explained by the feeding of the main vortex from the main flow (quantity of fluid ending in the main vortex), which is reduced when the secondary vortex arises.
\citet{younis_topological_2014} linked, for stable HSV, this feeding and the vortex size, but a relation between the feeding and the vortex velocity has not been found yet.
(ii) The delay $\Delta t$ between the motion of the main and the secondary vortices is strongly linked to the periodic behavior of the HSV, and makes its complex dynamics possible.

\subsection{Note on the vortex breaking}
\label{sec:note-vortex-breaking}
The breaking regime particular dynamics can also be approached using this model.
For high delays $\Delta t$ and high $V_1$ advection velocities, the maximum distance $\Delta x$ between $V_1$ and $V_2$ is expected to increase.
If $\Delta x$ exceeds a certain value ($\approx R_1 + R_2$ according to the present observations), the main flow passes between the two vortices, cancelling the interaction between $V_1$ and $V_2$.
The general vortex motion then results in a competition between the wall influence $v_{wall}$ and the flow influence, reduced to a simple boundary layer (without effect of $V_2$ through $\Delta y$).
The resulting new equilibrium position is observed to be stable and located in the vicinity of the obstacle (where the flow influence is small enough to be balanced by the wall influence).
The breaking vortex rapidly reaches this position and disappears due to the strong stretching in the near obstacle zone increasing its diffusion.

\section{Horseshoe vortex properties}
\label{sec:hors-vort-prop}
This last section presents the evolution of the HSV geometrical characteristics with the dimensionless parameters of the flow and an in-depth comparison of these results with the well-documented immersed configuration.

\subsection[Separation distance]{Separation distance $\lambda$}
\label{sec:separation-position}
The separation distance $\lambda$ (see figure~\ref{fig:INTRO-Hsv-diagram}) is a crucial parameter for the HSV, as it governs the shear layer shape and the HSV streamwise dimension.
Using all PIV and trajectography measurements of the HSV, the following correlation was obtained:
\begin{equation}
  \frac{\lambda}{W} = 1.91
  \label{eq:lambda_models_fit1}
\end{equation}
with a $R^2 = 0.95$ computed on the dimensional correlation: $\lambda=1.91W$.
The boundary layer separation position greatly depends on the adverse pressure gradient~\citep{lighthill_boundary_1963}, which depends on the obstacle width, according to 2D potential flow computations, and explains this result.
It is interesting to see that the boundary layer separation position does not depend on $\delta$, the boundary layer thickness measured before placing the obstacle.

\begin{figure}
  \centering
  \includegraphics[width=.75\figwidth]{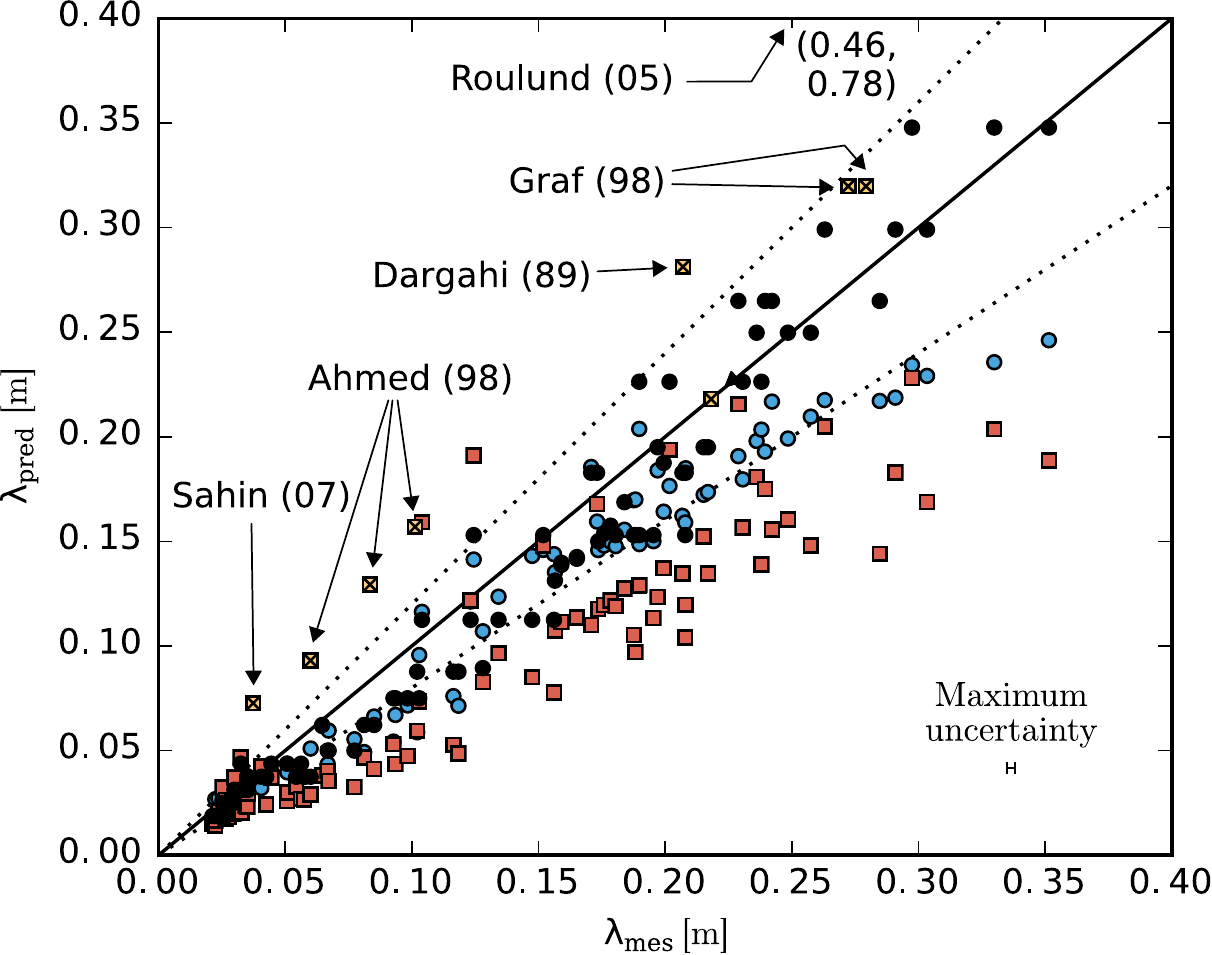}
  \caption{%
    Correlations quality for separation distance ($\lambda$) prediction.
    black filled circle symbols stands for equation~\ref{eq:lambda_models_fit1} applied on present data, with dotted lines for $20\%$ confidence interval,
    hollow (red) squares the correlation of~\citet{belik_secondary_1973} (equation~\ref{eq:lambda_models_belik}) applied on the present data and
    hollow (blue) circles represent the correlation of~\citet{baker_position_1985} (equation~\ref{eq:lambda_models_baker}) applied on the present data.
    Crossed-squares (yellow) represent equation~\ref{eq:lambda_models_fit1} applied on data from literature on emerging obstacles with turbulent boundary layer.
    Displayed maximum uncertainty (caption) only takes into account measurement uncertainties on PIV and trajectographies (see section~\ref{sec:hors-prop-meas}).
  }
  \label{fig:TYPO-Lambda-light}
\end{figure}

\citet{belik_secondary_1973} and \citet{baker_position_1985} proposed two correlations for the separation distance $\lambda$ for immersed cylindrical obstacles:
\begin{equation}
  \frac{\lambda_{Belik}}{D} = 0.5 + 35.5Re_{D}^{-0.424}
  \label{eq:lambda_models_belik}
\end{equation}
\begin{equation}
  \frac{\lambda_{Baker}}{D} = 0.5 + 0.338Re_{\delta^*}^{0.48}  \left(\frac{\delta^*}{D}\right)^{0.48} \tanh\left( \frac{3h}{D}\right)
  \label{eq:lambda_models_baker}
\end{equation}
with $D$ the obstacle diameter.
Application of these three correlations (equations \ref{eq:lambda_models_fit1}, \ref{eq:lambda_models_belik} and \ref{eq:lambda_models_baker}) are compared in figure~\ref{fig:TYPO-Lambda-light} for present data and data from the literature.
Both literature correlations for immersed obstacles underestimate the separation distance $\lambda$, revealing that this distance is greater for \emph{emerging} obstacles than for \emph{immersed} ones.
This difference can be explained by the fact that flow cannot pass above emerging obstacles, resulting in a higher pressure gradient than for immersed obstacles, and then, a precocious boundary layer separation.
\citet{sadeque_flow_2008} already noticed that the obstacle emergence increases the separation distance.
Application of equation~\ref{eq:lambda_models_fit1} on data from literature with emerging obstacles and turbulent boundary layers (figure \ref{fig:TYPO-Lambda-light}) overestimates the separation distance $\lambda$.
This can be explained by the fact that turbulent boundary layers are known to separate for higher pressure gradient than laminar ones.

The apparent simplicity of correlation~\ref{eq:lambda_models_fit1} compared to equations~\ref{eq:lambda_models_baker} and~\ref{eq:lambda_models_belik} is discussed here:
(i)~The boundary layer thickness $\delta$ is an important parameter for immersed obstacles, as it governs the position of the stagnation point on the face of the obstacle, and so, the down-flow and the adverse pressure gradient.
For emerging obstacles, the down-flow is stronger~\citep{sadeque_flow_2008} and the stagnation point position always high on the obstacle face, resulting in a strong adverse pressure gradient that is not influenced by the boundary layer thickness $\delta$.
(ii)~Correlations from literature are made more complex by the addition of $0.5D$ due to the method from~\citet{ballio_survey_1998} (see section~\ref{sec:hors-prop-meas}), and, for equation \ref{eq:lambda_models_baker}, by the last term ($\tanh\left(3h/2D\right)$) added afterwards to take into account the obstacle height.

\subsection{Vortex position}
\label{sec:vortex-position}
The present section aims at proposing empirical correlations for the main vortex position along $x$ and $y$ \textit{i.e.}\ the mean vortex position for the stable and oscillating regime configurations and the position of the vortex maximum circulation for the merging, diffusing and breaking regimes configurations.
They read:
\begin{equation}
  \frac{x_{1}}{W} = 1.01
  \label{eq:vort_pos_models_x_fit}
\end{equation}
\begin{equation}
  \frac{y_{1}}{\delta} = 0.2784 + 0.0229 \left(\frac{W}{h}\right)^2
  \label{eq:vort_pos_models_y_fit}
\end{equation}
with $R^2$ of respectively $0.96$ and $0.93$ on the dimensional forms of the correlations.

The main vortex position corresponds to the downstream end of the shear layer where the down-flow forces the reattachment of the boundary layer.
Consequently, the main vortex position should be linked to the boundary layer thickness $\delta$ (governing the shear layer shape) and the obstacle width $W$ (governing the down-flow), which explains the correlations \ref{eq:vort_pos_models_x_fit} and \ref{eq:vort_pos_models_y_fit}.

\citet{baker_position_1985} and~\citet{lin_characteristics_2008} correlations for vortex position in the case of immersed cylindrical obstacles read:
\begin{equation}
  \frac{x_{1, Baker}}{D} = 0.5 + 0.013Re_{\delta^*}^{0.67} \tanh\left(\frac{3h}{D}\right)
  \label{eq:vort_pos_models_x_baker}
\end{equation}
\begin{equation}
  \frac{x_{1, Lin}}{\delta} = 0.5\frac{D}{\delta} + 0.518\left(\frac{h}{\delta}\right)^{0.87}\left(\frac{h}{D}\right)^{-0.34}
  \label{eq:vort_pos_models_x_lin}
\end{equation}
\begin{equation}
  \frac{y_{1, Lin}}{\delta} = 0.207\left(\frac{h}{\delta}\right)^{0.6}\left(\frac{h}{D}\right)^{-0.77}
  \label{eq:vort_pos_models_y_lin}
\end{equation}

\begin{figure}
  \centering
  \includegraphics[width=.75\figwidth]{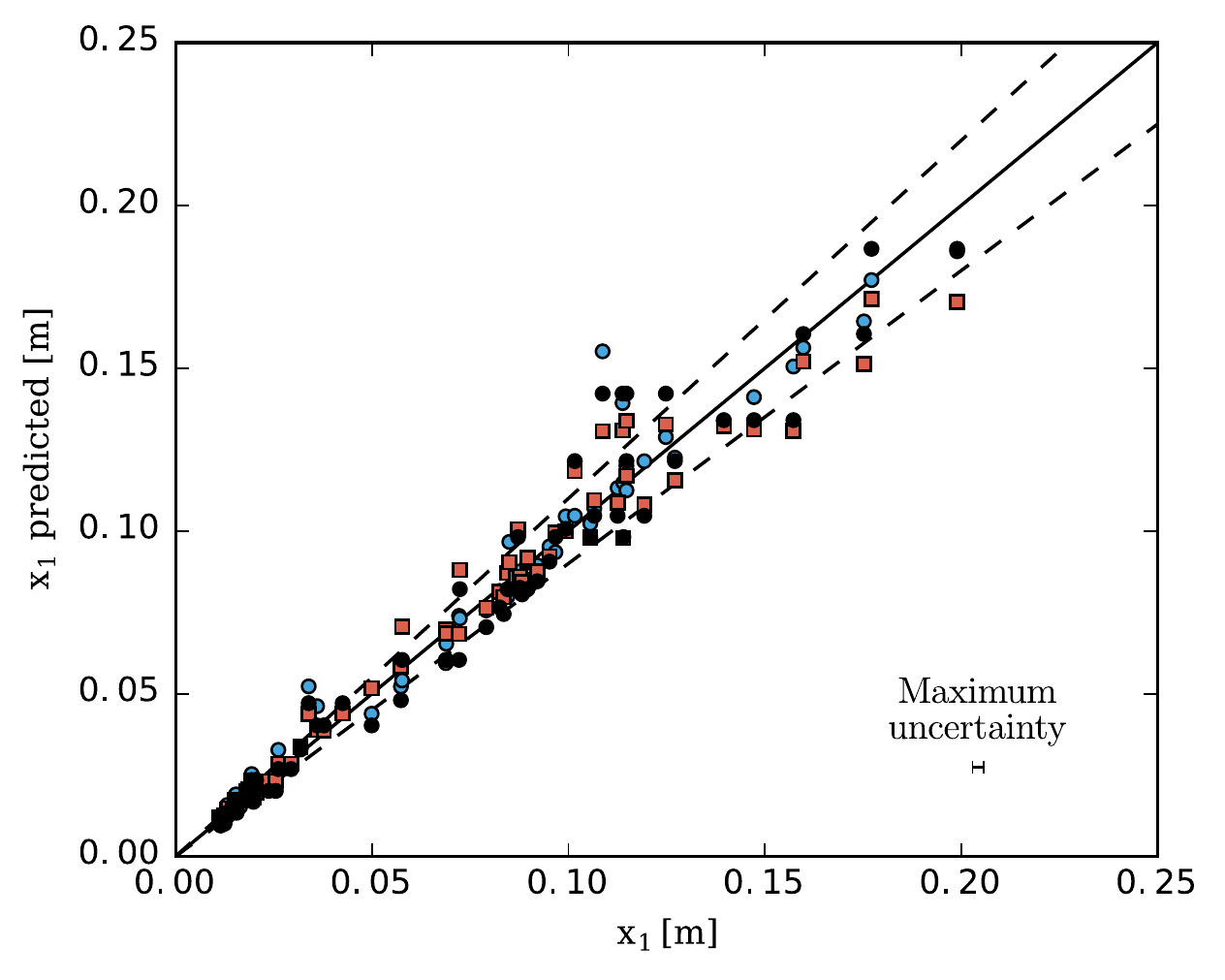}
  \caption{%
    Application of correlations for streamwise location of the main vortex ($x_1$):
    black filled circles: equation~\ref{eq:vort_pos_models_x_fit} with dashed line for 10\% confidence interval,
    hollow (blue) circles: equation~\ref{eq:vort_pos_models_x_baker} from~\citet{baker_position_1985} and
    hollow (red) squares: equation~\ref{eq:vort_pos_models_x_lin} from~\citet{lin_characteristics_2008}.
    Displayed maximum uncertainty (caption) only takes into account measurement uncertainties on PIV and trajectographies (see section~\ref{sec:hors-prop-meas}).
  }
  \label{fig:TYPO-Vort-pos-x}
\end{figure}

\begin{figure}
  \centering
  \includegraphics[width=.75\figwidth]{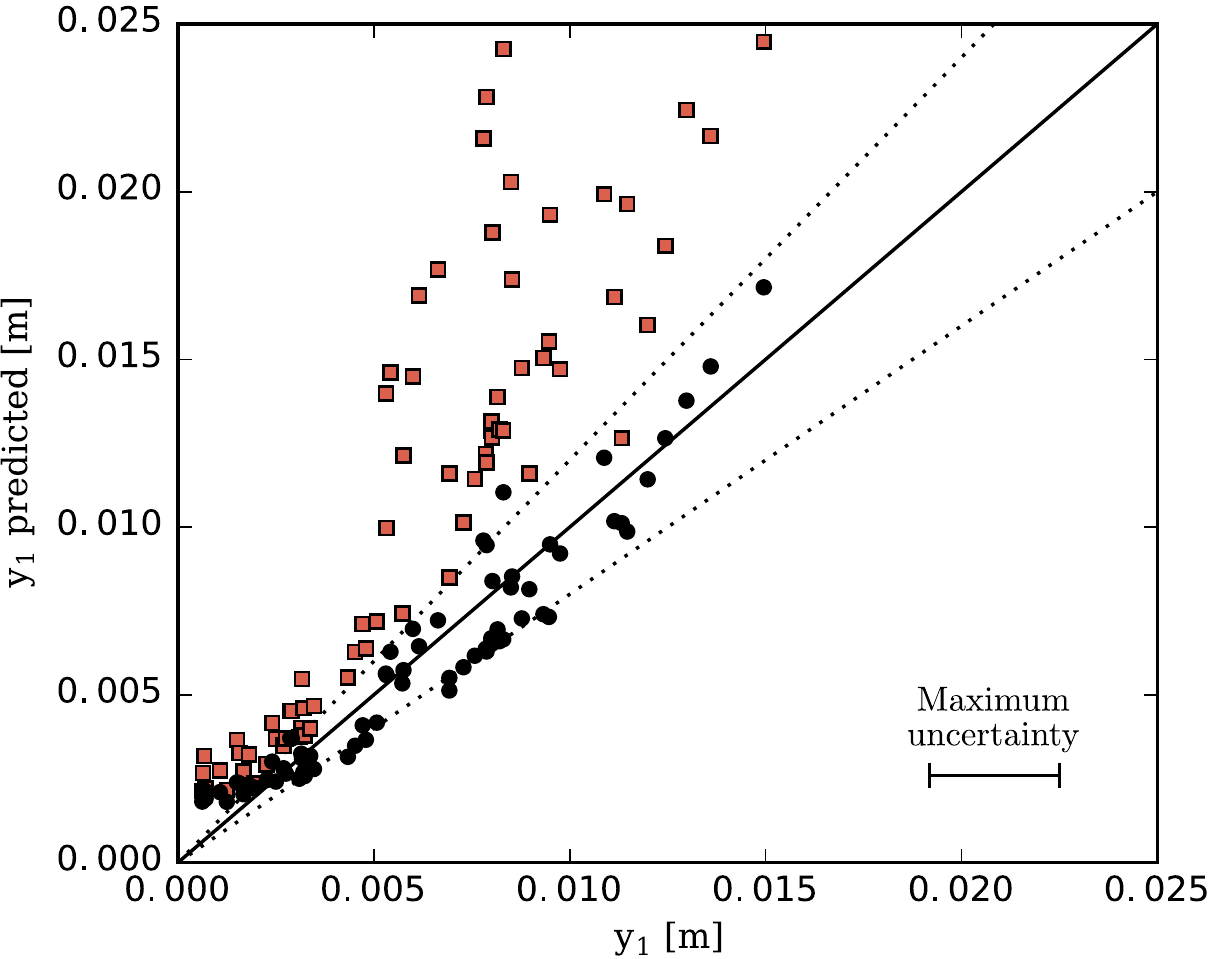}
  \caption{%
    Application of correlations for the vertical location of the vortex ($y_{1}$):
    black filled circles: equation~\ref{eq:vort_pos_models_y_fit} with dashed line for 20\% confidence interval,
    hollow (red) squares: equation~\ref{eq:vort_pos_models_y_lin} from~\citet{lin_characteristics_2008}.
    Displayed maximum uncertainty only takes into account measurement uncertainties on PIV or trajectographies (see section~\ref{sec:hors-prop-meas}).
  }
  \label{fig:TYPO-Vort-pos-y}
\end{figure}

Application of these correlations on the present data is plotted in figure~\ref{fig:TYPO-Vort-pos-x} and~\ref{fig:TYPO-Vort-pos-y}.
Equations~\ref{eq:vort_pos_models_x_baker} and~\ref{eq:vort_pos_models_x_lin} fairly fit the present data, showing that, contrary to the separation distance $\lambda$, the streamwise position of the main vortex $x_1$ does not differ for immersed and emerging obstacles
(and that it is not governed by the adverse pressure gradient).
Moreover, figure~\ref{fig:TYPO-Vort-pos-x} gives confidence to the equivalent diameter method described in section~\ref{sec:comp-betw-vari} and used herein.

Results for $y_1$ in figure~\ref{fig:TYPO-Vort-pos-y} show more dispersion, due to the higher relative uncertainty in the measurement of the vortex vertical location.
Correlations from the literature overestimate $y_{1}$.
This difference may be linked to the free-surface, that confines the shear layer vertically.

\subsection{Frequency}
\label{sec:frequency}
The frequency $f$ associated with the HSV vortex motion (for oscillating, diffusing and complex regimes) is of great importance to understand the HSV dynamics.
No correlation could be found for Strouhal numbers generally used in the literature ($St_\delta$ and $St_W$) nor for the Strouhal number based on the separation distance ($St_\lambda$) or on the distance between the main and the secondary vortices ( and $St_{\Delta x}$).
The frequency $f$ was however found to be mainly dependent on the bulk velocity $u_D$ as:
\begin{equation}
  f = 211u_D^{2.33}
  \label{eq:frequency} \\
\end{equation}
with $R^2=0.96$.
The main conclusion of this correlation is that the obstacle width $W$ (see also figure~\ref{fig:TRANS-O-M}c) and the boundary layer thickness $\delta$ have no influence on the HSV frequency.

\citet{thomas_unsteady_1987} measured the oscillation frequency for immersed obstacles in a wind tunnel with increasing obstacle width and observed a linear correlation (for Reynolds $\Rey_W=u_DW/\nu$ ranging from $2320$ to $10800$):
\begin{equation}
  St_W = 2.47\times 10^{-5} \Rey_W
  \label{eq:frequency_thomas_1}
\end{equation}
\begin{equation}
  \frac{fW}{u_D} = 2.47\times 10^{-5} \frac{u_D W}{\nu}
  \label{eq:frequency_thomas_1.5}
\end{equation}
that is:
\begin{equation}
  f = 1.8 u_D^2
  \label{eq:frequency_thomas_2}
\end{equation}
which confirms that the obstacle transverse dimension $W$ has no influence on the oscillating frequency.

As seen in section~\ref{sec:vort-dynam-repr}, a correlation can still be found with the average delay $\Delta t$ between the motion of $V_1$ and $V_2$ (equation~\ref{eq:corr-delay-freq}).

\subsection{vortices number}
\label{sec:vortices-number}
At a given time, the number of vortices is detected using critical points on PIV measurements, and using particle trajectories on trajectography measurements.
The average number of vortices (not including counter-rotating vortices $V_{ci}$) varies from $1$ to $3.5$ on the observed coherent regimes configurations.

The evolution of the vortices number should be strongly correlated to the shear layer shape.
Indeed, vortices follow each other along the shear layer at a distance from each other roughly equal to the sum of the vortices radii (which is correlated to their elevation $y$).
In such case, the vortices number should increase with the streamwise extension of the shear layer (equal to $\lambda - x_1$) and decrease with the shear layer elevation (because of the subsequent decrease of the vortices radius).

\begin{figure}
  \centering
  \includegraphics[width=.75\figwidth]{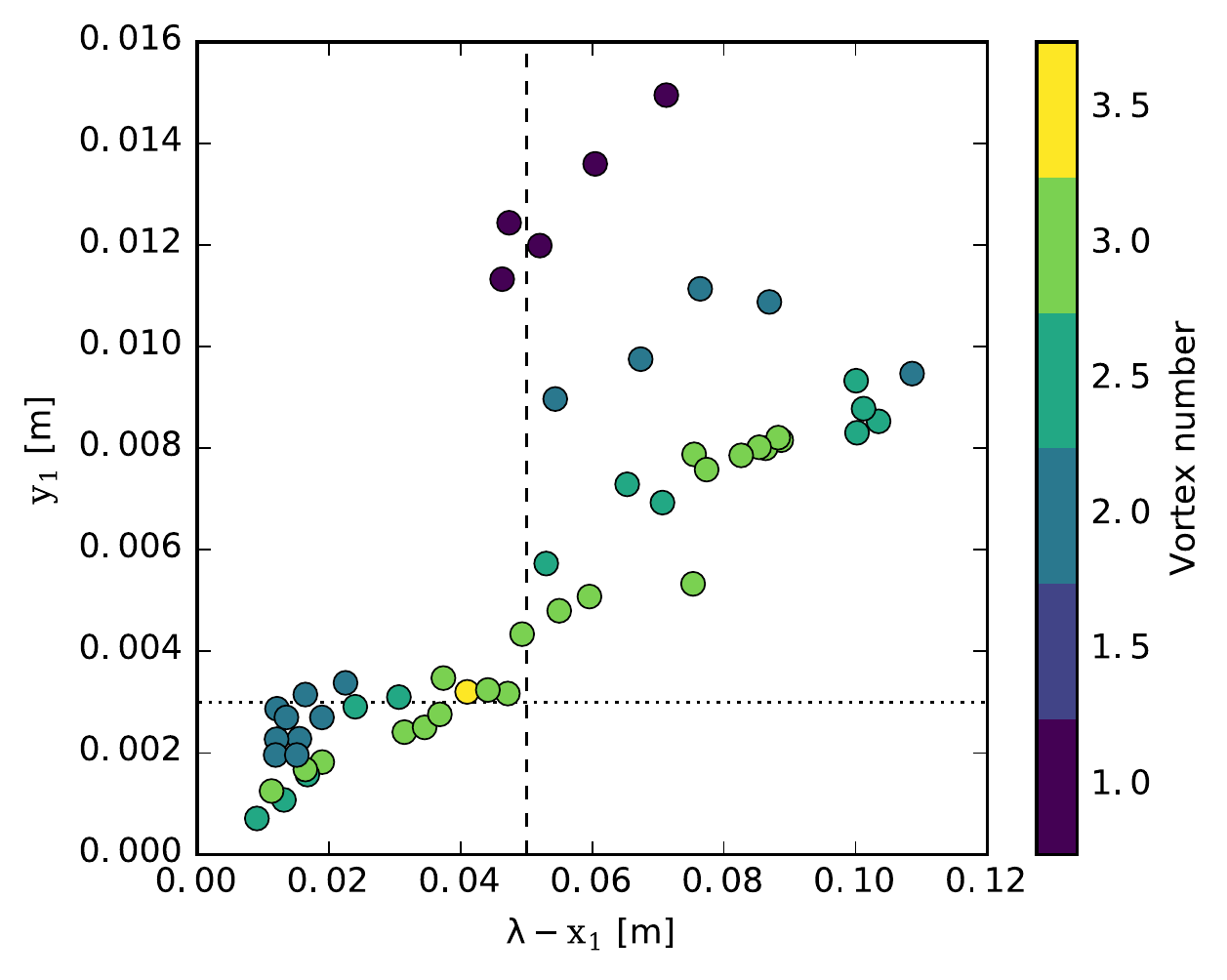}
  \caption{%
    Average vortices number evolution with the main vortex altitude $y_{1}$ and the shear layer streamwise extension: $\lambda - x_{1}$ for the coherent regime configurations.
    Non-integer values for the vortex number are associated to HSV with non-constant number of vortex.
    The visible continuous evolution (despite the non-homogeneous mapping), indicates that the vortex number is, as expected, governed by the shear-layer shape.
  }
  \label{fig:TYPO-vortex-number}
\end{figure}

Figure~\ref{fig:TYPO-vortex-number}, showing the evolution of the measured vortex number with these parameters, confirms this statement:
(i)~for a constant $\lambda - x_1$ (dashed line at $\lambda - x_1 = 0.05$ for instance), the vortices number decreases as $y_1$ increases and
(ii)~for a constant $y_1$ (dotted line at $y_1=0.003$ for instance), the vortices number increases as $\lambda - x_1$ increases.

\begin{figure}
  \centering
  \includegraphics[width=.75\figwidth]{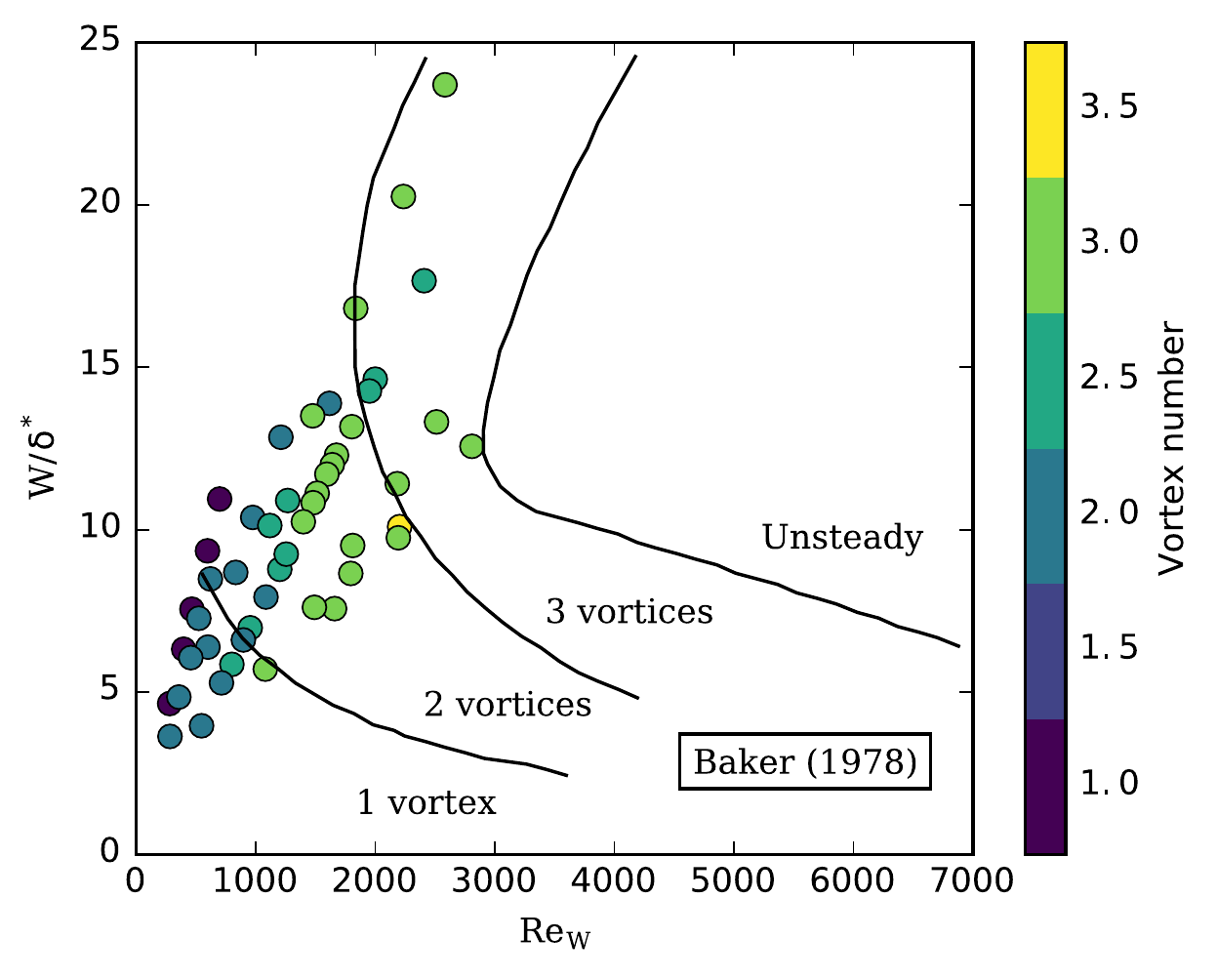}
  \caption{%
    Comparison of the HSV number of vortices with the observations of~\citet{baker_laminar_1978}.
    Circles represent data from the present study and are coloured according to the observed vortices number.
    $\delta^*$ is the boundary layer momentum thickness.
    Black lines represent the domain of constant vortices number observed by~\citet{baker_laminar_1978} (1 vortex, 2 vortices, 3 vortices and unsteady system).
    The bad agreement illustrates the differences between the immersed and emerging configurations in terms of dynamics.
  }
  \label{fig:TYPO-Vortex-number-baker}
\end{figure}

Comparison of the present data with the observations of~\citet{baker_laminar_1978} for immersed obstacles is presented in figure~\ref{fig:TYPO-Vortex-number-baker} and reveals that the vortices number (and so the shear layer shape evolution) differs greatly for the two emerging and immersed obstacles configurations.


\subsection{Conclusion on the impact of the emergence}
\label{sec:concl-impact-emerg}
The main difference between emerging and immersed obstacle configurations is the impossibility for the flow to pass over the obstacle in the latter case.
This has three main consequences:
(i)~the adverse pressure gradient is stronger in the case of an emerging obstacle, leading to a precocious boundary layer separation (longer $\lambda$, see figure~\ref{fig:TYPO-Lambda-light}).
(ii)~For an emerging obstacle, the whole flow facing the obstacle is deflected by the obstacle while for an immersed obstacle, the position of the flow stagnation point elevation along the upstream obstacle face (dependent on the boundary layer thickness) governs the quantity of fluid to be deflected.
This lead to a bigger dependency to the boundary layer thickness $\delta$ in the case of immersed obstacles.
(iii)~The difference of shear layer shape leads to different vortex number.
Nevertheless, the two configurations (immersed and emerging) share similarities in terms of~:
(i)~main vortex distance to the obstacle,
(ii)~observed frequency dependency to the bulk velocity $u_D$, and
(iii)~HSV vortices dynamics regimes.

\section{Conclusion}
\label{sec:conclusion}
Trajectographies and PIV measurements were performed to investigate the HSV developing at the toe of a rectangular obstacle emerging from a laminar free-surface flow.
In this context, vortex tracking methods based on critical points detection was shown to be valuable tools to extract and summarize the HSV vortices motion patterns.

The HSV dynamics was categorized using a typology adapted from the existing ones for immersed obstacle, by adding the newly observed complex regime, separating the coherent and the irregular evolutions (leading to a two-dimensional typology) and giving clear definitions of the typology regimes.
The observations of the complex regime and the investigation of the transitions between the different regimes showed the importance of considering the HSV as a dynamical system built on quasi-similar phases and undergoing a chaotic transition.
The coherent regime evolution was shown to depend on the three dimensionless parameters ($\Rey_h$, $h/\delta$ and $W/h$), in a continuous way from stable to complex regimes.
The irregular evolution is more complex, as it is linked to local vortex instabilities, but showed a main dependence to the Reynolds number $\Rey_h$.

A model for the main and secondary vortices motion was proposed and succeeded in reproducing the dynamics of the stable, oscillating and merging regimes.
This result allowed to identify two mechanisms of importance for the HSV dynamics.
Firstly the secondary vortex position has an influence on the main vortex velocity, certainly by modifying the quantity of fluid feeding it.
And secondly, the phase shift existing between the main and secondary vortices motion allows the apparition of a periodic motion in the present model.

Besides, the extraction of the HSV geometrical parameters allowed to highlight their correlations with the dimensionless parameters of the flow in the case of an emerging obstacle, showing that:
(i)~the separation distance $\lambda$ is mainly linked to the adverse pressure gradient,
(ii)~the main vortex position depends on the shear layer shape and on the downflow strength,
(iii)~the number of vortices composing the HSV is governed by the shear layer shape, and
(iv)~the effect of the free-surface confinement on the HSV is mainly indirect, through the limitation of the boundary layer thickness $\delta$.

Finally, the comparison with existing results for immersed obstacles indicated that the emerging configuration exhibits a higher adverse pressure gradient, due to a more important blocking effect, and a stronger downflow, due to the impossibility for the flow to bypass the obstacle by the top.
This implies a higher separation distance $\lambda$, a lower main vortex altitude $y_1$, and consequently, a higher number of vortices in the HSV in the emerging obstacle configuration.
The modification of these basic HSV properties is supposed to result in strong modifications of the HSV dynamics, making the comparison of the typology evolution for immersed and emerging configurations challenging.

In light of these results, some questions arise and should be the subject of future works investigation on this topic.
Firstly, the impact of the obstacle elongation ($L/W$), kept very high in this study, on the HSV should be investigated.
This parameter modifies the adverse pressure gradient, and, for sufficiently low $L/W$, the wake should be able to influence the HSV dynamics.
Secondly, the model presented in section \ref{sec:mech-behind-hors} should be extended to reproduce the other coherent regimes (merging, diffusing, breaking and complex).
This will only be possible by including additional ingredients in the model, such as the vortices circulation variation during a phase or the vortex merging, which are both challenging to handle.
Thirdly, little is known on the effect of a free-surface confinement on a HSV taking birth from a turbulent boundary layer.
Notably, the bi-modal behavior (previously reported for non-confined flows) should be influenced by the vertical confinement, as the zero-flow mode will be strongly modified.
Fourthly, experiments or numerical simulations on flow around obstacles with rounded or streamlined upstream faces would allow to confirm the validity of the equivalent diameter method for more complex obstacle shapes.
Finally, in a more application-oriented point of view, the effect of the HSV regime on the thermal exchanges, obstacle drag coefficient, wall shear stress and the downstream boundary layer properties could be investigated.

\bibliographystyle{jfm}

\begin{thebibliography}{59}
\expandafter\ifx\csname natexlab\endcsname\relax\def\natexlab#1{#1}\fi
\def\au#1{#1} \def\ed#1{#1} \def\yr#1{#1}\def\at#1{#1}\def\jt#1{\textit{#1}}
  \def\bt#1{#1}\def\bvol#1{\textbf{#1}} \def\vol#1{#1} \def\pg#1{#1}
  \def\publ#1{#1}\def\arxiv#1{#1}\def\org#1{#1}\def\st#1{\textit{#1}}

\bibitem[Adrian \& Westerweel(2011)]{adrian_particle_2011}
{\sc \au{Adrian, Ronald~J.} \& \au{Westerweel, Jerry}} \yr{2011} {\em Particle
  {{Image Velocimetry}}\/}.  \publ{{Cambridge University Press}}.

\bibitem[Agui \& Andreopoulos(1992)]{agui_experimental_1992}
{\sc \au{Agui, J.~H.} \& \au{Andreopoulos, J.}} \yr{1992}  \at{Experimental
  {{Investigation}} of a {{Three}}-{{Dimensional Boundary Layer Flow}} in the
  {{Vicinity}} of an {{Upright Wall Mounted Cylinder}} ({{Data Bank
  Contribution}})}.  \jt{Journal of Fluids Engineering}  \bvol{114}~(4),
  \pg{566--576}, 00000.

\bibitem[Baker(1978)]{baker_laminar_1978}
{\sc \au{Baker, C.~J.}} \yr{1978}  \at{The laminar horseshoe vortex}.
  \jt{Journal of Fluid Mechanics}  \bvol{95}~(02),  \pg{347}.

\bibitem[Baker(1979)]{baker_vortex_1979}
{\sc \au{Baker, C.~J.}} \yr{1979}  \at{Vortex flow around the bases of
  obstacles}. Thesis, University of Cambridge.

\bibitem[Baker(1980)]{baker_turbulent_1980}
{\sc \au{Baker, C.~J.}} \yr{1980}  \at{The turbulent horseshoe vortex}.
  \jt{Journal of Wind Engineering and Industrial Aerodynamics}
  \bvol{6}~(1\textendash{}2),  \pg{9--23}, 00142.

\bibitem[Baker(1985)]{baker_position_1985}
{\sc \au{Baker, C.~J.}} \yr{1985}  \at{The position of points of maximum and
  minimum shear stress upstream of cylinders mounted normal to flat plates}.
  \jt{Journal of Wind Engineering and Industrial Aerodynamics}  \bvol{18}~(3),
  \pg{263--274}, 00020.

\bibitem[Baker(1991)]{baker_oscillation_1991}
{\sc \au{Baker, C.~J.}} \yr{1991}  \at{The oscillation of horseshoe vortex
  systems}.  \jt{ASME Transactions Journal of Fluids Engineering}  \bvol{113},
  \pg{489--495}, 00048.

\bibitem[Ballio {\em et~al.\/}(1998)Ballio, Bettoni \&
  Franzetti]{ballio_survey_1998}
{\sc \au{Ballio, F.}, \au{Bettoni, C.} \& \au{Franzetti, S.}} \yr{1998}  \at{A
  {{Survey}} of {{Time}}-{{Averaged Characteristics}} of {{Laminar}} and
  {{Turbulent Horseshoe Vortices}}}.  \jt{Journal of Fluids Engineering}
  \bvol{120}~(2),  \pg{233}, 00044.

\bibitem[Belik(1973)]{belik_secondary_1973}
{\sc \au{Belik, L}} \yr{1973}  \at{The secondary flow about circular cylinders
  mounted normal to a flat plate}.  \jt{Aeronautical Quarterly}  \bvol{24},
  \pg{47--54}.

\bibitem[Bhattacharya {\em et~al.\/}(2006)Bhattacharya, Manoharan, Govindarajan
  \& Narasimha]{bhattacharya_critical_2006}
{\sc \au{Bhattacharya, Pinaki}, \au{Manoharan, M.~P.}, \au{Govindarajan, Rama}
  \& \au{Narasimha, R.}} \yr{2006}  \at{The critical {{Reynolds}} number of a
  laminar mixing layer}.  \jt{arXiv:physics/0604009} ,  \arxiv{arXiv:
  physics/0604009}.

\bibitem[Dargahi(1989)]{dargahi_turbulent_1989}
{\sc \au{Dargahi, B.}} \yr{1989}  \at{The turbulent flow field around a
  circular cylinder}.  \jt{Experiments in Fluids}  \bvol{8}~(1-2),  \pg{1--12},
  00107.

\bibitem[Depardon {\em et~al.\/}(2007)Depardon, Lasserre, Brizzi \&
  Bor{\'e}e]{depardon_automated_2007}
{\sc \au{Depardon, S.}, \au{Lasserre, J.~J.}, \au{Brizzi, L.~E.} \&
  \au{Bor{\'e}e, J.}} \yr{2007}  \at{Automated topology classification method
  for instantaneous velocity fields}.  \jt{Experiments in Fluids}
  \bvol{42}~(5),  \pg{697--710}, 00018.

\bibitem[Devenport \& Simpson(1990)]{devenport_time-depeiident_1990}
{\sc \au{Devenport, William~J.} \& \au{Simpson, Roger~L.}} \yr{1990}
  \at{Time-depeiident and time-averaged turbulence structure near the nose of a
  wing-body junction}.  \jt{Journal of Fluid Mechanics}  \bvol{210},
  \pg{23--55}.

\bibitem[Dimotakis(2000)]{dimotakis_mixing_2000}
{\sc \au{Dimotakis, Paul~E.}} \yr{2000}  \at{The mixing transition in turbulent
  flows}.  \jt{Journal of Fluid Mechanics}  \bvol{409},  \pg{69--98}.

\bibitem[Doligalski {\em et~al.\/}(1994)Doligalski, Smith \&
  Walker]{doligalski_vortex_1994}
{\sc \au{Doligalski, T.~L.}, \au{Smith, C.~R.} \& \au{Walker, J. D.~A.}}
  \yr{1994}  \at{Vortex {{Interactions}} with {{Walls}}}.  \jt{Annual Review of
  Fluid Mechanics}  \bvol{26}~(1),  \pg{573--616}.

\bibitem[Dritschel \& Waugh(1992)]{dritschel_quantification_1992}
{\sc \au{Dritschel, D.~G.} \& \au{Waugh, D.~W.}} \yr{1992}  \at{Quantification
  of the inelastic interaction of unequal vortices in two-dimensional vortex
  dynamics}.  \jt{Physics of Fluids A: Fluid Dynamics}  \bvol{4}~(8),
  \pg{1737}.

\bibitem[Eckerle \& Awad(1991)]{eckerle_effect_1991}
{\sc \au{Eckerle, W.~A.} \& \au{Awad, J.~K.}} \yr{1991}  \at{Effect of
  {{Freestream Velocity}} on the {{Three}}-{{Dimensional Separated Flow
  Region}} in {{Front}} of a {{Cylinder}}}.  \jt{Journal of Fluids Engineering}
   \bvol{113}~(1),  \pg{37--44}.

\bibitem[Eckerle \& Langston(1987)]{eckerle_horseshoe_1987}
{\sc \au{Eckerle, W.~A.} \& \au{Langston, L.~S.}} \yr{1987}  \at{Horseshoe
  {{Vortex Formation Around}} a {{Cylinder}}}.  \jt{Journal of Turbomachinery}
  \bvol{109}~(2),  \pg{278}, 00067.

\bibitem[Effenberger \& Weiskopf(2010)]{effenberger_finding_2010}
{\sc \au{Effenberger, F.} \& \au{Weiskopf, D.}} \yr{2010}  \at{Finding and
  classifying critical points of {{2D}} vector fields: A cell-oriented approach
  using group theory}.  \jt{Computing and Visualization in Science}
  \bvol{13}~(8),  \pg{377--396}.

\bibitem[Escauriaza \& Sotiropoulos(2011)]{escauriaza_reynolds_2011}
{\sc \au{Escauriaza, C.} \& \au{Sotiropoulos, F.}} \yr{2011}  \at{Reynolds
  {{Number Effects}} on the {{Coherent Dynamics}} of the {{Turbulent Horseshoe
  Vortex System}}}.  \jt{Flow, Turbulence and Combustion}  \bvol{86}~(2),
  \pg{231--262}, 00006.

\bibitem[Euler \& Herget(2012)]{euler_controls_2012}
{\sc \au{Euler, Thomas} \& \au{Herget, J{\"u}rgen}} \yr{2012}  \at{Controls on
  local scour and deposition induced by obstacles in fluvial environments}.
  \jt{CATENA}  \bvol{91},  \pg{35--46}.

\bibitem[Floryan(1986)]{floryan_gortler_1986}
{\sc \au{Floryan, J.~M.}} \yr{1986}  \at{G{\"o}rtler instability of boundary
  layers over concave and convex walls}.  \jt{Physics of Fluids}
  \bvol{29}~(8),  \pg{2380}.

\bibitem[Graf \& Yulistiyanto(1998)]{graf_experiments_1998}
{\sc \au{Graf, W.H.} \& \au{Yulistiyanto, B.}} \yr{1998}  \at{Experiments on
  flow around a cylinder; the velocity and vorticity fields}.  \jt{Journal of
  Hydraulic Research}  \bvol{36}~(4),  \pg{637--654}, 00063.

\bibitem[Graftieaux {\em et~al.\/}(2001)Graftieaux, Michard \&
  Grosjean]{graftieaux_combining_2001}
{\sc \au{Graftieaux, L.}, \au{Michard, M.} \& \au{Grosjean, N.}} \yr{2001}
  \at{Combining {{PIV}}, {{POD}} and vortex identification algorithms for the
  study of unsteady turbulent swirling flows}.  \jt{Measurement Science and
  Technology}  \bvol{12}~(9),  \pg{1422}, 00198.

\bibitem[Greco(1990)]{greco_flow_1990}
{\sc \au{Greco, J.J.}} \yr{1990}  \at{The {{Flow Structure}} in the
  {{Vicinity}} of a {{Cylinder}}-{{Flat Plate Junction}}: {{Flow Regimes}},
  {{Periodicity}}, and {{Vortex Interac}}- tions}. PhD thesis, Department of
  Mechanical Engineering and Mechanics, Lehigh University, Lehigh.

\bibitem[Hunt {\em et~al.\/}(1978)Hunt, Abell, Peterka \&
  Woo]{hunt_kinematical_1978}
{\sc \au{Hunt, J. C.~R.}, \au{Abell, C.J.}, \au{Peterka, J.~A.} \& \au{Woo,
  H.}} \yr{1978}  \at{Kinematical studies of the flows around free or
  surface-mounted obstacles; applying topology to flow visualization}.
  \jt{Journal of Fluid Mechanics}  \bvol{86}~(01),  \pg{179}, 00454.

\bibitem[Jeong \& Hussain(1995)]{jeong_identification_1995}
{\sc \au{Jeong, J.} \& \au{Hussain, F.}} \yr{1995}  \at{On the identification
  of a vortex}.  \jt{Journal of Fluid Mechanics}  \bvol{285},  \pg{69--94}.

\bibitem[Johnson \& Ting(2003)]{johnson_measurements_2003}
{\sc \au{Johnson, K.} \& \au{Ting, F.}} \yr{2003}  \at{Measurements of {{Water
  Surface Profile}} and {{Velocity Field}} at a {{Circular Pier}}}.
  \jt{Journal of Engineering Mechanics}  \bvol{129}~(5),  \pg{502--513}, 00006.

\bibitem[Josserand \& Rossi(2007)]{josserand_merging_2007}
{\sc \au{Josserand, C.} \& \au{Rossi, M.}} \yr{2007}  \at{The merging of two
  co-rotating vortices: A numerical study}.  \jt{European Journal of Mechanics
  - B/Fluids}  \bvol{26}~(6),  \pg{779--794}.

\bibitem[Kelso \& Smits(1995)]{kelso_horseshoe_1995}
{\sc \au{Kelso, R.~M.} \& \au{Smits, A.~J.}} \yr{1995}  \at{Horseshoe vortex
  systems resulting from the interaction between a laminar boundary layer and a
  transverse jet}.  \jt{Physics of Fluids}  \bvol{7}~(1),  \pg{153}.

\bibitem[Kerswell(2002)]{kerswell_elliptical_2002}
{\sc \au{Kerswell, Richard~R.}} \yr{2002}  \at{Elliptical instability}.
  \jt{Annual review of fluid mechanics}  \bvol{34}~(1),  \pg{83--113}.

\bibitem[Khan \& Ahmed(2005)]{khan_topological_2005}
{\sc \au{Khan, M.~J.} \& \au{Ahmed, A.}} \yr{2005}  \at{Topological model of
  flow regimes in the plane of symmetry of a surface-mounted obstacle}.
  \jt{Physics of Fluids}  \bvol{17}~(4),  \pg{045101}.

\bibitem[Khan {\em et~al.\/}(1995)Khan, Ahmed \& Trosper]{khan_dynamics_1995}
{\sc \au{Khan, M.~J.}, \au{Ahmed, A.} \& \au{Trosper, J.~R.}} \yr{1995}
  \at{Dynamics of the juncture vortex}.  \jt{AIAA Journal}  \bvol{33}~(7),
  \pg{1273--1278}.

\bibitem[Kol{\'a}{\v r}(2007)]{kolar_vortex_2007}
{\sc \au{Kol{\'a}{\v r}, V.}} \yr{2007}  \at{Vortex identification: {{New}}
  requirements and limitations}.  \jt{International Journal of Heat and Fluid
  Flow}  \bvol{28}~(4),  \pg{638--652}.

\bibitem[Larousse {\em et~al.\/}(1993)Larousse, Martinuzzi \&
  Tropea]{larousse_flow_1993}
{\sc \au{Larousse, A.}, \au{Martinuzzi, R.} \& \au{Tropea, C.}} \yr{1993}
  \at{Flow {{Around Surface}}-{{Mounted}}, {{Three}}-{{Dimensional
  Obstacles}}}.  \bt{In {\em Turbulent {{Shear Flows}} 8\/} (ed. \ed{Franz
  Durst, Rainer Friedrich, Brian~E. Launder, Frank~W. Schmidt, Ulrich Schumann
  \& James~H. Whitelaw})},  \pg{pp. 127--139}.  \publ{{Springer Berlin
  Heidelberg}}.

\bibitem[Lighthill(1963)]{lighthill_boundary_1963}
{\sc \au{Lighthill, M.~J.}} \yr{1963} {\em Boundary Layer Theory\/}.
  \publ{{Oxford University Press London}}.

\bibitem[Lin {\em et~al.\/}(2002)Lin, Chiu \& Shieh]{lin_characteristics_2002}
{\sc \au{Lin, C.}, \au{Chiu, P.-H.} \& \au{Shieh, S.-J.}} \yr{2002}
  \at{Characteristics of horseshoe vortex system near a vertical
  plate\textendash{}base plate juncture}.  \jt{Experimental Thermal and Fluid
  Science}  \bvol{27}~(1),  \pg{25--46}, 00030.

\bibitem[Lin {\em et~al.\/}(2008)Lin, Ho \& Dey]{lin_characteristics_2008}
{\sc \au{Lin, C.}, \au{Ho, T.~C.} \& \au{Dey, S.}} \yr{2008}
  \at{Characteristics of {{Steady Horseshoe Vortex System}} near {{Junction}}
  of {{Square Cylinder}} and {{Base Plate}}}.  \jt{Journal of Engineering
  Mechanics}  \bvol{134}~(2),  \pg{184--197}, 00006.

\bibitem[Lin {\em et~al.\/}(2003)Lin, Lai \& Chang]{lin_simultaneous_2003}
{\sc \au{Lin, C.}, \au{Lai, W.} \& \au{Chang, K.}} \yr{2003}  \at{Simultaneous
  {{Particle Image Velocimetry}} and {{Laser Doppler Velocimetry Measurements}}
  of {{Periodical Oscillatory Horseshoe Vortex System}} near {{Square
  Cylinder}}-{{Base Plate Juncture}}}.  \jt{Journal of Engineering Mechanics}
  \bvol{129}~(10),  \pg{1173--1188}, 00025.

\bibitem[Loucks \& Wallace(2012)]{loucks_velocity_2012}
{\sc \au{Loucks, R.~B.} \& \au{Wallace, J.~M.}} \yr{2012}  \at{Velocity and
  velocity gradient based properties of a turbulent plane mixing layer}.
  \jt{Journal of Fluid Mechanics}  \bvol{699},  \pg{280--319}.

\bibitem[Meunier {\em et~al.\/}(2005)Meunier, Le~Diz{\`e}s \&
  Leweke]{meunier_physics_2005}
{\sc \au{Meunier, P.}, \au{Le~Diz{\`e}s, S.} \& \au{Leweke, T.}} \yr{2005}
  \at{Physics of vortex merging}.  \jt{Comptes Rendus Physique}
  \bvol{6}~(4-5),  \pg{431--450}.

\bibitem[Mignot {\em et~al.\/}(2016)Mignot, Cai, Launay, Riviere \&
  Escauriaza]{mignot_coherent_2016}
{\sc \au{Mignot, E.}, \au{Cai, W.}, \au{Launay, G.}, \au{Riviere, N.} \&
  \au{Escauriaza, C.}} \yr{2016}  \at{Coherent turbulent structures at the
  mixing-interface of a square open-channel lateral cavity}.  \jt{Physics of
  Fluids}  \bvol{28}~(4),  \pg{045--104}.

\bibitem[Ozturk {\em et~al.\/}(2008)Ozturk, Akkoca \& Sahin]{ozturk_flow_2008}
{\sc \au{Ozturk, N.~A.}, \au{Akkoca, A.} \& \au{Sahin, B.}} \yr{2008}  \at{Flow
  details of a circular cylinder mounted on a flat plate}.  \jt{Journal of
  Hydraulic Research}  \bvol{46}~(3),  \pg{344--355}.

\bibitem[Paik {\em et~al.\/}(2007)Paik, Escauriaza \&
  Sotiropoulos]{paik_bimodal_2007}
{\sc \au{Paik, J.}, \au{Escauriaza, C.} \& \au{Sotiropoulos, F.}} \yr{2007}
  \at{On the bimodal dynamics of the turbulent horseshoe vortex system in a
  wing-body junction}.  \jt{Physics of Fluids (1994-present)}  \bvol{19}~(4),
  \pg{045107}.

\bibitem[Peltier {\em et~al.\/}(2014)Peltier, Erpicum, Archambeau, Pirotton \&
  Dewals]{peltier_meandering_2014}
{\sc \au{Peltier, Y.}, \au{Erpicum, S.}, \au{Archambeau, P.}, \au{Pirotton, M.}
  \& \au{Dewals, B.}} \yr{2014}  \at{Meandering jets in shallow rectangular
  reservoirs: {{POD}} analysis and identification of coherent structures}.
  \jt{Experiments in Fluids}  \bvol{55}~(6).

\bibitem[Riviere {\em et~al.\/}(2011)Riviere, La{\"\i}ly, Mignot \&
  Doppler]{riviere_supercritical_2011}
{\sc \au{Riviere, N.}, \au{La{\"\i}ly, A.-G.}, \au{Mignot, E.} \& \au{Doppler,
  D.}} \yr{2011}  \at{Supercritical {{Flowaround}} and {{Beneath}} a {{Fixed
  Obstacle}}} .

\bibitem[Roulund {\em et~al.\/}(2005)Roulund, Sumer, Fredsoe \&
  Michelsen]{roulund_numerical_2005}
{\sc \au{Roulund, A.}, \au{Sumer, B.~M.}, \au{Fredsoe, J.} \& \au{Michelsen,
  J.}} \yr{2005}  \at{Numerical and experimental investigation of flow and
  scour around a circular pile}.  \jt{Journal of Fluid Mechanics}  \bvol{534},
  \pg{351--401}, 00147.

\bibitem[Sabatino \& Smith(2008)]{sabatino_boundary_2008}
{\sc \au{Sabatino, D.~R.} \& \au{Smith, C.~R.}} \yr{2008}  \at{Boundary {{Layer
  Influence}} on the {{Unsteady Horseshoe Vortex Flow}} and {{Surface Heat
  Transfer}}}.  \jt{Journal of Turbomachinery}  \bvol{131}~(1),
  \pg{011015--011015}, 00000.

\bibitem[Sadeque {\em et~al.\/}(2008)Sadeque, Rajaratnam \&
  Loewen]{sadeque_flow_2008}
{\sc \au{Sadeque, M.}, \au{Rajaratnam, N.} \& \au{Loewen, M.}} \yr{2008}
  \at{Flow around {{Cylinders}} in {{Open Channels}}}.  \jt{Journal of
  Engineering Mechanics}  \bvol{134}~(1),  \pg{60--71}, 00018.

\bibitem[Schwind(1962)]{schwind_three_1962}
{\sc \au{Schwind, R.~G.}} \yr{1962} {\em The Three Dimensional Boundary Layer
  near a Strut\/}.  \publ{{Massachusetts Institute of Technology}}.

\bibitem[Seal {\em et~al.\/}(1995)Seal, Smith, Akin \&
  Rockwell]{seal_quantitative_1995}
{\sc \au{Seal, C.V.}, \au{Smith, C.R.}, \au{Akin, O.} \& \au{Rockwell, D.}}
  \yr{1995}  \at{Quantitative characteristics of a laminar, unsteady necklace
  vortex system at a rectangular block-flat plate juncture}.  \jt{Journal of
  Fluid Mechanics}  \bvol{286},  \pg{117--135}.

\bibitem[Seal {\em et~al.\/}(1997)Seal, Smith \& Rockwell]{seal_dynamics_1997}
{\sc \au{Seal, C.~V.}, \au{Smith, C.~R.} \& \au{Rockwell, D.}} \yr{1997}
  \at{Dynamics of the {{Vorticity Distribution}} in {{Endwall Junctions}}}.
  \jt{AIAA Journal}  \bvol{35}~(6),  \pg{1041--1047}.

\bibitem[Shavit {\em et~al.\/}(2006)Shavit, Lowe \&
  Steinbuck]{shavit_intensity_2006}
{\sc \au{Shavit, U.}, \au{Lowe, R.~J.} \& \au{Steinbuck, J.~V.}} \yr{2006}
  \at{Intensity {{Capping}}: A simple method to improve cross-correlation
  {{PIV}} results}.  \jt{Experiments in Fluids}  \bvol{42}~(2),  \pg{225--240}.

\bibitem[Simpson(2001)]{simpson_junction_2001}
{\sc \au{Simpson, R.~L.}} \yr{2001}  \at{Junction {{Flows}}}.  \jt{Annual
  Review of Fluid Mechanics}  \bvol{33}~(1),  \pg{415--443}.

\bibitem[Thomas(1987)]{thomas_unsteady_1987}
{\sc \au{Thomas, A. S.~W.}} \yr{1987}  \at{The unsteady characteristics of
  laminar juncture flow}.  \jt{Physics of Fluids (1958-1988)}  \bvol{30}~(2),
  \pg{283--285}.

\bibitem[Trieling {\em et~al.\/}(1998)Trieling, Linssen \&
  Van~Heijst]{trieling_monopolar_1998}
{\sc \au{Trieling, R.~R.}, \au{Linssen, A.~H.} \& \au{Van~Heijst, G. J.~F.}}
  \yr{1998}  \at{Monopolar vortices in an irrotational annular shear flow}.
  \jt{Journal of Fluid Mechanics}  \bvol{360},  \pg{273--294}.

\bibitem[Tropea {\em et~al.\/}(2007)Tropea, Yarin \&
  Foss]{tropea_springer_2007}
{\sc \au{Tropea, C.}, \au{Yarin, A.~L.} \& \au{Foss, J.~F.}}, ed. \yr{2007}
  {\em Springer {{Handbook}} of {{Experimental Fluid Mechanics}}\/}, 2007th
  edn.  \publ{New York, NY: {Springer}}.

\bibitem[Wygnanski \& Fiedler(1970)]{wygnanski_two-dimensional_1970}
{\sc \au{Wygnanski, I.} \& \au{Fiedler, H.~E.}} \yr{1970}  \at{The
  two-dimensional mixing region}.  \jt{Journal of Fluid Mechanics}
  \bvol{41}~(02),  \pg{327--361}.

\bibitem[Younis {\em et~al.\/}(2014)Younis, Zhang, Hu \&
  Mehmood]{younis_topological_2014}
{\sc \au{Younis, M.~Y.}, \au{Zhang, H.}, \au{Hu, B.} \& \au{Mehmood, S.}}
  \yr{2014}  \at{Topological evolution of laminar juncture flows under
  different critical parameters}.  \jt{Science China Technological Sciences}
  \bvol{57}~(7),  \pg{1342--1351}.

\end{thebibliography}

\end{document}